\RequirePackage{fix-cm}
\documentclass[natbib]{svjour3}                    
\smartqed  
\usepackage[table]{xcolor} 
\usepackage{graphicx}
\usepackage{fixltx2e}
\usepackage{balance}
\usepackage{epstopdf}
\usepackage{wrapfig}
\usepackage[utf8]{inputenc}
\usepackage{clrscode}
\usepackage{listings}
\usepackage{paralist}
\usepackage{subfig}
\usepackage{tikz}
\usepackage{amsmath}
\newcommand{\distance}[4]{\overrightarrow{d}_{\hspace*{-.5ex}#1,#2}(#3,#4)}
\newcommand{\Frd}{Fr\'echet distance}
\def\checkmark{\tikz\fill[scale=0.4](0,.35) -- (.25,0) -- (1,.7) -- (.25,.15) -- cycle;}

\begin{document}

\title{A Comparison and Evaluation of Map Construction Algorithms Using Vehicle Tracking Data}

\author{Mahmuda Ahmed         \and
        Sophia Karagiorgou    \and
        Dieter Pfoser		  \and
        Carola Wenk
}

\institute{Mahmuda Ahmed \at
              University of Texas at San Antonio \\
              San Antonio, TX, USA\\
              \email{mahmudaahmed@gmail.com}                     
           \and
           Sophia Karagiorgou \at
              National Technical\\
              University of Athens, Greece \\
              \email{sokaragi@mail.ntua.gr}           
           \and
           Dieter Pfoser \at
              George Mason University \\
              Fairfax, VA, USA\\
              \email{dpfoser@gmu.edu}           
            \and
            Carola Wenk \at
              Tulane University \\
              New Orleans, LA, USA\\
              \email{cwenk@tulane.edu}           
}

\date{Received: date / Accepted: date}

\maketitle

\begin{abstract}
Map construction construction methods automatically produce and/or update street map datasets using vehicle tracking data. Enabled by the ubiquitous generation of geo-referenced tracking data, there has been a recent surge in map construction algorithms coming from different computer science domains. A cross-comparison of the various algorithms is still very rare, since (i) algorithms and constructed maps are generally not publicly available and (ii) there is no standard approach to assess the result quality, given the lack of benchmark data and quantitative evaluation methods.
This work represents a first comprehensive attempt to benchmark such map construction algorithms. We provide an evaluation and comparison of seven algorithms using four datasets and four different evaluation measures. 
%
In addition to this comprehensive comparison, we make our datasets, source code of map construction algorithms and evaluation measures publicly available on {\tt mapconstruction.org}. This site has been established as a repository for map construction data and algorithms and we invite other researchers to contribute by uploading code and benchmark data supporting their contributions to map construction algorithms. 




\keywords{tracking data \and map construction \and quality measures \and algorithms \and performance}
\end{abstract} 


\section{Introduction}
\label{sec:introduction}

\begin{figure}[htbp]
\begin{center}
	\subfloat[Vehicle tracking data -
Berlin.]{\label{fig:tracking_data}\includegraphics[width=0.47\columnwidth]{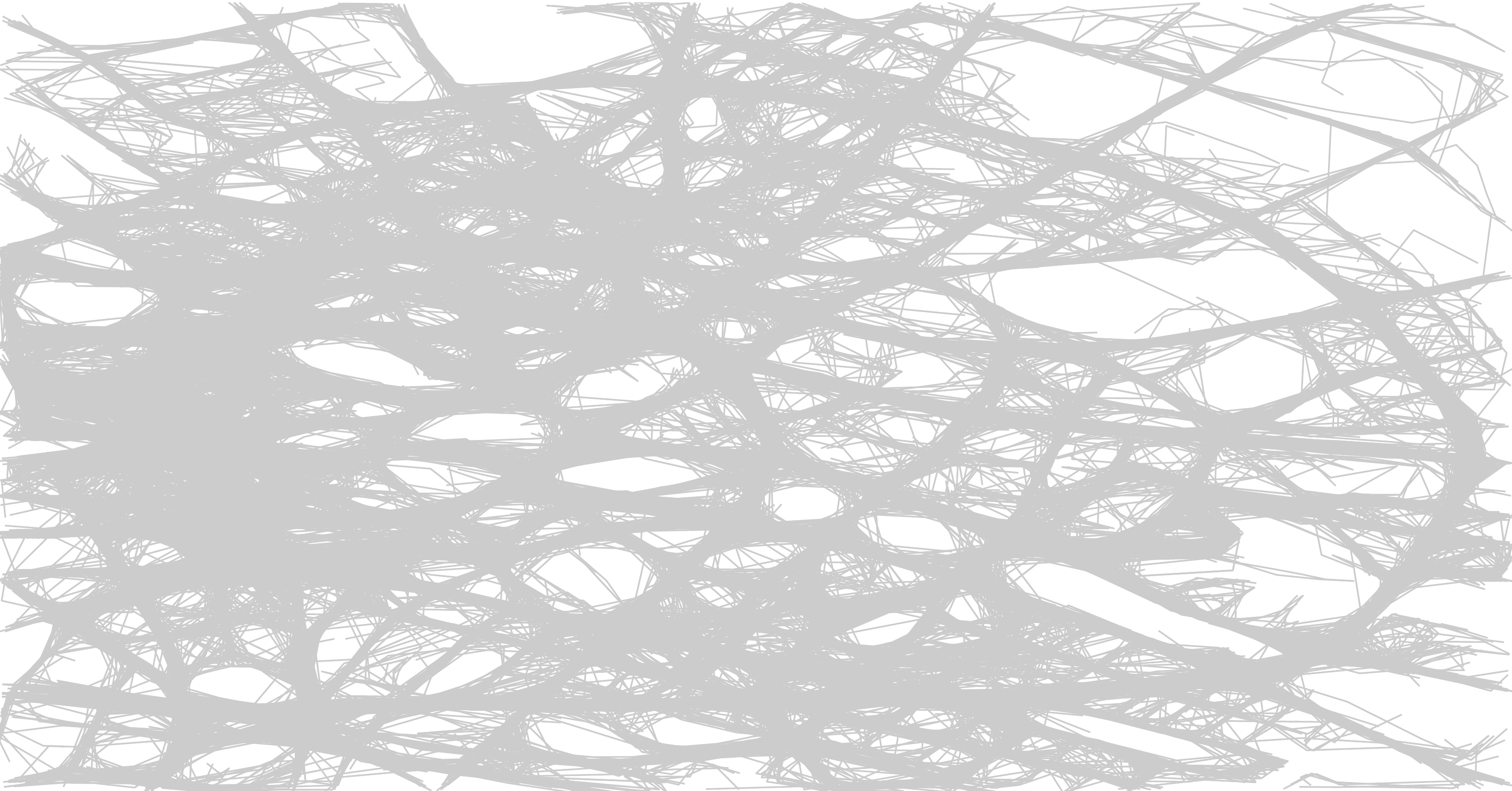}}
\hspace{5pt}
	\subfloat[Constructed map using~\cite{Karagiorgou:2012:VTD:2424321.2424334}(in black) overlayed on ground-truth (in grey).]{\includegraphics[width=0.47\columnwidth]{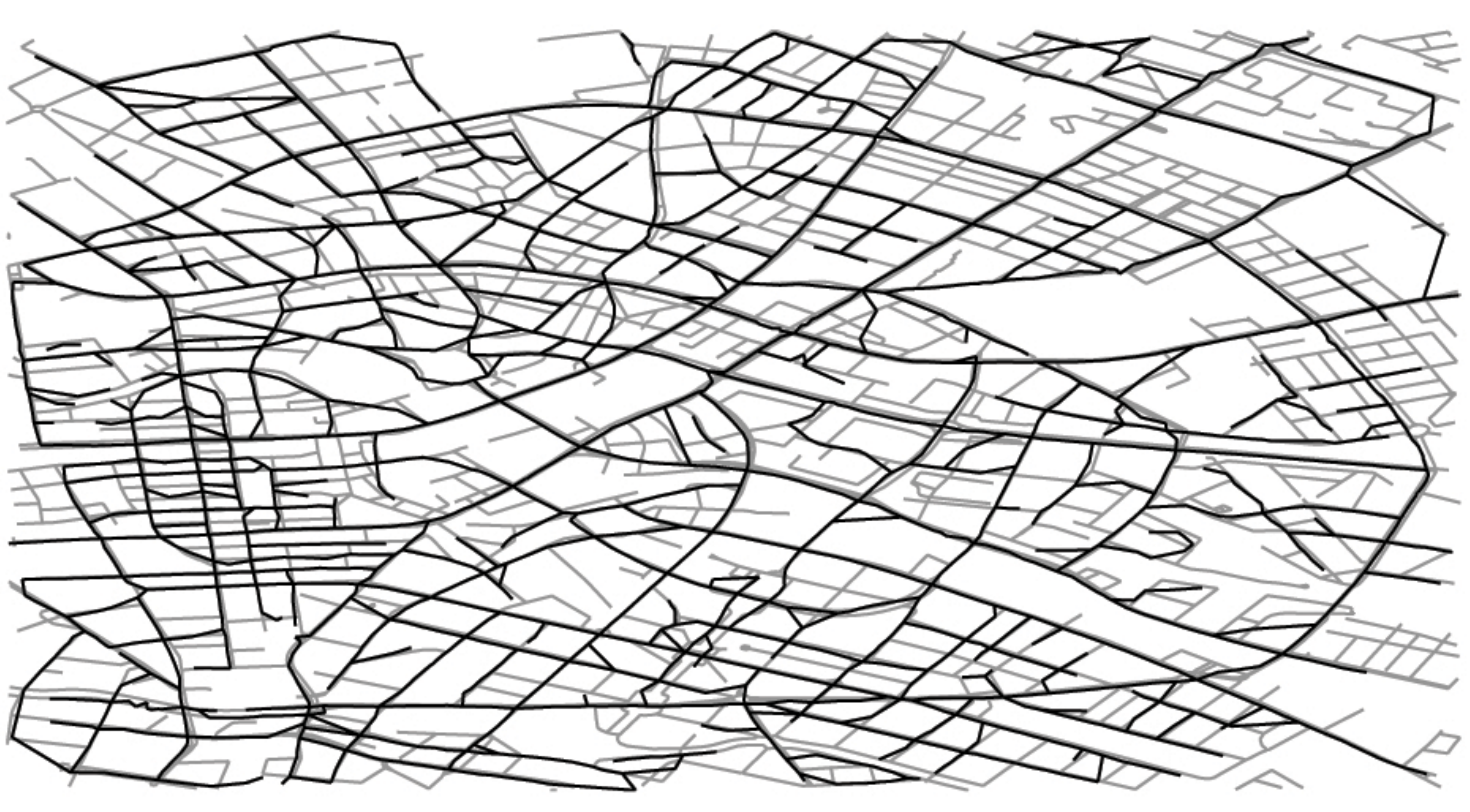}} 
\end{center}
\vspace{-10pt}
\caption{Vehicle tracking data vs constructed map overlayed on ground-truth.}
\label{fig:example}
\end{figure}

Street maps and transportation networks are of fundamental importance in a wealth of applications. In the past, the production of street maps required expensive field surveying and labor-intensive postprocessing. Proprietary data vendors such as Navteq (now Nokia), TeleAtlas (now TomTom) and Google therefore dominated the market. 
Over the last years, Volunteered Geographic Information (VGI) \cite{goodchild07} efforts such as OpenStreetMap (OSM) \cite{hw-osmugsm-08, osm:lk} have complemented commercial map datasets. They provide map coverage especially in areas which are of less commercial interest. VGI efforts however still require dedicated users to author maps using specialized software tools.  
Lately, on the other hand, the commoditization of GPS technology and integration in mobile phones coupled with the advent of low-cost fleet management and positioning software has triggered the generation of vast amounts of tracking data. As a size indicator one can consider the contribution of tracking data in OpenStreetMap, which is steadily increasing in size and currently amounts to 2.6 trillion points \cite{osmplanet}. 
Besides the use of such data in traffic assessment and forecasting \cite{efentakis13b}, i.e., map-matching vehicle trajectories to road networks to obtain travel times \cite{Brakatsoulas05onmap-matching}, there has been a recent surge of actual \emph{map construction algorithms} that derive not only travel time attributes but actual road network geometries from tracking data, e.g., \cite{Aanjaneya:2011,Agamennoni:2011:RIP:2218592.2218957,csm_esa2012,be-irmgp-12,Biagioni:2012:MIF:2424321.2424333,bruntrup:2005:imgg,Cao:2009:GTR:1653771.1653776,chen:2008:rdmg,chen:2010:rnr,Davies:2006:SDR:1175887.1176088,edelkamp:2003:rpmi,fathi:2010:dri,DBLP:conf/nips/GeSBW11,DBLP:conf/igarss/GuoIK07,jang:2010:mgs,Karagiorgou:2012:VTD:2424321.2424334,Liu:2012:MLS:2339530.2339637,schroedl:2004:mgtm,shi:2009:agm,sten:2011:mga,Worrall:2007:automatedprocess,DBLP:conf/gis/ZhangTS10}. Among those only a few algorithms give theoretical quality guarantees \cite{Aanjaneya:2011,csm_esa2012,chen:2010:rnr}.
An example of a constructed map is given in Figure~\ref{fig:example}, which shows (a) the vehicle trajectories collected for Berlin in grey color and (b) the respective constructed map, shown in black color, using the algorithm of \cite{Karagiorgou:2012:VTD:2424321.2424334} with an OpenStreetMap background map, shown in grey color.

A major challenge in the research community is to compare the performance and to evaluate the quality of the various map construction algorithms. 
Visual inspection remains the most common evaluation approach throughout the literature and only a few recent papers incorporate quantitative distance measures \cite{aw-SIGSPATIAL-13, be-irmgp-12, Biagioni:2012:MIF:2424321.2424333, Karagiorgou:2012:VTD:2424321.2424334, Liu:2012:MLS:2339530.2339637}. However, the cross-comparison of different algorithms remains rare, since algorithms and constructed maps are generally not publicly available. Also, there is a lack of benchmark data, and the quantitative evaluation with suitable distance measures is in its infancy.
A cultural shift has recently been triggered by Biagioni and Eriksson \cite{be-irmgp-12}: In addition to providing an extensive survey of eleven map construction algorithms, they have performed a quantitative evaluation of three representative map construction algorithms. And they have made their implementations of these algorithms, as well as their dataset, publicly available.
The present work {\em complements and significantly expands these benchmarking efforts to provide an evaluation and comparison of more map construction algorithms on more diverse datasets using various quality measures suitable for different applications.}
Such an effort can only be sustained in a culture of sharing that makes data, methods and source code publicly available.
%

In this work, we evaluate and compare seven map construction algorithms using four benchmark tracking datasets and four different distance measures. 
The \emph{algorithms} we compare represent the state-of-the-art over the past several years and constitute representatives of different map construction algorithm classes. The algorithms we evaluate include the recent algorithms by Ahmed and Wenk \cite{csm_esa2012}, by Ge et al. \cite{DBLP:conf/nips/GeSBW11}, and by Karagiorgou and Pfoser \cite{Karagiorgou:2012:VTD:2424321.2424334}, in addition to the algorithms by Cao and Krumm \cite{Cao:2009:GTR:1653771.1653776}, Davies et al. \cite{Davies:2006:SDR:1175887.1176088}, Edelkamp and Schr\"odl \cite{edelkamp:2003:rpmi}, and Biagioni and Eriksson \cite{Biagioni:2012:MIF:2424321.2424333}. 
Among those, the algorithms by \cite{Cao:2009:GTR:1653771.1653776}, \cite{Davies:2006:SDR:1175887.1176088} and \cite{edelkamp:2003:rpmi} were previously compared by Biagioni and Eriksson \cite{be-irmgp-12}. We have used their publicly available implementations of the algorithms by \cite{Cao:2009:GTR:1653771.1653776}, \cite{Davies:2006:SDR:1175887.1176088, edelkamp:2003:rpmi} and by \cite{Biagioni:2012:MIF:2424321.2424333}, and the authors of \cite{DBLP:conf/nips/GeSBW11} ran their algorithm for us. Our own implementations of the algorithms by \cite{csm_esa2012,Karagiorgou:2012:VTD:2424321.2424334} we have also made publicly available, see below.

The \emph{four distance measures used to assess the constructed map quality} comprise two novel distance measures that have not been used for comparative evaluations of map construction before and that work with unmodified and unbiased ground-truth maps: the Directed Hausdorff distance \cite{ag-dgsmi-99} and the path-based distance measure presented by Ahmed et al. \cite{aw-SIGSPATIAL-13}.
We also use a distance measure based on shortest paths by Karagiorgou and Pfoser \cite{Karagiorgou:2012:VTD:2424321.2424334} and the graph-sampling-based distance measure by Biagioni and Eriksson \cite{Biagioni:2012:MIF:2424321.2424333}. The implementation of the latter distance measure  \cite{Biagioni:2012:MIF:2424321.2424333} has been made available to us by the authors.

The \emph{tracking datasets} include the \emph{Chicago} dataset provided by Biagioni and Eriksson \cite{be-irmgp-12,Biagioni:2012:MIF:2424321.2424333}, and three additional tracking datasets: two from \emph{Athens}, Greece and one from \emph{Berlin}, Germany (see detail in Section \ref{sec:sec_data}). They are available together with unmodified ground-truth maps obtained from OpenStreetMap. We use different datasets because they cover diverse roads (i.e. highways, secondary roads), different sampling rates and different scale.

In addition to providing the largest comprehensive comparison of map construction algorithms, we make our three new benchmark datasets, the map construction algorithms and outputs by Ahmed and Wenk \cite{csm_esa2012} and by Karagiorgou and Pfoser \cite{Karagiorgou:2012:VTD:2424321.2424334}, as well as the metric code for computing the three distance measures: the Directed Hausdorff distance \cite{ag-dgsmi-99}, the path-based distance \cite{aw-SIGSPATIAL-13} and shortest path based measure~\cite{Karagiorgou:2012:VTD:2424321.2424334} publicly available on the internet at \mbox{\tt mapconstruction.org}. We have established this web site as a repository for map construction data and algorithms, and we invite other researchers to contribute by uploading code and benchmark data supporting their map construction algorithms. We expect that such a central repository will encourage a culture of sharing and will enable the development of improved map construction algorithms.

Our main goal with this work is to provide a common platform to do comparative analysis of map construction algorithms. As different distance measures capture different features of a constructed map, it is hard to combine them into a single score and rank the algorithms based on that. Also, which algorithm is the best highly depends on the quality of the input data and for what purpose the map will be used. For example, for the \emph{Chicago} dataset the KDE-based algorithm by Davies et al. \cite{Davies:2006:SDR:1175887.1176088} generates a very good-quality map in terms of spatial distance to the ground-truth map (captured using path-based and Directed Hausdorff distance), but if the user is interested in maps with good coverage (captured by shortest path based and graph-sampling based distance measure) this algorithm will not be the best choice as it ignores tracks in sparse areas as outliers/noise. So, we leave it to the user to pick the distance measure that suits his/her needs best.

The outline of the paper is as follows. 
Section~\ref{sec:sec_mca} surveys map construction algorithms by introducing categories for types of algorithms and gives more details on the algorithms that we will use in our evaluation. 
Section~\ref{sec:sec_qmes} discusses quality measures that will allow us to assess the quality of the constructed maps. The tracking datasets that we provide for evaluation purposes are briefly discussed in Section~\ref{sec:sec_data}. The datasets are available for download and also include the respective ground-truth map data. A comprehensive performance study comparing the various algorithms across datasets is given in Section~\ref{sec:sec_exp}. Finally, Section~\ref{sec:conclusions} provides conclusions and directions for future work.  


\section{Map Construction Algorithms}
\label{sec:sec_mca}

We assume that the input is given as a set of {\em tracks}, where each track is a sequence of {\em measurements}. Each measurement consists of a point (latitude/longitude or $\left(x,y\right)$-coordinates after suitable projection), a time stamp, and optionally additional information such as vehicle heading or speed. The desired output is to construct a {\em street map}. There are many possible models for street maps, mostly depending on the desired application and granularity. For example, an intersection can be modeled as a single vertex embedded as a point in the plane, or it could be a set of vertices, possibly annotated with turn restrictions, or it could be a region. An edge can be modeled as an abstract connection between vertices, as a curve embedded in the plane, as a set of curves to model multiple lanes, and an edge might be directed to model one-way streets. We will focus on the most basic model of a street map as an undirected geometric graph, where each vertex is embedded as a point in the plane and each edge is a polygonal curve that connects two vertices. All map construction algorithms in the literature follow this basic model, i.e. $\left(x,y\right)$-coordinates, time stamp.  In addition, some algorithms enhance this basic model by deriving some additional information (such as mean speed, directions, turn restrictions, number of lanes), which often is computed in an additional post-processing step, e.g.~\cite{Biagioni:2012:MIF:2424321.2424333, Cao:2009:GTR:1653771.1653776, Davies:2006:SDR:1175887.1176088, edelkamp:2003:rpmi, schroedl:2004:mgtm}.

\subsection{Related Work}
There exist several different approaches in the literature for constructing street maps from tracking data. These can be organized into the following categories: Point clustering (this includes $k$-means algorithms and Kernel Density Estimation (KDE) as described in Biagioni and Eriksson \cite{Biagioni:2012:MIF:2424321.2424333}), incremental track insertion, and intersection linking.

\subsubsection{Point Clustering} 
Algorithms in this general category assume the input consists of a set of points which are then clustered in various different ways to obtain street segments which finally connect to a street map. The input point set either comprises the set of all raw input measurements, or a dense sample of all input tracks. Here, the input tracks are assumed to be continuous curves obtained from interpolating (usually piecewise-linearly) between measurements. 

One type of approach, speerheaded by Edelkamp and Schr\"odl \cite{edelkamp:2003:rpmi}, employs the $k$-means algorithm to cluster the input point set, using distance measures (e.g., Euclidean distance) and possibly also vehicle heading of the measurement, as a condition to introduce seeds at fixed distances along a path. Their map construction algorithm incorporates new algorithms for road segmentation, map-matching, and lane clustering. In \cite{schroedl:2004:mgtm} this approach was used to refine an existing map rather than building it entirely from scratch. 
In their short paper \cite{DBLP:conf/igarss/GuoIK07}, Guo et al. make use of statistical analysis of GPS tracks, assuming that the GPS data follows a symmetric 2D Gaussian distribution. This assumption may become unrealistic, especially in error-prone environments. Worrall et al. \cite{Worrall:2007:automatedprocess} compute point clusters based on location and heading, and in a second step link these clusters together using non-linear least-squares fitting. They emphasize compression of the input tracks to infer a digitized road map and present their results only for small datasets. They are mostly concerned with topological elements and not with connected way points.
%
%
 Agamennoni et al. \cite{Agamennoni:2011:RIP:2218592.2218957} presented a 
machine-learning method to consistently build a representation of the map mostly in dynamic environments such as open-pit mines. They focus on estimating a set of principal curves from the input traces to represent the constructed map.
Liu et al. \cite{Liu:2012:MLS:2339530.2339637} first cluster line segments based on proximity and direction, and then use the resulting point clusters and fit polylines to them, to extract road segments. 
In our comparisons, we use the algorithm by Edelkamp and Schr\"odl \cite{edelkamp:2003:rpmi}.

Another approach employs KDE methods to first transform the input point set to a density-based discretized image. Most of the KDE algorithms function well either when the data is frequently sampled (i.e., once per second) \cite{chen:2008:rdmg}, or when there is a lot of data redundancy~\cite{Biagioni:2012:MIF:2424321.2424333, sten:2011:mga, shi:2009:agm, Davies:2006:SDR:1175887.1176088}. A similar approach to \cite{Biagioni:2012:MIF:2424321.2424333} is presented in Liu et al.\ \cite{Liu:2012:MLS:2339530.2339637}.
Generally, KDE algorithms have a hard time overcoming the problem of noisy samples when they accumulate in an area. Recently, Wang et al. \cite{2013:csu:wlw} addressed the problem of map updates by applying their approach to OpenStreetMap data using a KDE-based approach.
In our comparisons, we use the algorithms by Davies et al.\ \cite{Davies:2006:SDR:1175887.1176088} and by Biagioni and Eriksson \cite{Biagioni:2012:MIF:2424321.2424333}, which are both KDE-based but use very different approaches to extract the map from the kernel density estimate.
 
In the computational geometry community, map construction algorithms have been proposed that cluster the input points using local neighborhood properties by employing Voronoi diagrams, Delaunay triangulations~\cite{chen:2010:rnr,DBLP:conf/nips/GeSBW11}, or other neighborhood complexes such as the Vietoris-Rips complex \cite{Aanjaneya:2011}. All these algorithms assume a densely sampled input point set, and provide theoretical quality guarantees for the constructed output map, under certain assumptions on the underlying street map and the input tracks.
Aanjaneya et al. \cite{Aanjaneya:2011} view street maps as metric graphs, and they focus on computing the combinatorial structure by computing an almost isometric space with lower complexity, but they do not compute an explicit embedding of vertices and edges. 
Chen et al. \cite{chen:2010:rnr} focus on detecting ``good'' street portions in the map and connect them subsequently. The theoretical quality guarantees, however, assume dense point sample coverage and error bounds, and make assumptions on the road geometry.
In our comparisons, we use the algorithm by Ge and Wang \cite{DBLP:conf/nips/GeSBW11}. 

\subsubsection{Incremental Track Insertion}
Algorithms in this category construct a street map by incrementally inserting tracks into an initially empty map~\cite{Niehofer:2009:GCM:1590964.1591648}, often making use of map-matching ideas \cite{quddus:2007:cmma}. Distance measures and vehicle headings are also used to perform additions and deletions during the incremental construction of the map.
One of the first algorithms in this category \cite{rogers:1999:mgd} clusters the tracks merely to refine an existing map and not to compute it from scratch. Cao and Krumm \cite{Cao:2009:GTR:1653771.1653776} first introduce a clarification step in which they modify the input tracks by applying physical attraction to group similar input tracks together. Then they incrementally insert each track by using local criteria such as distance and direction. 
Bruntrup et al.~\cite{bruntrup:2005:imgg} propose a spatial-clustering based algorithm that requires high quality tracking data (sampling rate and positional accuracy).
The work in \cite{DBLP:conf/gis/ZhangTS10} discusses a map update algorithm based on spatial similarity. It uses a method similar to GPS trace merging to continuously refine existing road maps.
Ahmed and Wenk \cite{csm_esa2012} present an incremental method that employs the \Frd\ to partially match the tracks to the map. 
In our comparisons, we use the algorithms by Cao and Krumm \cite{Cao:2009:GTR:1653771.1653776} and by Ahmed and Wenk  \cite{csm_esa2012} which use very different approaches for incremental track insertion.

\subsubsection{Intersection Linking}
While related to point clustering, the intersection linking approach is to first detect the intersection vertices of the street map, and in a second step link those intersections together by identifying suitable street segments. 
Fathi and Krumm \cite{fathi:2010:dri} provide an approach that detects intersections by using a prototypical detector trained on ground truth data from an existing map. While a map is finally derived, their approach works best for well aligned maps and it uses frequently sampled data of 1s or 5s. The method by Karagiorgou and Pfoser \cite{Karagiorgou:2012:VTD:2424321.2424334} relies on detecting changes in the direction of movement to infer intersection nodes, and then ``bundling'' the trajectories around them to create the map edges.
In our comparisons, we use the algorithm by Karagiorgou and Pfoser \cite{Karagiorgou:2012:VTD:2424321.2424334}.

\subsection{Compared Algorithms}
\label{subsec:subsec_ca}
Here we give some more details on the map construction algorithms that we compare in Section \ref{sec:sec_exp}. 
The algorithms categories are also provided in Table~\ref{tab:tab_alg}.

\begin{table}[htbp]\scriptsize
\centering
\begin{tabular}{|c||c c c|}
\hline
		 &Point&Incremental Track&Intersection\\
Algorithm&Clustering&Insertion&Linking\\
\hline
\hline
Ahmed and Wenk \cite{csm_esa2012}&&\checkmark&\\
Biagioni and Eriksson \cite{Biagioni:2012:MIF:2424321.2424333}&\checkmark&&\\
Cao and Krumm \cite{Cao:2009:GTR:1653771.1653776}&&\checkmark&\\
Davies et al. \cite{Davies:2006:SDR:1175887.1176088}&\checkmark&&\\
Edelkamp and Schr\"odl \cite{edelkamp:2003:rpmi}&\checkmark&&\\
Ge et al. \cite{DBLP:conf/nips/GeSBW11}&\checkmark&&\\
Karagiorgou and Pfoser \cite{Karagiorgou:2012:VTD:2424321.2424334}&&&\checkmark\\
\hline
\end{tabular}
\caption{Algorithms categories.}
\label{tab:tab_alg}
\end{table}

\subsubsection{Ahmed and Wenk \protect\cite{csm_esa2012}}
\label{subsub:subsub_aw}
\begin{figure}[btph]
 \begin{center}
	 \subfloat[Existing graph and trajectory to be added]{\label{subfig:mahmuda_1} \includegraphics[width=0.3\columnwidth]{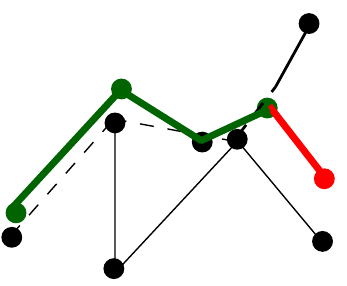}}	 
	 \subfloat[Merged graph]{\label{subfig:mahmuda_2} \includegraphics[width=0.30\columnwidth]{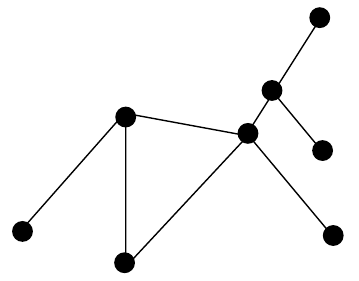}}		  
 \end{center}
\caption{Incremental track insertion algorithm (images from presentation of\cite{csm_esa2012})}
\label{fig:mahmuda}
\end{figure}

The algorithm by Ahmed and Wenk \cite{csm_esa2012} is a simple and practical incremental track insertion algorithm. The insertion of one track proceeds in three steps. The first step performs a partial map-matching of the track to the partially constructed map in order to identify matched portions and unmatched portions. Figure~\ref{subfig:mahmuda_1} gives an example of a track with its matched portions shown in dark green and its unmatched portions shown in red. 
This partial map-matching is based on a variant of the \Frd. In the second step, the unmatched portions of the track are then inserted into the partially constructed map, which requires creating new vertices and creating and splitting edges. In a third step, the already existing edges in the map that are covered by the matched portions of the trajectory, are updated using a minimum-link algorithm to compute a new representative edge (cf. Figure~\ref{subfig:mahmuda_2}). This last step is only needed to provide a guaranteed bound on the complexity of the output map; in the implementation of this algorithm that we use in Section \ref{sec:sec_exp}, this last step has been omitted. Ahmed and Wenk also give theoretical quality guarantees for the output map computed by their algorithm, which include a one-to-one correspondence between well-separated ``good'' portions of the underlying map and the output map, with a guaranteed \Frd\ between those portions. 

\subsubsection{Biagioni and Eriksson \protect\cite{Biagioni:2012:MIF:2424321.2424333}}
\begin{figure}[htbp]
 \begin{center} 
	 \subfloat[High threshold]{\label{subfig:biagioni_1} \includegraphics[width=0.31\columnwidth]{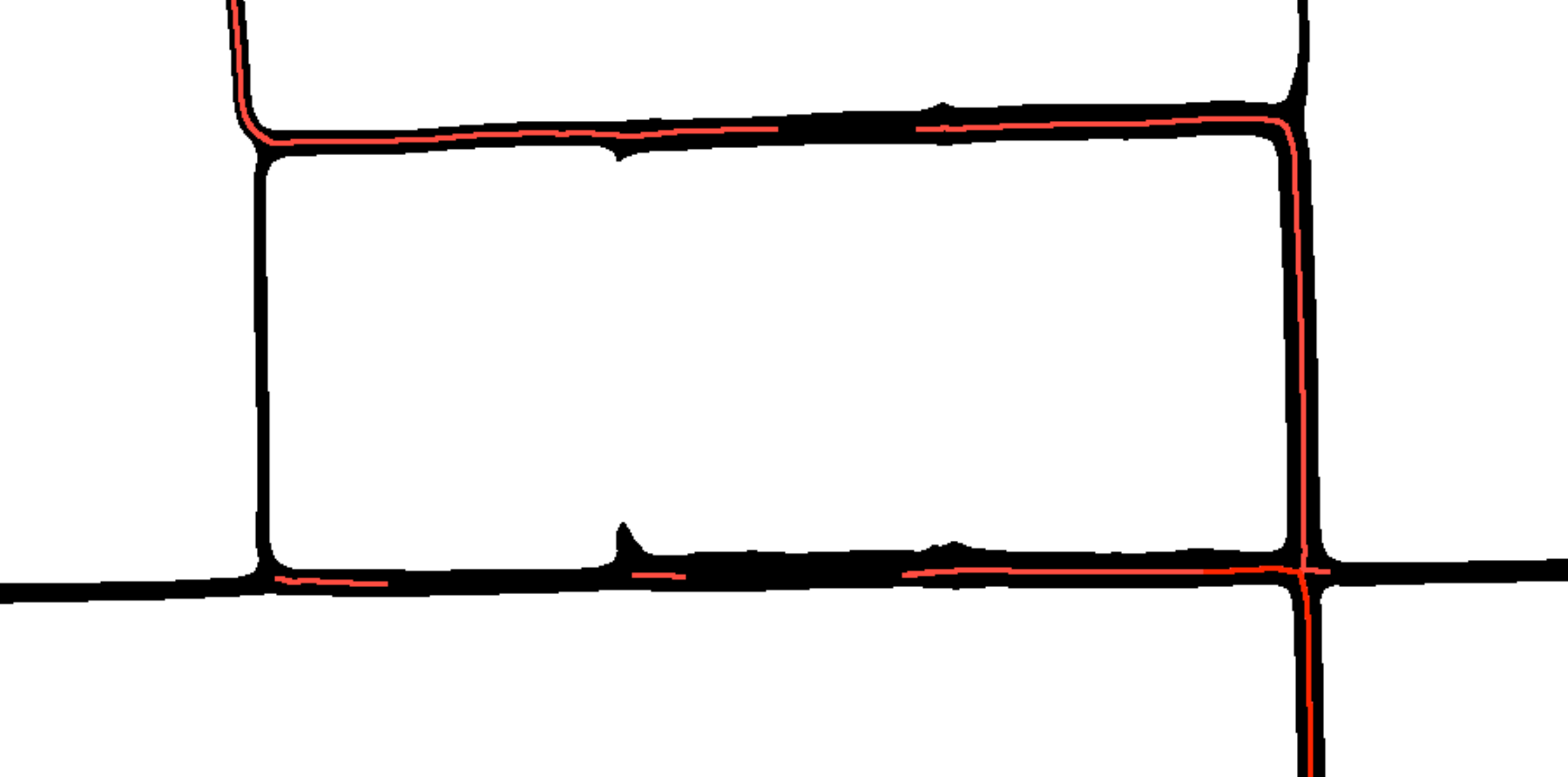}}	 
	 \subfloat[Medium threshold]{\label{subfig:biagioni_2} \includegraphics[width=0.31\columnwidth]{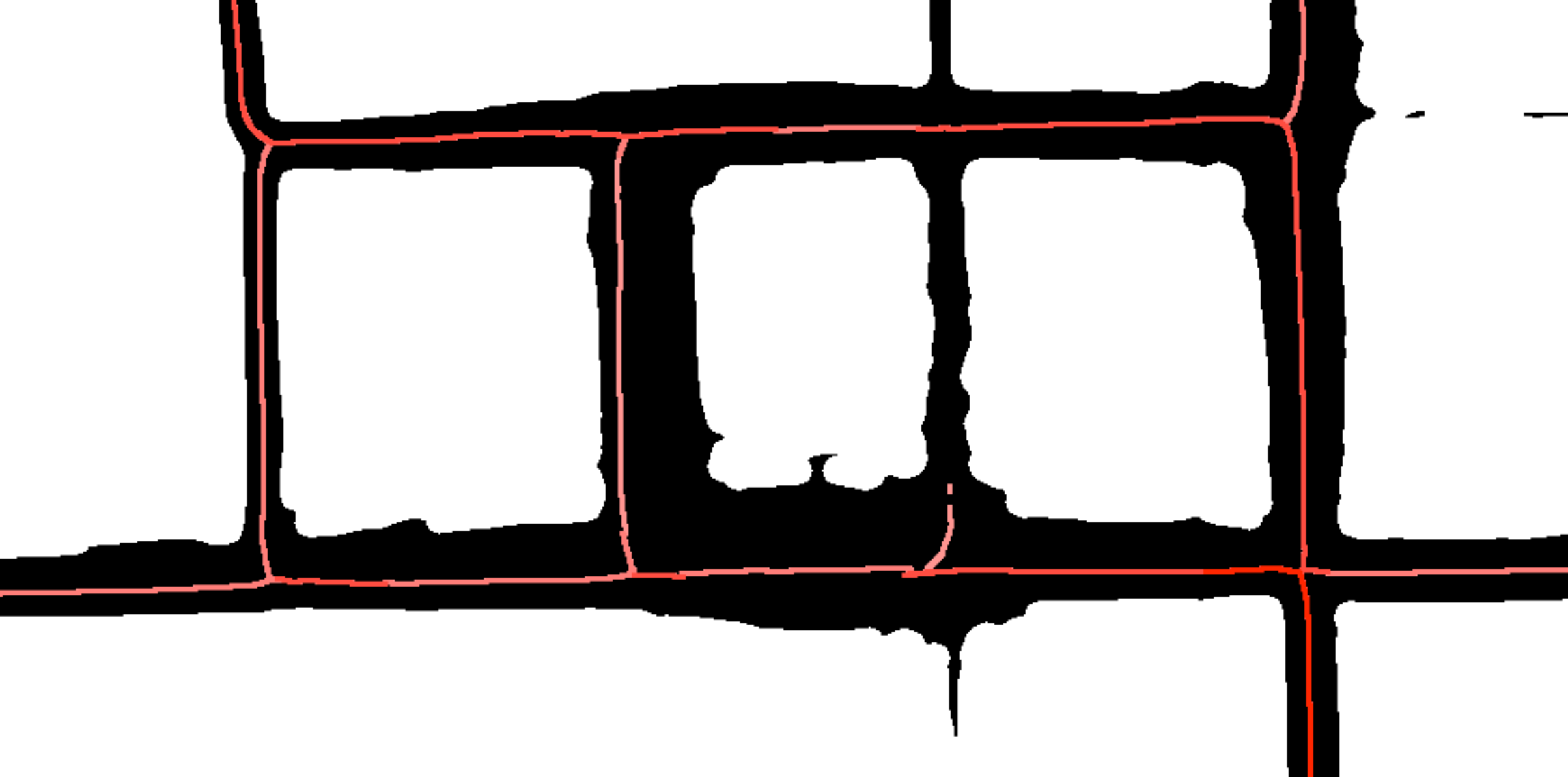}}		 
	 \subfloat[Low threshold]{\label{subfig:biagioni_3} \includegraphics[width=0.31\columnwidth]{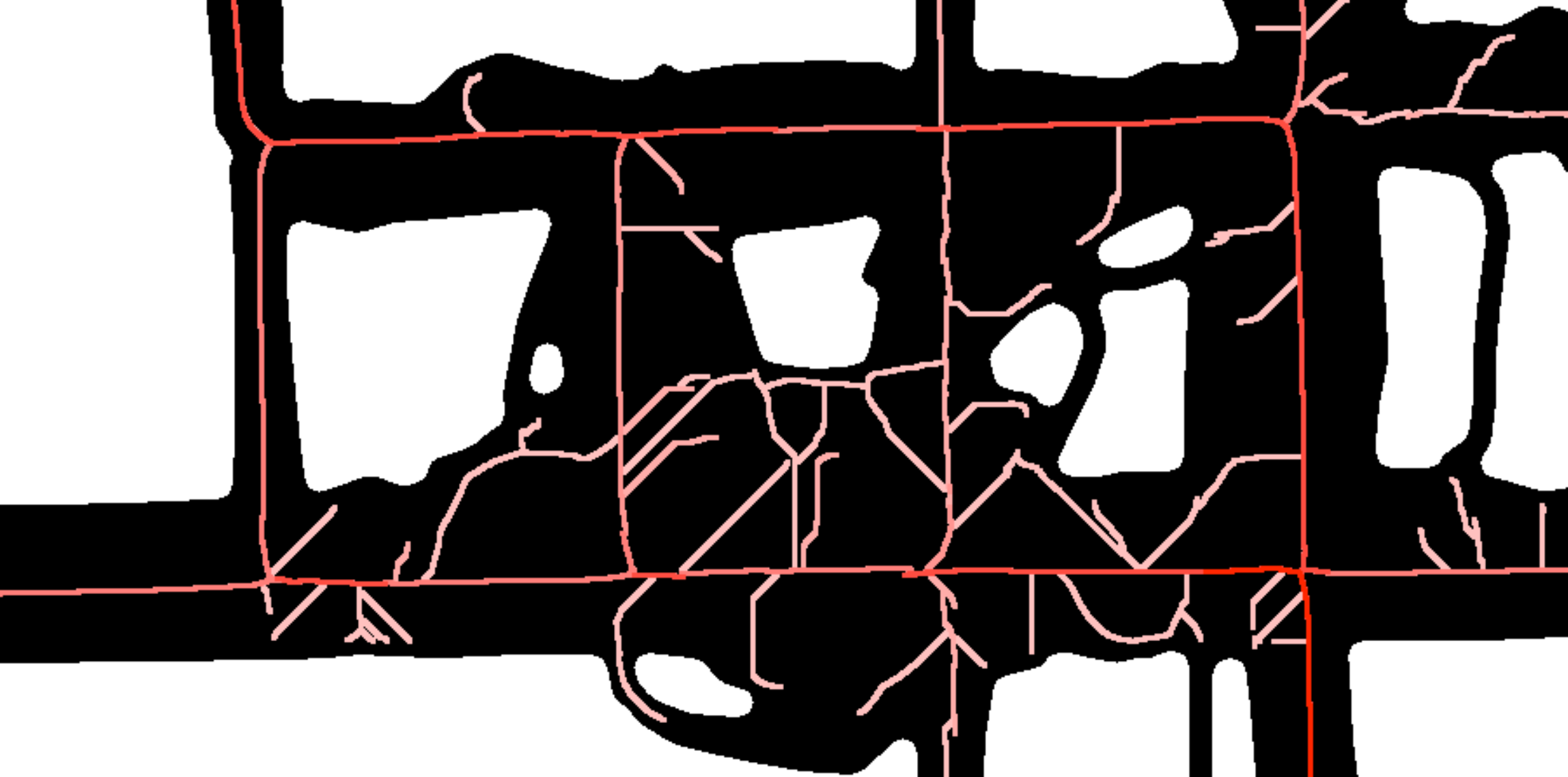}}		  
 \end{center}
\caption{KDE-based map construction using threshold ranges (images from presentation of  \cite{Biagioni:2012:MIF:2424321.2424333})}
\label{fig:biagioni}
\end{figure}

Biagioni and Eriksson \cite{Biagioni:2012:MIF:2424321.2424333} describe a point clustering-based algorithm that uses KDE methods. Their algorithm proceeds in using KDE with various thresholds to compute successive versions of a skeleton map. They annotate the map by performing a map-matching pass of the input tracks with the skeleton map. 
Figure~\ref{fig:biagioni} gives three example stages of the skeleton construction process using high to low KDE thresholds.

\subsubsection{Cao and Krumm \protect\cite{Cao:2009:GTR:1653771.1653776}}
This incremental track insertion approach proceeds in two stages. In the first stage, simulation of physical attraction is used to modify the input tracks to group portions of the tracks that are similar together. This results in a cleaner data set in which track clusters are more pronounced and different lanes are more separated. Then, this much cleaner data is used as the input for a fairly simple incremental track insertion algorithm. This algorithm makes local decisions based on distance and direction to insert an edge or vertex and either merge the vertex into an existing edge, or add a new edge and vertex. 

\begin{figure}[htbp]
 \begin{center}
	 \subfloat[Three input trajectories]{\label{subfig:cao_1} \includegraphics[width=0.45\columnwidth]{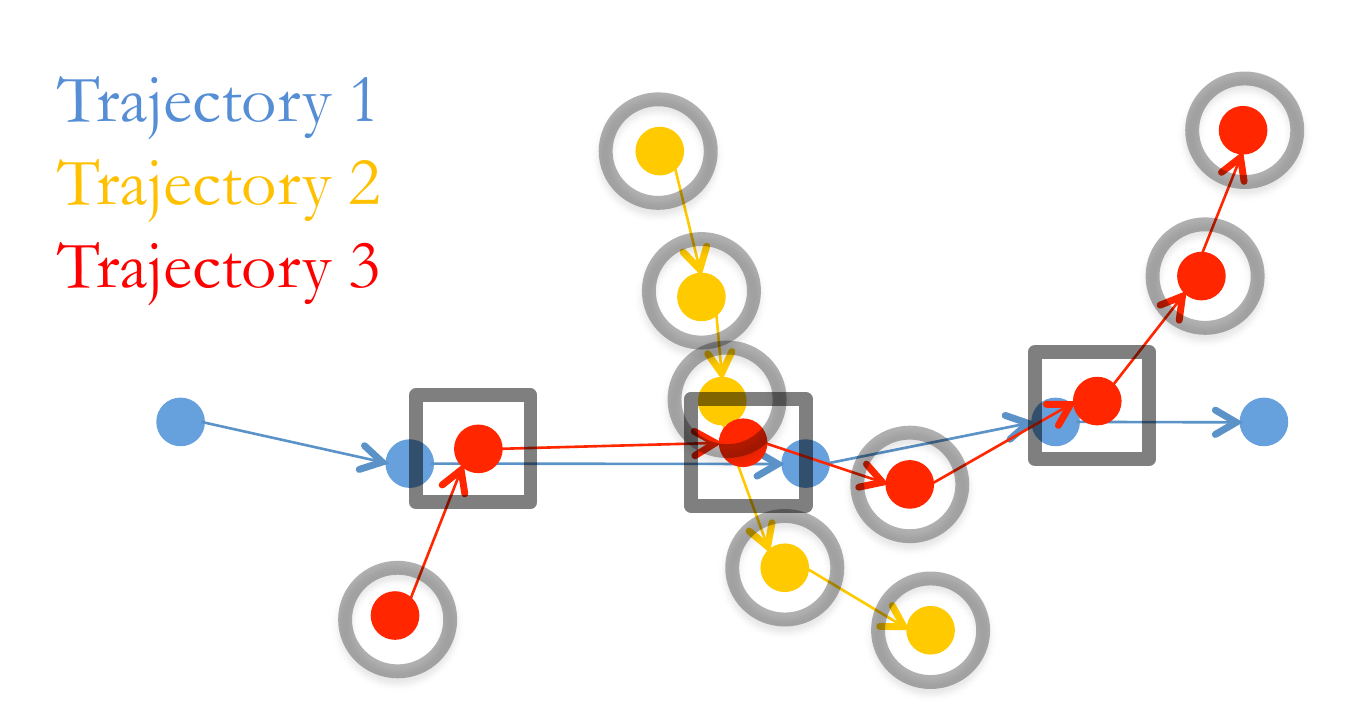}}	 
	 \subfloat[Merged graph]{\label{subfig:cao_2} \includegraphics[width=0.45\columnwidth]{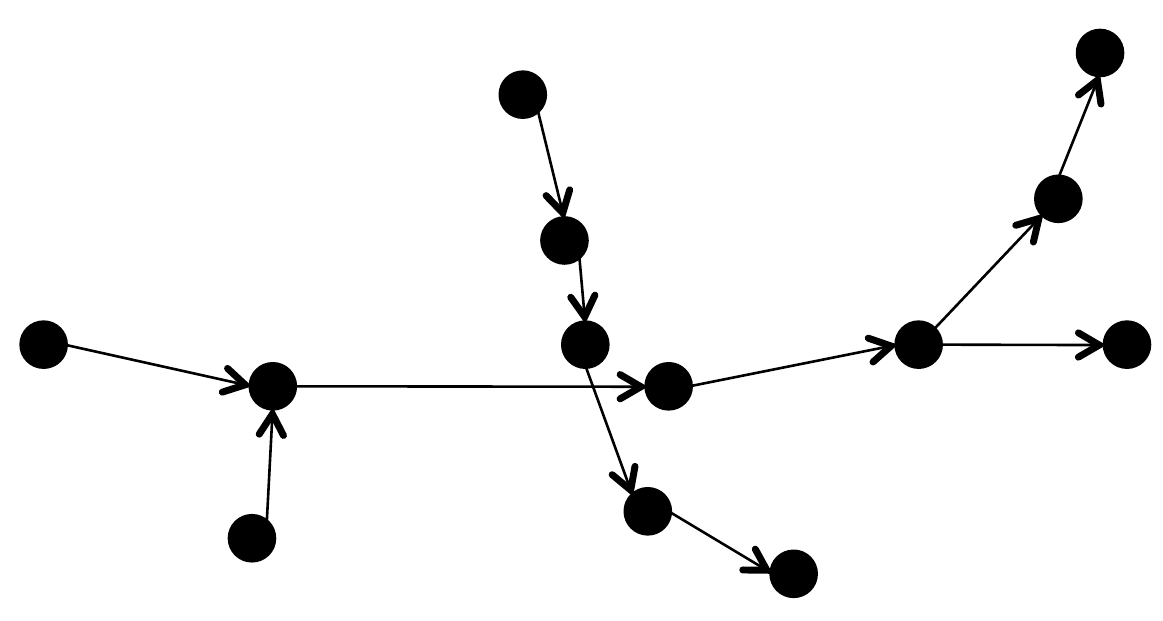}}		  
 \end{center}
\caption{The incremental track insertion algorithm - adapted from \cite{Cao:2009:GTR:1653771.1653776}}
\label{fig:cao}
\end{figure}

Figure~\ref{fig:cao} gives a respective map construction example. The three trajectories of Figure~\ref{subfig:cao_1} are used to incrementally build the graph in Figure~\ref{subfig:cao_2} by (i) either merging nodes to existing nodes if the distances are small and the directions of the traces match (nodes in boxes), or (ii) by creating new nodes and edges otherwise (nodes in circles). 

\subsubsection{Davies et al. \protect\cite{Davies:2006:SDR:1175887.1176088}}
\begin{figure}[htbp]
 \begin{center} 
	 \subfloat[Blurred trajectory histogram]{\label{subfig:kde_1} \includegraphics[width=0.31\columnwidth]{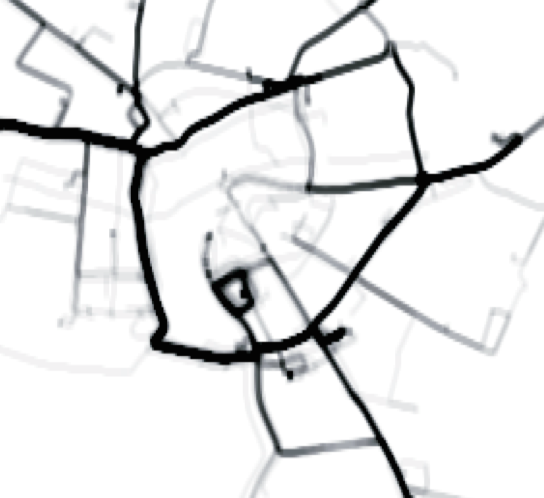}}	 
	 \subfloat[Contours]{\label{subfig:kde_2} \includegraphics[width=0.31\columnwidth]{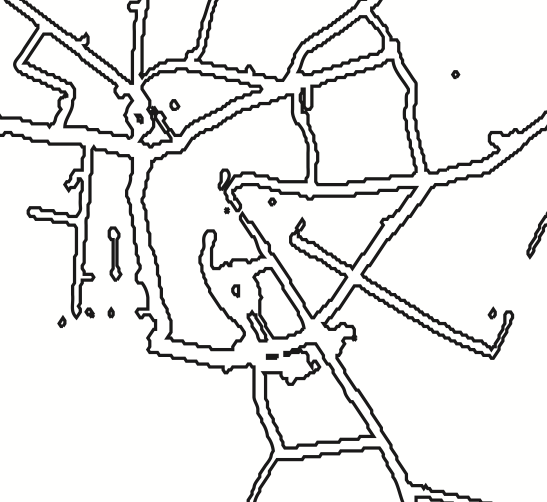}}		 
	 \subfloat[Centerlines, graph]{\label{subfig:kde_3} \includegraphics[width=0.31\columnwidth]{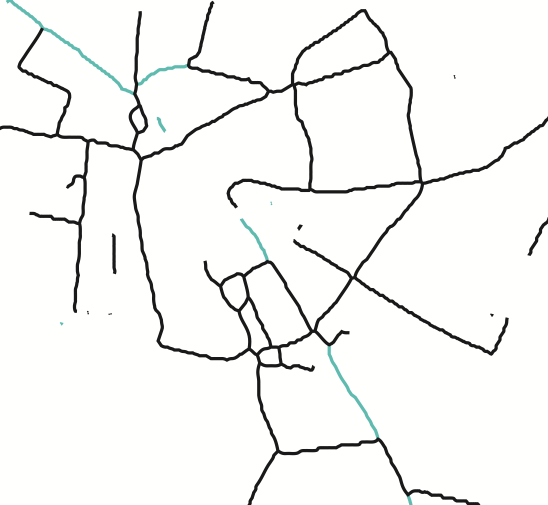}}		 
 \end{center}
\caption{Clustering-based map construction algorithm (images from \cite{Davies:2006:SDR:1175887.1176088})}
\label{fig:kde}
\end{figure}

This is a classical KDE-based map construction algorithm. It first computes for each grid cell the density of tracks that pass through it (cf. the example of Figure~\ref{subfig:kde_1}). Then it computes the contour of the resulting bit map (Figure~\ref{subfig:kde_2}), and then it uses the Voronoi diagram of the contour to compute a center line representation, followed by additional cleanup (Figure~\ref{subfig:kde_3}).

\subsubsection{Edelkamp and Schr\"odl \protect\cite{edelkamp:2003:rpmi}}

Edelkamp and Schr\"odl \cite{edelkamp:2003:rpmi} were the first to propose a map construction approach based on the $k$-means method. Their point clustering algorithm creates road segments based on tracking data, represents the center line of the road using a fitted spline and performs lane finding. The lanes are found by clustering tracks based on their distance from the road center line.

\begin{figure}[htbp]
 \begin{center} 
	 \subfloat[Input trajectories, clusters, and segments]{\label{subfig:edelkamp_1} \includegraphics[width=0.45\columnwidth]{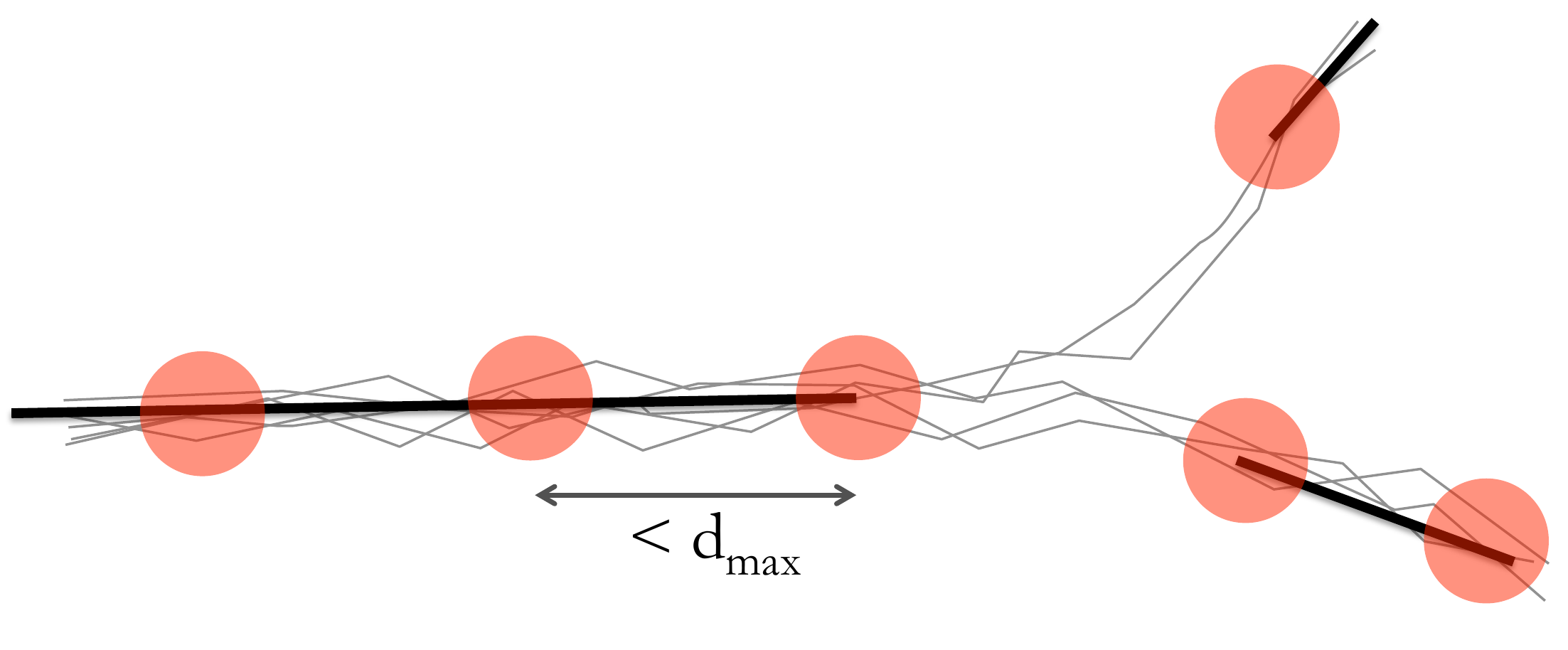}}	 
	 \subfloat[Centerlines, refined graph]{\label{subfig:edelkamp_2} \includegraphics[width=0.45\columnwidth]{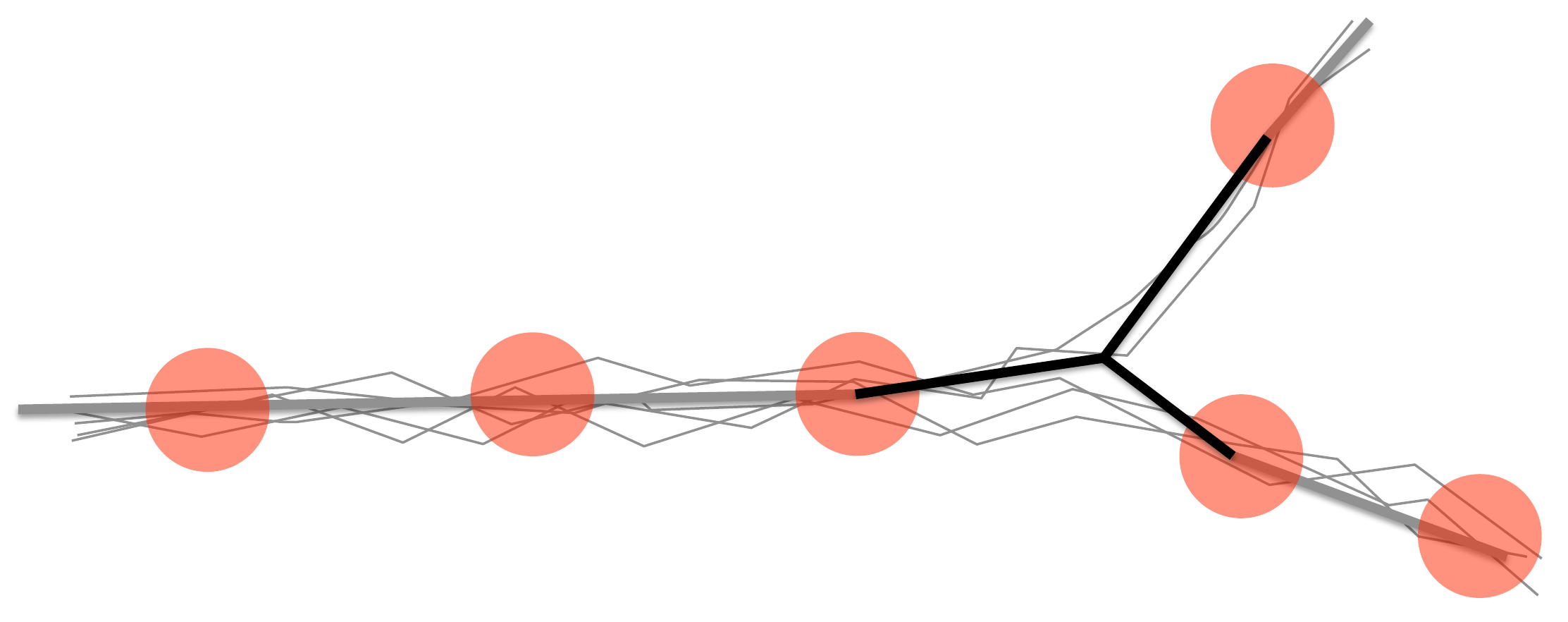}}		 
 \end{center}
\caption{Clustering-based map construction algorithm - adapted from \cite{edelkamp:2003:rpmi}}
\label{fig:edelkamp}
\end{figure}

\subsubsection{Ge et al. \protect\cite{DBLP:conf/nips/GeSBW11}}
\label{subsub:subsub_ge}
\begin{figure}[htbp]
 \begin{center}
	 \subfloat[Input points and initial graph]{\label{subfig:ge_1} \includegraphics[width=0.3\columnwidth]{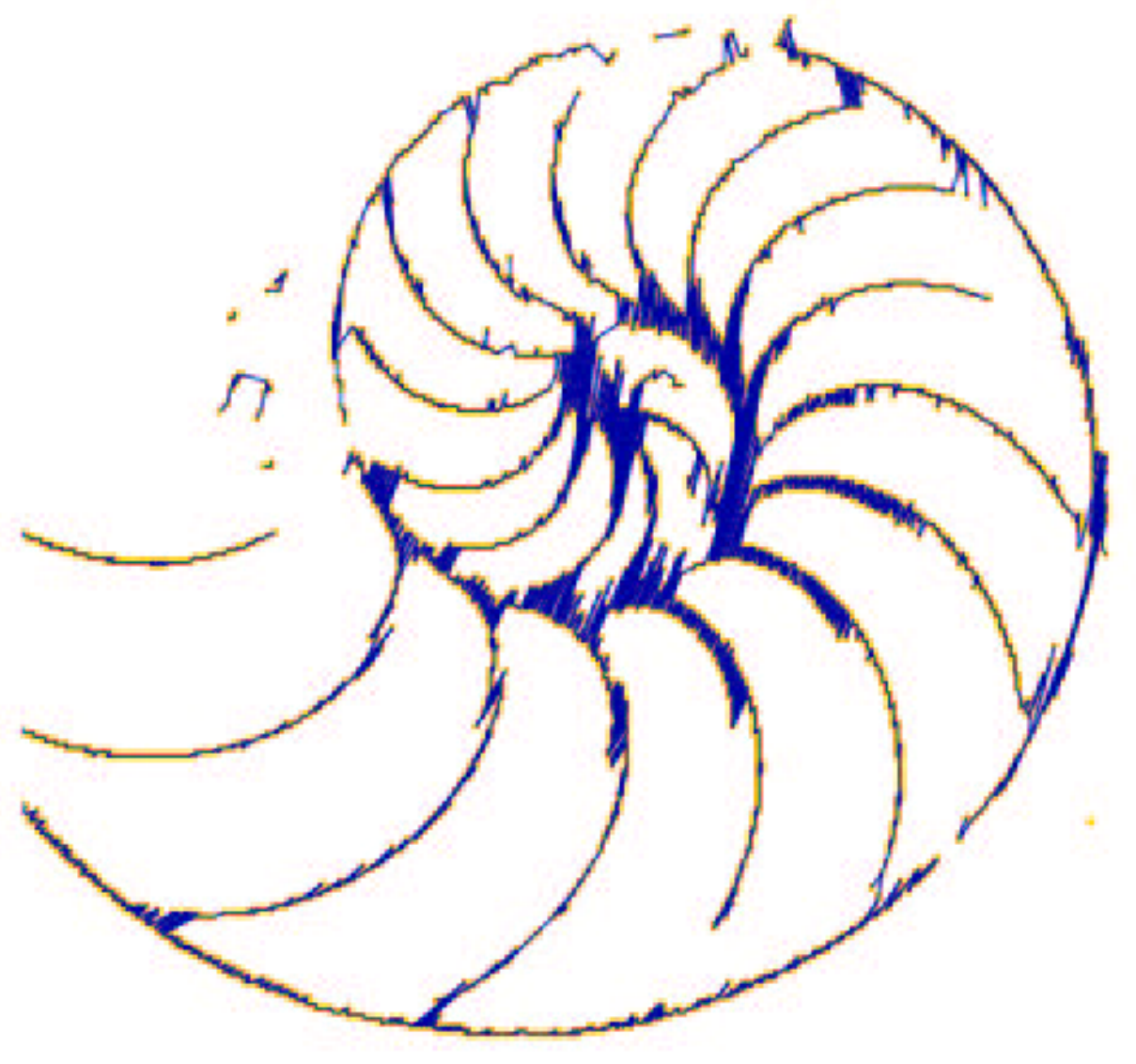}}	 
	 \subfloat[Graph after smoothing]{\label{subfig:ge_2} \includegraphics[width=0.3\columnwidth]{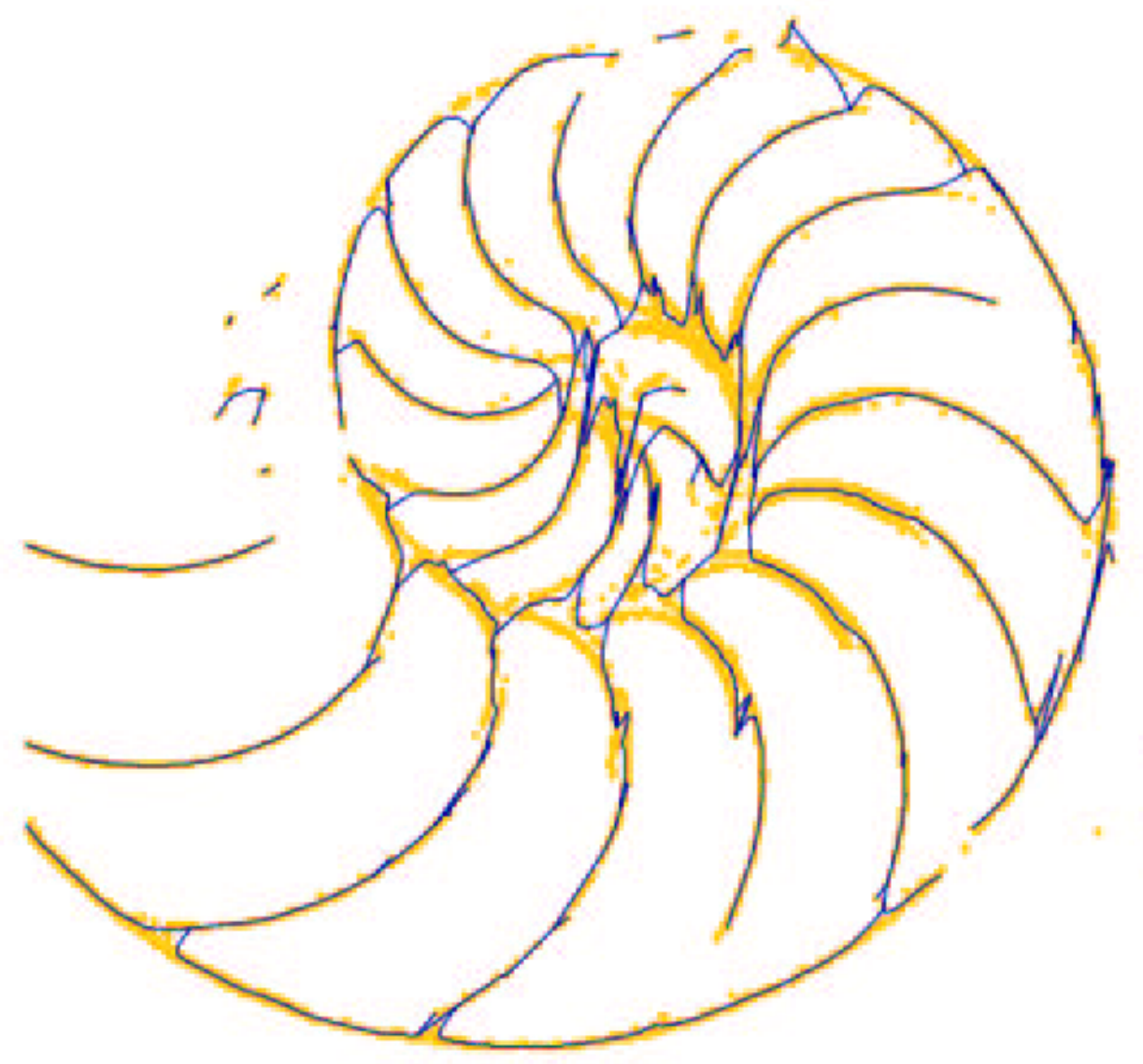}}		 
 \end{center}
\caption{Reeb graph based map construction (images from \cite{DBLP:conf/nips/GeSBW11})}
\label{fig:ge}
\end{figure}

This algorithm is a point clustering approach that applies topological tools to extract the underlying graph structure. The main idea of this algorithm is to decompose the input data set into sets each corresponding to a single branch in the underlying graph. The authors assume that the input point set is densely sampled, and their algorithm only needs a distance matrix or proximity graph of the point set as input. 
Then they define a function on the proximity graph, which assigns to every point in the graph its geodesic distance to an arbitrary base point. They employ the Reeb graph to model the connected components of the level set of the inverse of this function.
Finally, there is a canonical way
to measure importance of features in the Reeb graph, which allows them to easily simplify the resulting graph.
%
They provide runtime guarantees as well as partial quality guarantees for correspondences of cycles. 
An embedding for the edges is then obtained by using a principal curve algorithm \cite{Kegl:2000:LDP:333666.333696} that fits a curve to the points contributing to the edge.
Figure~\ref{fig:ge} gives an example of a constructed graph based on a point cloud shown as light (yellow) dots. 

\subsubsection{Karagiorgou and Pfoser \protect\cite{Karagiorgou:2012:VTD:2424321.2424334}}
\label{subsub:subsub_kp}
\begin{figure}[htbp]
 \begin{center}
	 \subfloat[Intersection nodes]{\label{subfig:intersections_sophia} \includegraphics[width=0.5\columnwidth]{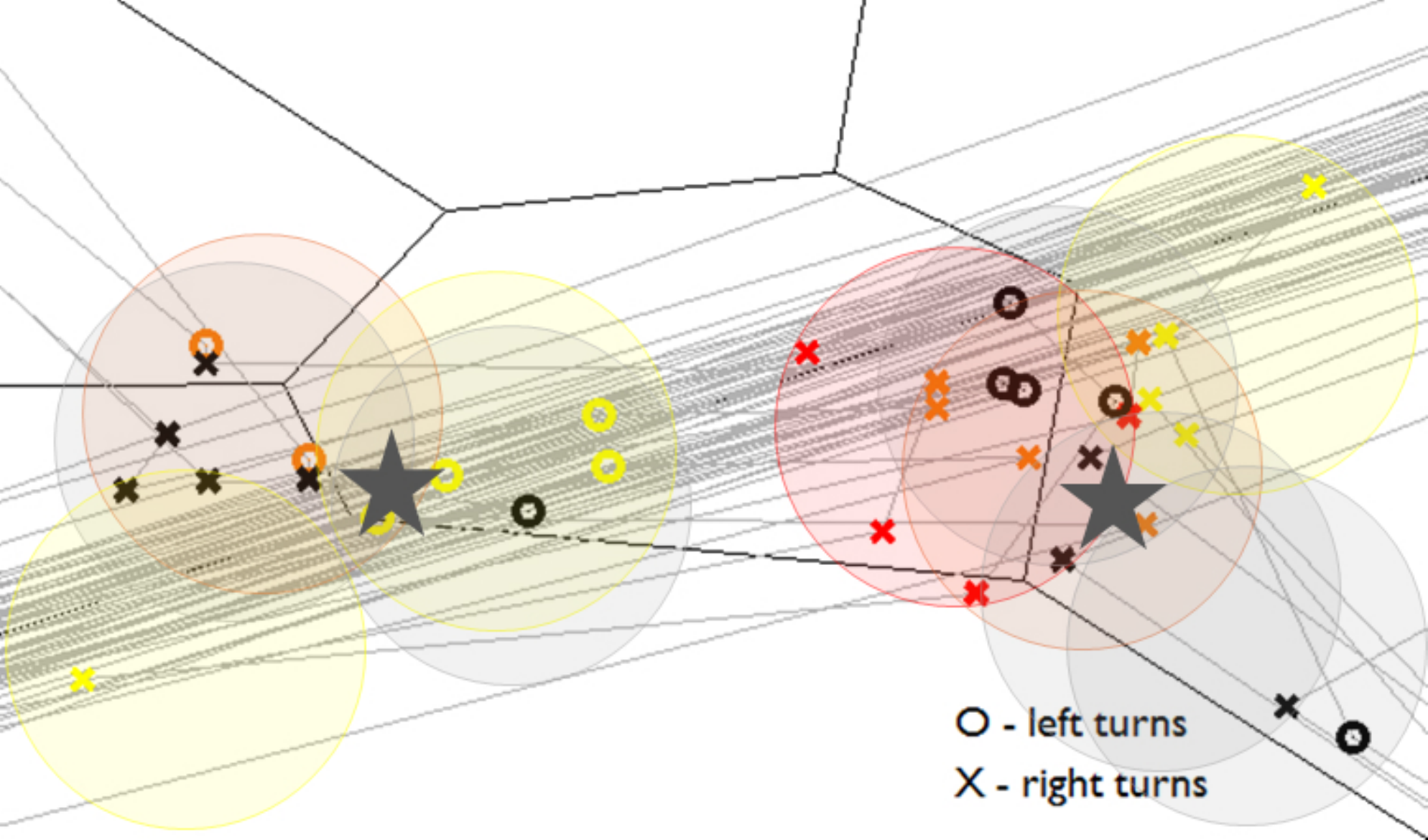}}	 
	 \subfloat[Compacting links]{\label{subfig:links_sophia} \includegraphics[width=0.4\columnwidth]{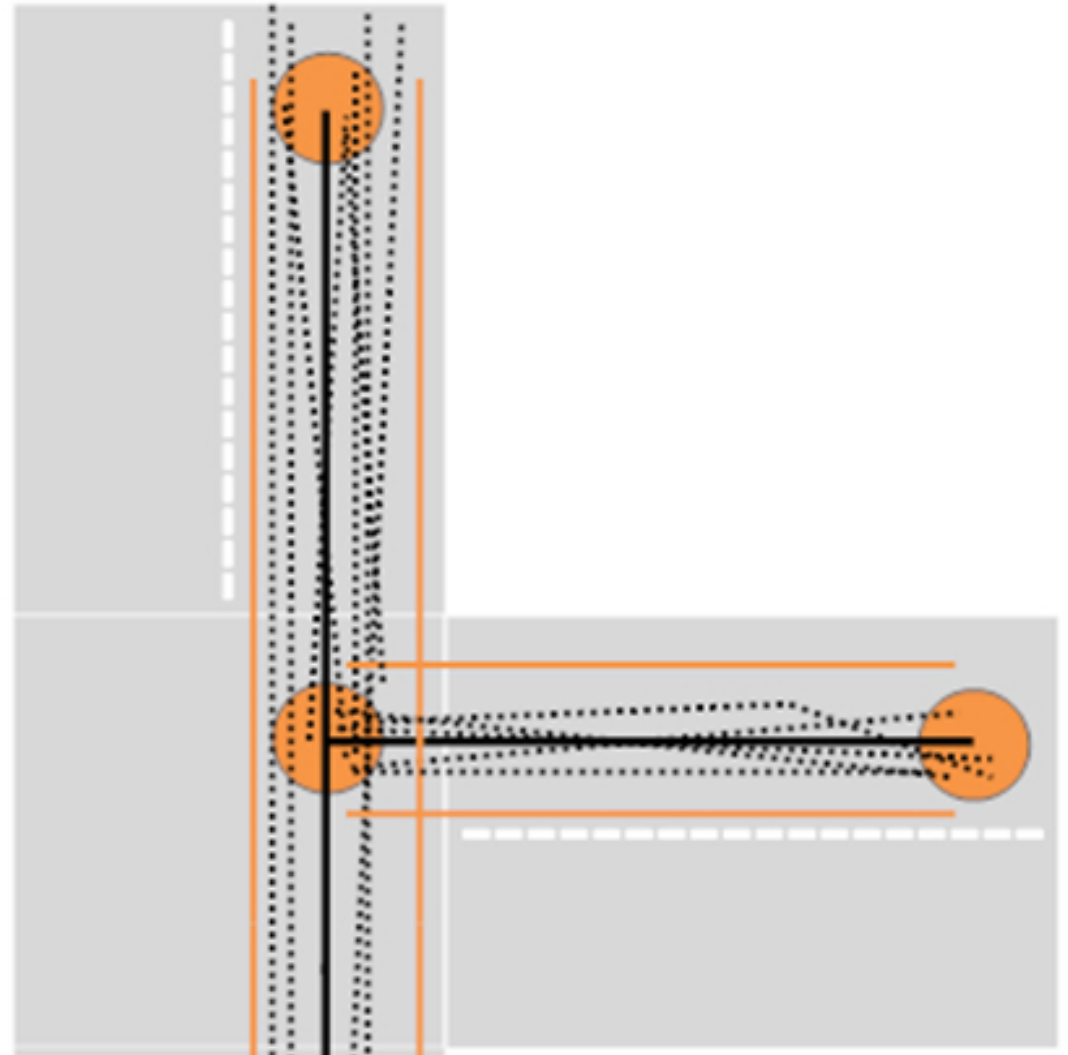}}		 
 \end{center}
\caption{The \emph{TraceBundle} algorithm \cite{Karagiorgou:2012:VTD:2424321.2424334}}
\label{fig:tracebundle}
\end{figure}

This intersection-linking map construction algorithm is a heuristic approach that ``bundles'' trajectories around intersection nodes. 
The main contribution of this \emph{TraceBundle} algorithm is its methodology to derive intersection nodes. The basic heuristic relies on detecting changes in movement and then clustering ``similar'' nodes. A change in direction and speed is considered a turn indicator. Clustering these turns based on (i) spatial proximity and (ii) turn type results in turn clusters. The centroid location of each of these turn clusters represents an intersection node. Links, and consequently the entire geometry of the map, are generated by connecting the intersection nodes with trajectories, and compacting the trajectories. Figure~\ref{fig:tracebundle} presents the steps of this algorithm. Figure~\ref{subfig:intersections_sophia} shows the constructed intersection nodes as gray stars from turn clusters (x and o markers) and Figure~\ref{subfig:links_sophia} shows as black lines the created links after compacting the  trajectories. 

\section{Quality Measures for Map Comparison}
\label{sec:sec_qmes}

There are two key ingredients for evaluating the quality of a constructed map: (1) the availability of an adequate ground-truth map $G$ as part of the benchmark data, and (2) a quality measure used to evaluate the similarity between the constructed map $C$ and the ground-truth map $G$.


There are essentially two cases of what can be considered as a ground-truth map $G$. Ideally, $G$ is the underlying map consisting of all streets, and only those streets, that have been traversed by the entities that generated the set of input tracks. If such a $G$ was available, then a suitable quality measure would compare $C$ to all of $G$ and the ideal would be for $C$ to equal $G$. However, in practice, it is hard to obtain an unbiased ground-truth map that exactly corresponds to the coverage of the tracking data. 
This non-trivial task has been addressed in the past by pruning the ground-truth either manually, by proximity to the tracking data, or by map-matching the tracking data to the map \cite{be-irmgp-12,Biagioni:2012:MIF:2424321.2424333,Karagiorgou:2012:VTD:2424321.2424334,Liu:2012:MLS:2339530.2339637}. By using graph topologies resulting from human judgment or from the cropping behaviors of the different pruning algorithms, clearly all these approaches introduce an undesired bias.

Actually, it is much easier to obtain a ground-truth map that covers a superset of all the streets covered by the input tracks, e.g., street maps taken by proprietary vendors or OpenStreetMap. 
Therefore, if $G$ is a superset, then the quality measure attempts to {\em partially} match $C$ to $G$.
Of course, another possible scenario is that $C$ contains additional streets that are not present in either variation of $G$. 

\subsection{Related Work}
In the graph theory literature, there are various distance measures for comparing two abstract graphs, that do not necessarily have a geometric embedding \cite{cfsv-tygmpr-04,JGT:JGT3190030202,JGT:JGT3190010410}. Most closely related to street map comparison are the subgraph isomorphism problem and the maximum common isomorphic subgraph problem, both of which are NP-complete. These, however, rely on one-to-one mappings of graphs or subgraphs, and they do not take any geometric embedding into account. Graph edit distance \cite{gxtl-sged-10,Zeng:2009:CSA:1687627.1687631} is a way to allow noise by seeking a sequence of edit operations to transform one graph into the other, however it is NP-hard as well. Cheong et al. \cite{CheongGKSS09} consider a graph edit distance for geometric graphs (embedded in two different coordinate systems, however), and also show that it is NP-hard to compute. 

For comparing street maps, distance measures based on \emph{point sets} and distance measures based on \emph{sets of paths} have been proposed. 
\emph{Point set-based distance measures} treat each graph as the set of points in the plane that is covered by all its vertices and edges. The idea is then to compute a distance between the two point sets. A straightforward distance measure for point sets are the directed and undirected Hausdorff distances \cite{ag-dgsmi-99}. The main drawback of such an approach is that it does not use the topological structure of the graph. Biagioni and Eriksson \cite{be-irmgp-12,Liu:2012:MLS:2339530.2339637}, use two distance measures that essentially both use a variant of a partial one-to-one bottleneck matching that is based on sampling both graphs densely.
The two distance measures compare the total number of matched sample points to the total number of sample points in the graph, thus providing a measure of how much of the graph has been matched. They do require though to have as input a ground-truth graph that closely resembles the underlying map and not a superset.

For \emph{path-based distance measures} on the other hand, the underlying idea is to represent the graphs by sets of paths, and then define a distance measure based on distances between the paths. This captures some of the topological information in the graphs, and paths are of importance for street maps in particular since the latter are often used for routing applications for which similar connectivity is desirable. 
Mondzech and Sester~\cite{MondzechS11} use shortest paths to compare the suitability of two road networks for pedestrian navigation by considering basic properties such as respective path length. 
Karagiorgou and Pfoser \cite{Karagiorgou:2012:VTD:2424321.2424334} also use shortest paths, but to actually assess the similarity of road network graphs. Computing random sets of start and end nodes, the computed paths are compared using Discrete \Frd\ and the Average Vertical distance. Using those sets of distances, a global network similarity measure is derived.
%
In another effort, Ahmed and Wenk \cite{aw-SIGSPATIAL-13} cover the networks to be compared with paths of $k$ link-length and map-match the paths to the other graph using the \Frd. They are the first to introduce the concept {\em local signature} to identify {\em how} and {\em where} two graphs differ.  

\subsection{Quality Measures used for Comparison}
\label{subsec:subsec_qmu}

Here we give some more details on the quality measures that we use in Section \ref{sec:sec_exp} to compare the different road network construction algorithms. Note that in our experiments the ground-truth $G$ is an unmodified street map from OpenStreetMap and thus expected to be a superset of the underlying graph. We use the Directed Hausdorff distance \cite{ag-dgsmi-99}, the path-based distance measure presented by Ahmed et al. \cite{aw-SIGSPATIAL-13}, the distance measure based on shortest paths by Karagiorgou and Pfoser \cite{Karagiorgou:2012:VTD:2424321.2424334} and graph-sampling based distance measure 
by Biagioni and Eriksson \cite{be-irmgp-12}. The first two measures have not been used for comparative evaluations of road network constructions before.

\subsubsection{Directed Hausdorff Distance \protect\cite{ag-dgsmi-99}}
The {\em directed Hausdorff distance}
of two sets of points $A, B$ is defined as $\overrightarrow{d}(A,B) = \max_{a \in A}\min_{b \in B} d(a,b)$. Here,
$d(a,b)$ is usually the Euclidean distance between two points $a$ and $b$. Intuitively, the directed Hausdorff distance assigns to every point in $a$ its nearest neighbor $b\in B$ and takes the maximum of all distances between assigned points.
In order to compare two graphs, we identify each graph as the set of points that is covered by all its vertices and edges. 
If the directed Hausdorff distance from graph $C$ to graph $G$ is at most $\varepsilon$, this means that for every point on any edge or vertex of $C$ there is a point on $G$ at distance at most $\varepsilon$. 
Or equivalently, every point of $C$ is contained in the Minkowski sum of $G$ with a disk of radius $\varepsilon$; the Minkowski sum intuitively ``fattens'' $G$ by ``drawing'' each of its edges with a thick circular pen.  
This distance measure gives a notion about spatial distance for graphs. 
If $C$ is the constructed graph and $G$ is the ground-truth, the lower the distance from $C$ to $G$, the closer the graph $C$ to $G$.
%


\subsubsection{Path-Based Distance \protect\cite{aw-SIGSPATIAL-13}}
The path-based map distance considers graphs as sets of paths. The distance between two sets of paths is then computed in the Hausdorff setting, while the \Frd\, which is a natural distance measure for curves that takes monotonicity and continuity into account, is used to compute the distance between two paths. 

For curves $f, g$, the \Frd\ is defined as

\begin{equation}
	\delta_{F}(f,g) = \inf_{\substack{\alpha,\beta:[0,1]\rightarrow [0,1]}}
	\max_{\substack{t\in\left[ 0,1\right] }}d(f(\alpha(t)),g(\beta(t))), 
\end{equation}
where $\alpha, \beta$ range over continuous, surjective and non decreasing
repara\-metrizations.

A common intuition is to explain it as the minimum leash length required such that a man and dog can walk on the two curves from beginning to end in a monotonic way.

Under this scope, let $C$ and $G$ be two planar geometric graphs, and let $\pi_C$ be a set of paths generated from $C$, and $\pi_G$ be a set of paths generated from $G$. The {\em path-based distance} is defined as:
\begin{equation}
\distance{C}{G}{\pi_C}{\pi_G} = \max_{p_C \in \pi_C}\min_{p_G \in \pi_G}\delta_F(p_C,p_G)
\end{equation}

Ideally, $\pi_C$ and $\pi_G$ should be the set of all paths in $C$ and $G$, which however has exponential size. In \cite{aw-SIGSPATIAL-13} they showed that $\distance{C}{G}{\Pi_C}{\Pi_G}$ can be approximated 
using $\distance{C}{G}{\Pi^{3}_C}{\Pi_G}$ in polynomial time using the map-matching algorithm of \cite{aerw-mpm-03}, under some assumptions on $C$. Here, $\Pi_C$ is the set of all paths and $\Pi^{3}_C$ is the set of all link-$3$ paths of $C$. 
A link-$k$ path consists of $k$ ``edges'', where vertices of degree two in the graph are not counted as vertices.
 Using this asymmetric distance measure $\distance{C}{G}{\Pi^{k}_C}{\Pi_G}$, which can be computed in polynomial time
for constant $k$, the following properties have been shown in \cite{aw-SIGSPATIAL-13}, under some assumptions on $C$:

\begin{enumerate}
\item{$k=1$: For each edge in $C$, there is a path in $G$ which is within \Frd\ $\distance{C}{G}{\Pi^{1}_C}{\Pi_G}$.}
\item{$k=2$: For each vertex $v$ in $C$ there is a vertex in $G$ within bounded distance $\distance{C}{G}{\Pi^{2}_C}{\Pi_G})/\sin{\frac{\theta}{2}}$, where $\theta$ is the minimum incident angle at $v$ between its adjacent edges.}
\item{$k=3$: $\distance{C}{G}{\Pi^{3}_C}{\Pi_G}$ approximates $\distance{C}{G}{\Pi_C}{\Pi_G}$ within a factor of $1/\sin{\frac{\theta}{2}}$ if the vertices of $C$ are reasonably well separated and have degree $\neq 3$.
\footnote{The degree assumption is only a technical requirement for the theoretical quality guarantees, and the authors have shown \cite{aw-SIGSPATIAL-13} that similar approximation guarantees appear to hold in practice as well.}}
\end{enumerate}

Similar to Directed Hausdorff distance, the lower the value of $\distance{C}{G}{\Pi_C}{\Pi_G}$ the more closely the constructed map $C$ resembles the ground-truth map $G$.

The {\em local signature} of a vertex $v\in C$ is defined as $\Delta_{v}=\distance{C}{G}{{\Pi_C}_v}{\Pi_G}$ where ${\Pi_C}_v$ is a set of paths that contains $v$. In a similar way, the local signature of
an edge $e\in C$ is defined as $\Delta_{e}=\distance{C}{G}{{\Pi_C}_e}{\Pi_G}$ where ${\Pi_C}_e$ is a set of paths that contains $e$. Based on the value of these signatures one can identify which vertices or edges are very similar and which are not.

\subsubsection{Shortest Path Based Distance \protect\cite{Karagiorgou:2012:VTD:2424321.2424334}}
Karagiorgou et al.~\cite{Karagiorgou:2012:VTD:2424321.2424334} propose a measure that essentially samples each graph using random sets of shortest paths.
Given the constructed and ground-truth networks $C$ and $G$ respectively, a common set of node pairs (origin, destination) is selected in both, using the nearest neighbor search if necessary. For all node pairs, shortest paths are computed in both networks. The geometric difference/similarity between the respective shortest paths is used to assess the similarity between $C$ and $G$ and consequently as a means to assess the quality of the constructed network.
The Discrete \Frd\ and the Average Vertical distance are used to compare the shortest paths.
The rationale for using this approach is that measuring the similarity for sets of paths instead of individual links allows one to better reason about the connectivity of the generated network. The more ``similar'' the shortest paths in the constructed network are to the ground-truth network, the higher also the quality of the network.
The results of this shortest path based distance measure can be assessed by plotting the distance of all paths against each other, or by comparing average values for the entire set of paths. We employ both approaches in our experiments below.

\subsubsection{Graph-Sampling Based Distance \protect\cite{be-irmgp-12}}
Biagioni and Eriksson \cite{be-irmgp-12} introduce a graph-sampling based distance measure in order to evaluate geometry and topology of the constructed road networks represented by graphs. The main idea is as follows: starting from a random street location, explore the topology of the graphs by placing point samples on each graph outward within a maximum radius. This produces two sets of locations, which are essentially spatial samples of a local graph neighborhood. These two point sets are compared using one-to-one bottleneck matching and counting the unmatched points in each set. The sampling process is repeated for several seed locations.

For the bottleneck matching, the sample points on one graph can be considered as ``marbles'' and on the other graph as ``holes''. Intuitively, if a marble lands close to a hole it falls in, marbles that are too far from a hole remain where they land, and holes with no marbles nearby remain empty. If one of the graphs is the ground truth, this difference represents the accuracy of the other graph. Counting the number of unmatched marbles and empty holes quantifies the accuracy of the generated road network with respect to the ground truth according to two metrics. The first metric is the proportion of spurious marbles, $spurious =spurious\_ marbles /\left(spurious\_marbles + matched\_ marbles\right)$ and the second is the proportion of missing locations (empty holes), where $missing =$ $empty\_holes/\left(empty\_holes + matched\_holes\right)$.

To produce a combined performance measure from these two values, the well-known F-score is used, which is computed as follows:
\begin{equation}
F \mbox{-} score = 2 * \frac{precision * recall}{precision+recall} 
\end{equation}
where, $precision = 1 - spurious$ and $recall = 1-missing$.

The higher the F-score, the closer the match. Sampling the graphs locally is an important aspect of this approach as it provides the ability to capture the connectivity of the graphs at a very detailed level, 
allowing the topological similarity to be measured. Repeated local sampling at randomly chosen locations yields an accurate view of local geometry and topology throughout the graph.

A modified version is used in \cite{Biagioni:2012:MIF:2424321.2424333} where the method ignores parts of the road network where no correspondence could be found between generated and ground-truth networks, for our experiments we used this modified version.

\subsection{Comparison of Distance Measures}
\begin{figure}
\centering
\includegraphics[width=0.5\textwidth]{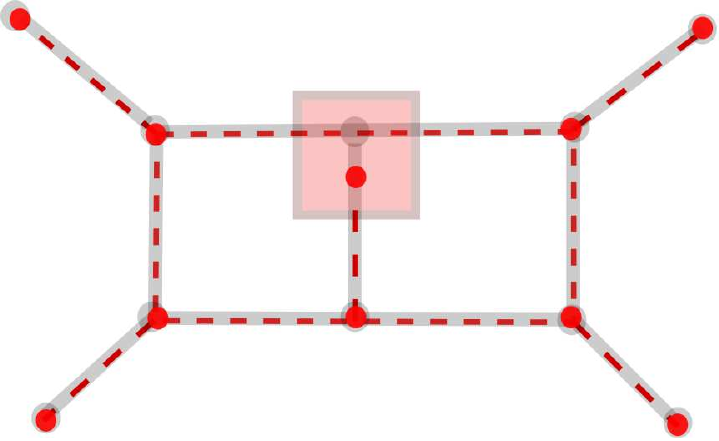}
\caption{Graph $G$ (dotted edges) overlayed on $H$ (gray). $G$ and $H$ differs in the shaded squared region. The distance measure in ~\cite{Biagioni:2012:MIF:2424321.2424333} fails to capture the broken connection in $G$, as there is always detour available to reach every edge and sample it.}
\label{fig:distance_illustration}
\end{figure}
All the distance measures described in Subsection~\ref{subsec:subsec_qmu} capture different properties of graphs.
Based on the desired kind of similarity, different distance measures could be employed.
For example, if one is interested in ensuring similar shortest paths in the two graphs, requiring that independent queries produce similar routes, then the  shortest path based measure would be the prefect choice~\cite{MondzechS11, Karagiorgou:2012:VTD:2424321.2424334} among 
all. If, however, one wants to know the spatial displacement between the two graphs without necessarily considering any kind of topology or path similarity, then the directed Hausdorff distance \cite{ag-dgsmi-99} would be the distance measure to choose.

On the other hand, the two distance measures described in~\cite{Biagioni:2012:MIF:2424321.2424333} and~\cite{aw-SIGSPATIAL-13} maximize the use of topology in comparing graphs. Using the concept of {\em local signature} described in~\cite{aw-SIGSPATIAL-13} one can visualize the exact differences in graphs using any of these two measures. Figure~\ref{fig:distance_illustration} shows an example where the graph sampling based distance~\cite{Biagioni:2012:MIF:2424321.2424333} fails to identify local differences (the dotted graph has a broken connection in the gray square region). As it samples small sub-graphs starting from a root location, it cannot capture this kind of broken connection when another connecting detour between the two parts is available in that small sub-graph. As 
the path-based distance~\cite{aw-SIGSPATIAL-13} exploits every adjacency transition around a vertex, it verifies all connectivities. 

Among these four measures only the graph sampling based distance~\cite{Biagioni:2012:MIF:2424321.2424333} ensures one-to-one correspondence. 
So, if one of the graphs has missing streets or extra edges, that is reflected in the overall score as well as in the local signatures.


\section{Datasets}
\label{sec:sec_data}

A basic means for assessing map construction algorithms is the underlying dataset comprising vehicle trajectories and ground-truth map datasets. The datasets are in a projected coordinate system (UTM, GGRS87). All the visualizations of the datasets are also available on the {\tt mapconstruction.org} web site. The statistics of the datasets are provided in Table~\ref{tab:dstatistics}.

\begin{figure}[htbp]
 \begin{center}
	 \subfloat[\emph{Athens large}]{\label{subfig:track_al} \includegraphics[width=0.5\columnwidth]{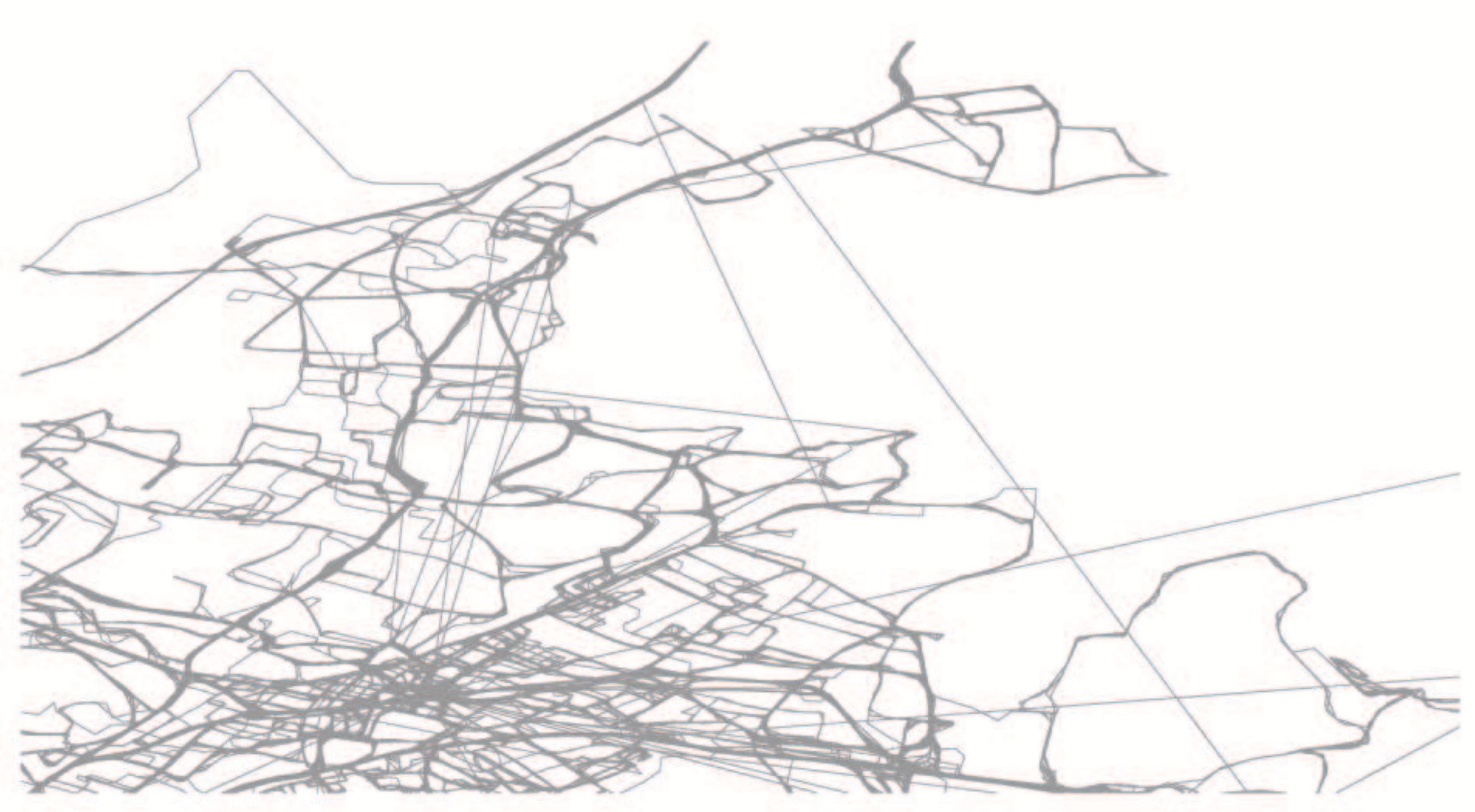}}	 
	 \subfloat[\emph{Athens small}]{\label{subfig:track_as} \includegraphics[width=0.5\columnwidth]{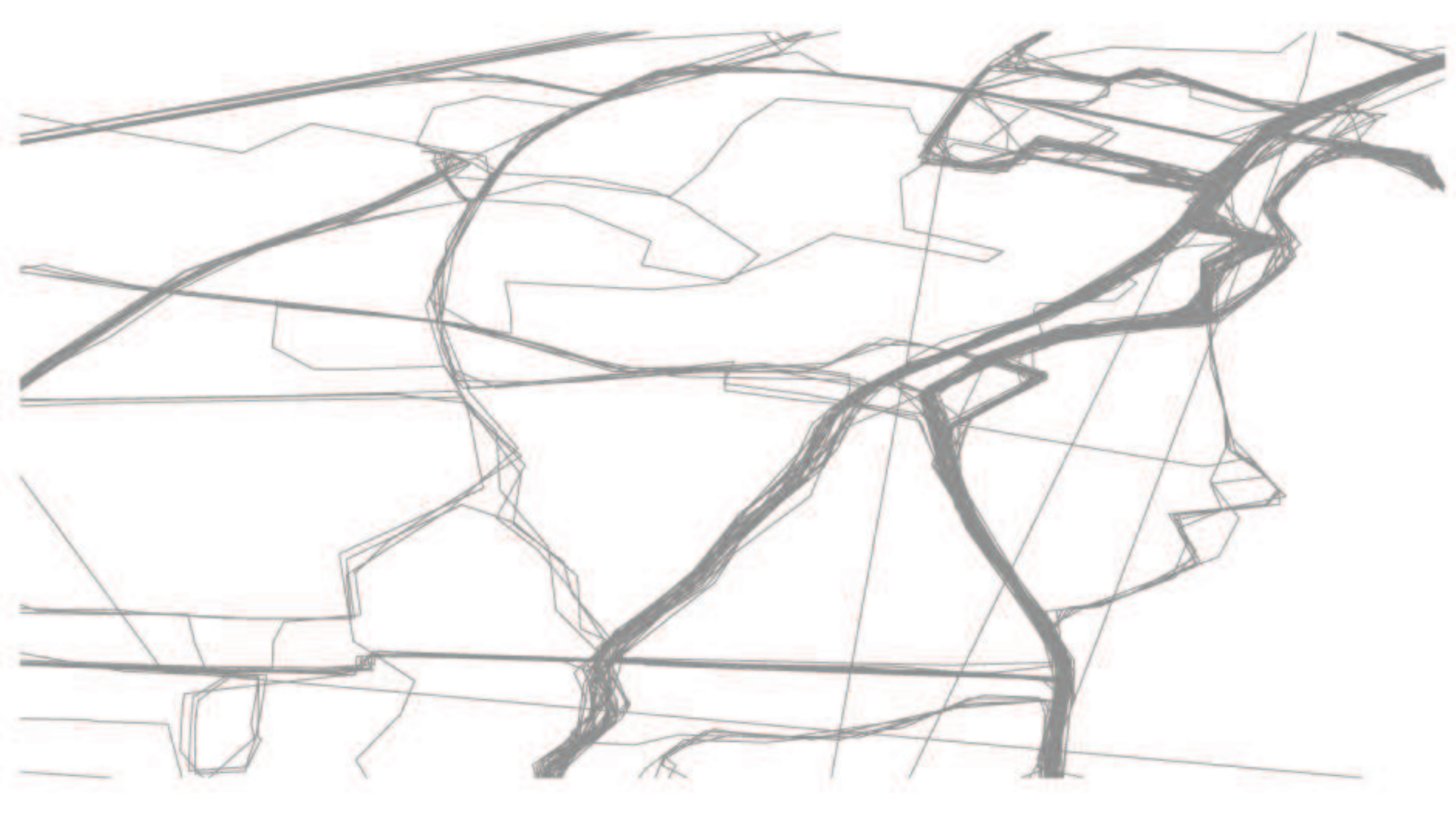}}	 
	 \\
	 \subfloat[\emph{Berlin}]{\label{subfig:track_b} \includegraphics[width=0.5\columnwidth]{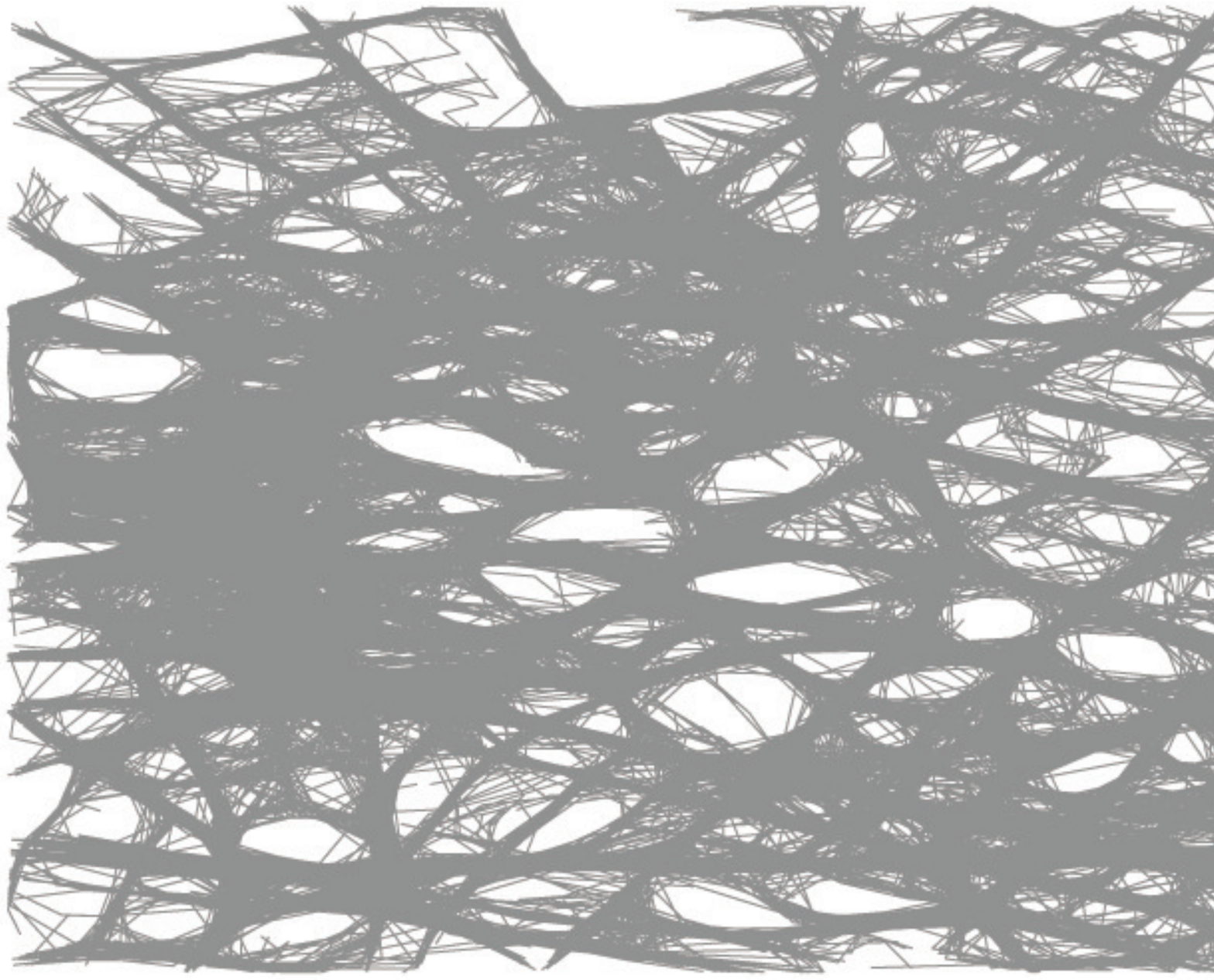}}	 
	 \subfloat[\emph{Chicago}]{\label{subfig:track_c} \includegraphics[width=0.5\columnwidth]{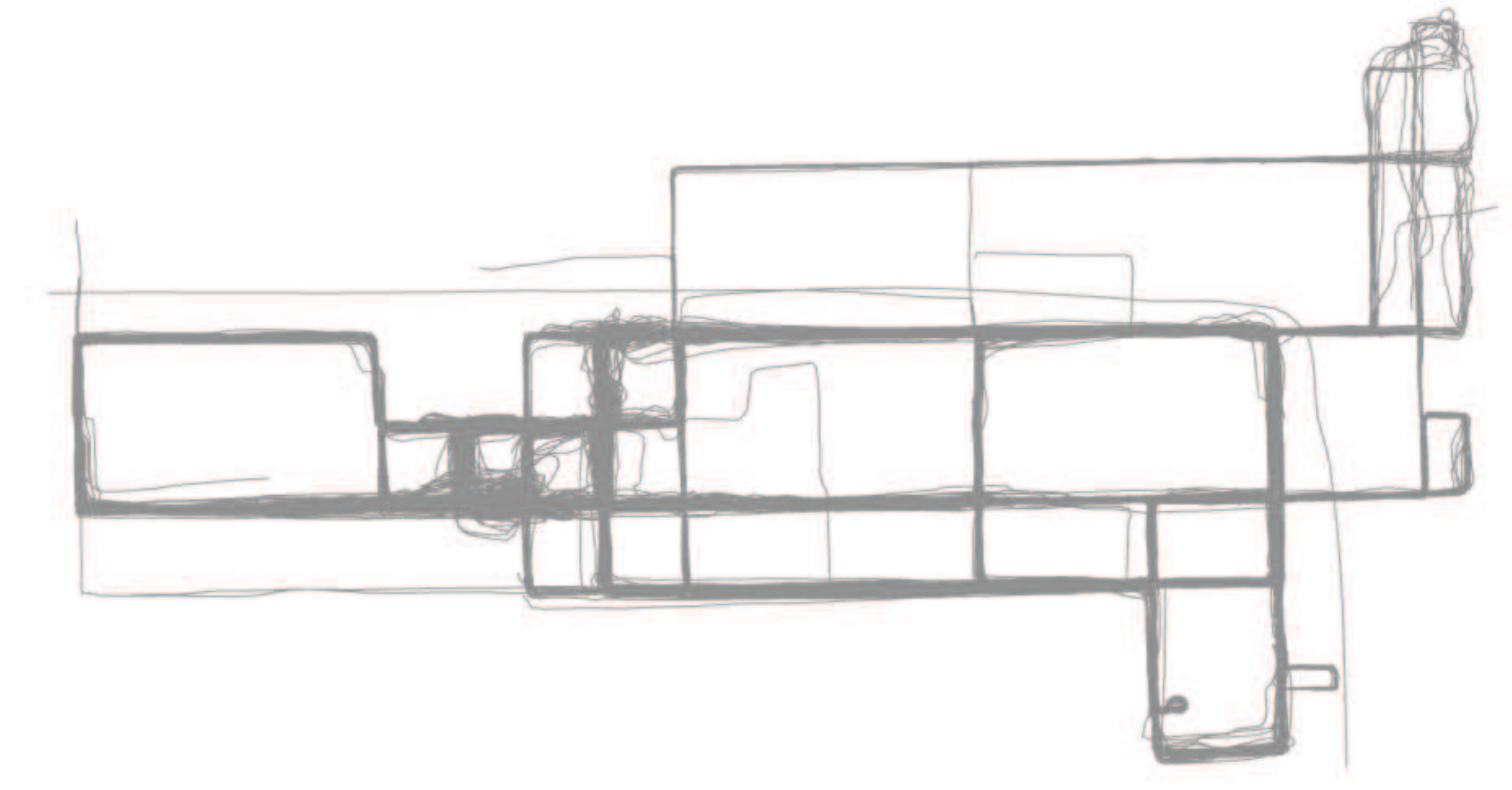}}
 \end{center}
\caption{Tracking data.}
\label{fig:tracking}
\end{figure}

Our experiments use several tracking datasets from different cities (Figure~\ref{fig:tracking}).
While other publicly available GPS-based vehicle tracking datasets exist, e.g., GeoLife \cite{journals/debu/ZhengXM10} and OpenStreetMap GPX track data \cite{osmgps}, 
the selected range covers the various types of existing datasets produced by different types of vehicles, at varying sampling rates and representing different network sizes. 

The \emph{Athens large} dataset consists of 511 trajectories with a total length of 6,781$km$ (average: 13.27$km$ and standard deviation: 10.79$km$) obtained from school buses covering an area of 12$km$ $\times$ 14$km$; the tracks range from 32 to 80 position samples, with a sampling rate of 20s to 30s (average: 30.14$s$ and standard deviation: 24.77$s$) and an average speed of 20.16$km/h$.  
The \emph{Athens small} dataset consists 129 tracks with a total length of 443$km$ (average: 3.82$km$ and standard deviation: 1.45$km$) obtained from school buses covering an area of 2.6$km$ $\times$ 6$km$; the tracks range from 13 to 47 position samples, with a sampling rate of 20$s$ to 30$s$ (average: 34.07$s$ and standard deviation: 31.92$s$) and an average speed of 19.55$km/h$. 
The \emph{Berlin} dataset consists of 26,831 tracks with a total length of 41,116$km$ (average: 1.53$km$ and standard deviation: 634.51$m$) obtained from a taxi fleet covering an area of 6$km$ $\times$ 6$km$; the tracks comprise from 22  up to 58 position samples, with a sampling rate of 15$s$ to 127$s$ (average: 41.98$s$ and standard deviation: 38.70$s$) and an average speed of 35.23$km/h$. 
%
The \emph{Chicago} dataset \cite{be-irmgp-12,Biagioni:2012:MIF:2424321.2424333} consists of 889 tracks with a total length of 2,869$km$ (average: 3.22$km$ and standard deviation: 894.28$m$) obtained from university shuttle buses covering an area of 7$km$ $\times$ 4.5$km$; the tracks range from 100 to 363 position samples, with a sampling rate of 1$s$ to 29$s$ (average: 3.61$s$ and standard deviation: 3.67$s$) and an average speed of 33.14$km/h$. 

For all cases, we consider as ground-truth map data the corresponding OpenStreetMap excerpt. 
%

%
\begin{table}
	[t] \centering \scriptsize 
	\begin{tabular}
		{|c||c|c|c|c|c|} \hline Tracking & Trajec- & Sampling & Trajectory & Speed \\
		Data & tories & rate ($s$) & length ($km$) & ($km$/$h$) \\
		\hline \hline \emph{Athens large} & 120 & 30.14 & 6,781 & 20.16 \\
		\hline \emph{Athens small} & 129  & 34.07 & 443 & 19.55 \\
		\hline \emph{Berlin} & 26,831 & 41.98 & 41,116 & 35.23 \\
		\hline \emph{Chicago} & 889 & 3.61 & 2869 & 33.14 \\
		\hline 
	\end{tabular}
	\\
	\vspace{10pt} 
	\begin{tabular}
		{|c||c|c|c|c|} \hline OSM Network & Vertices & Edges & Length ($km$) & Area ($km^2$) \\
		\hline \hline \emph{Athens large} & 32,212 & 39,699 & 2,000 & 12 $\times$ 14 \\
		\hline \emph{Athens small} & 2,694 & 3,436 & 193 & 2.6 $\times$ 6 \\
		\hline \emph{Berlin} & 5,894 & 6,839 & 360 & 6 $\times$ 6 \\
		\hline \emph{Chicago} & 9,429 & 11,801 & 61 & 7 $\times$ 4.5 \\
		\hline 
	\end{tabular}
	\caption{Statistics for datasets used.} \label{tab:dstatistics}  
\end{table}

\section{Experiments}
\label{sec:sec_exp}

What follows is a description of the map construction experiments that were conducted for the range of algorithms, datasets and evaluation measures, with the scope to \emph{assess the quality of the constructed maps}. 
The seven algorithms used in this experimentation are implemented in C, Java, Python and Matlab. The experiments for six algorithms have been performed by the authors and the implementations have been made available at the {\tt mapconstruction.org} web site. The authors of \cite{DBLP:conf/nips/GeSBW11} performed the experiments themselves, since we did not have access to their implementation. 
Given the implementations, (i) their difference in code base, (ii) their scope, i.e., to construct small-scale maps from GPS trajectories, and (iii) their quality, i.e., all are academic prototypes, we did not assess the characteristics of the algorithms themselves by means of, e.g., a performance study or theoretical analysis. However, to at least give an impression of their running times, for the \emph{Chicago} dataset the running times of the algorithms range from 10$min$ to 20$h$. For the larger \emph{Berlin} dataset, the running times range from 2$h$ to 4$days$.
Given the quality of the implementations, another problem we encountered was that some algorithms could not cope with the size of the input dataset (trajectories) resulting in runtime crashes. Hence, not all algorithms could be tested on the large datasets 
and results for all algorithms are only available for the smaller datasets, i.e., \emph{Athens small} and \emph{Chicago}.

\subsection{Constructed Maps}
\label{sub:constructed_maps}

\begin{figure}[htbp]
 \begin{center}
	 \subfloat[Ahmed - \emph{Chicago}]{\includegraphics[width=0.47\columnwidth]{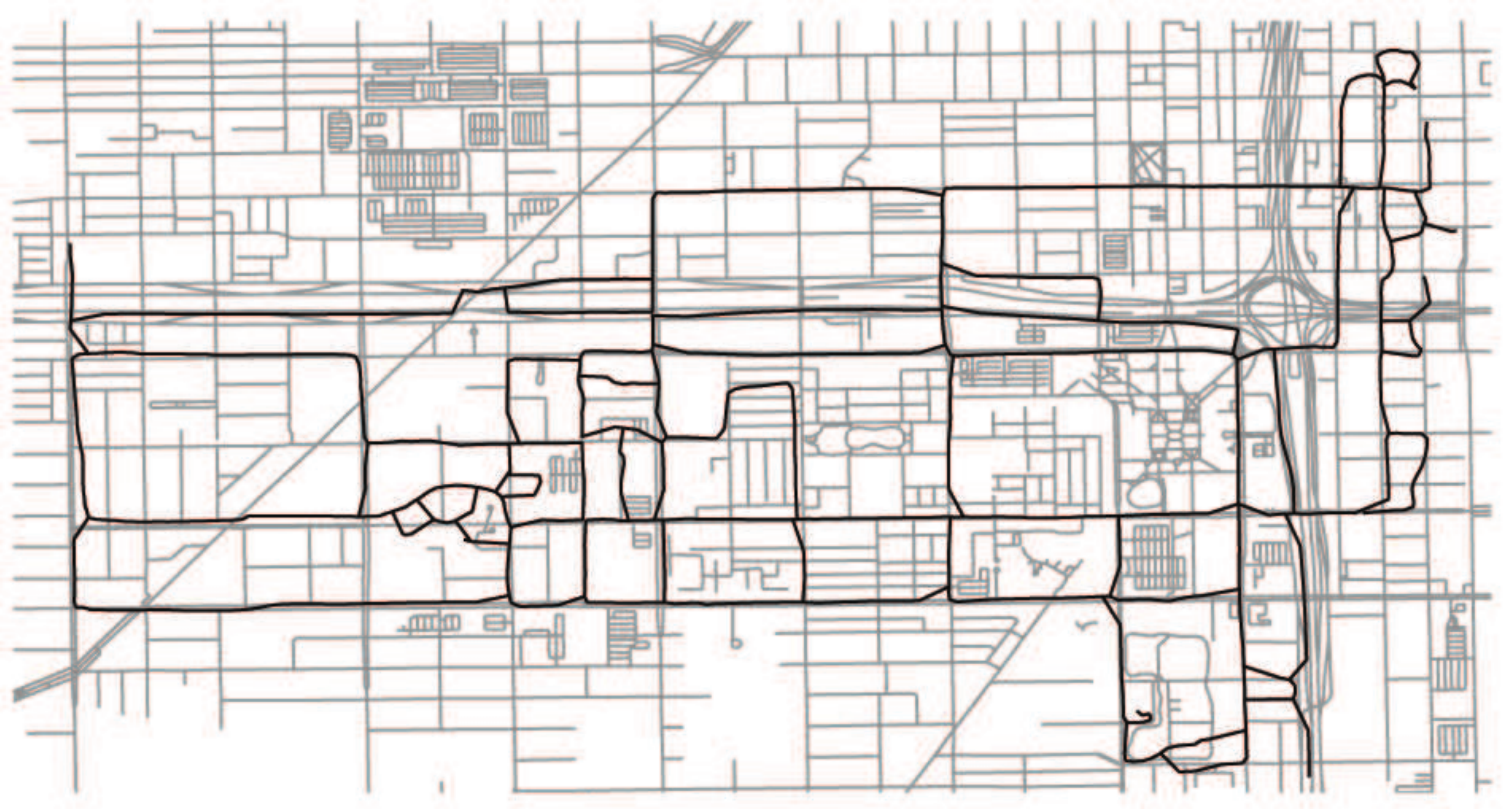}}
	 \subfloat[Biagioni - \emph{Chicago}]{\includegraphics[width=0.47\columnwidth]{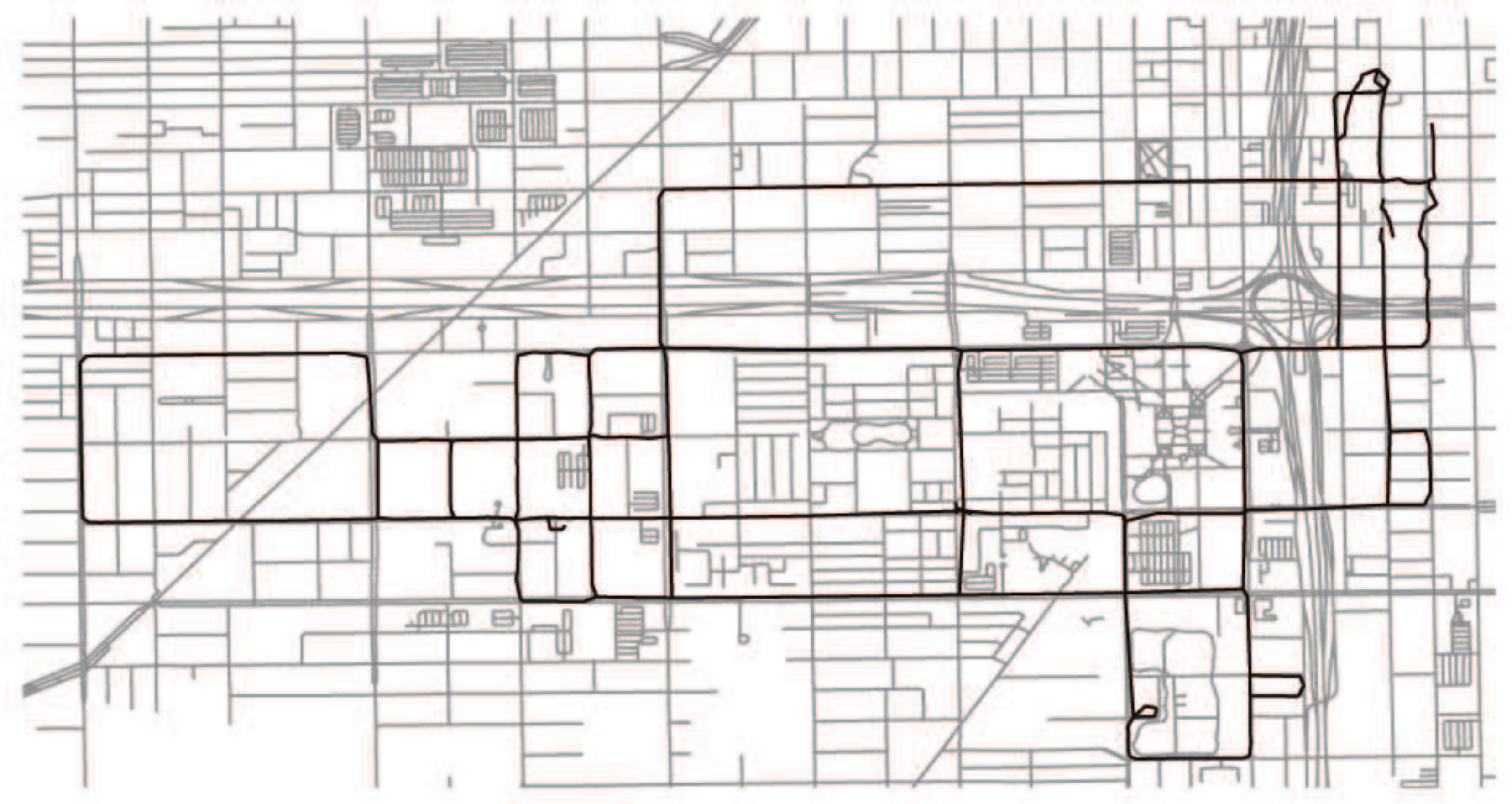}}
	 \\
	 \subfloat[Cao - \emph{Chicago}]{\includegraphics[width=0.47\columnwidth]{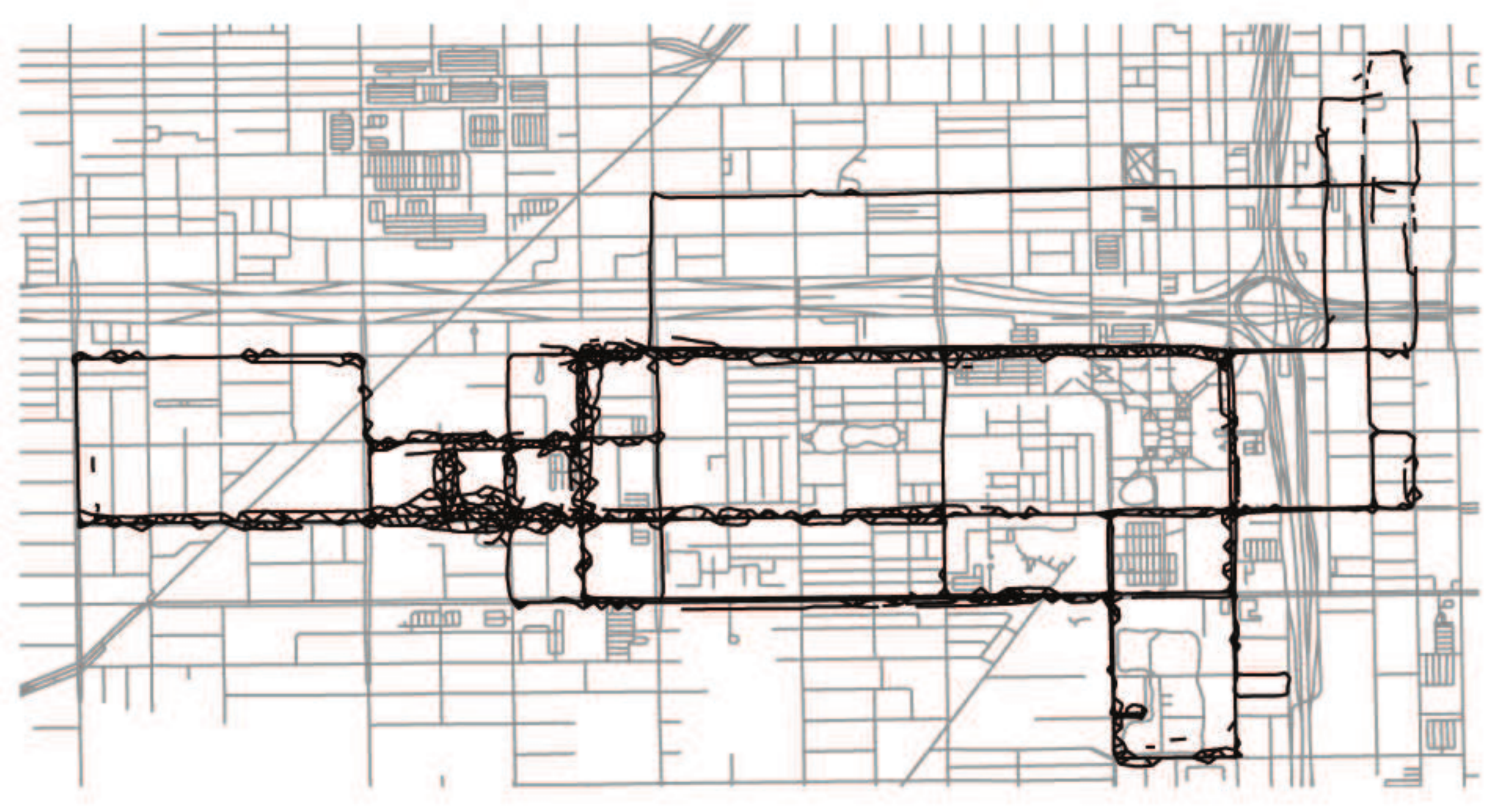}}
	 \subfloat[Davies - \emph{Chicago}]{\includegraphics[width=0.47\columnwidth]{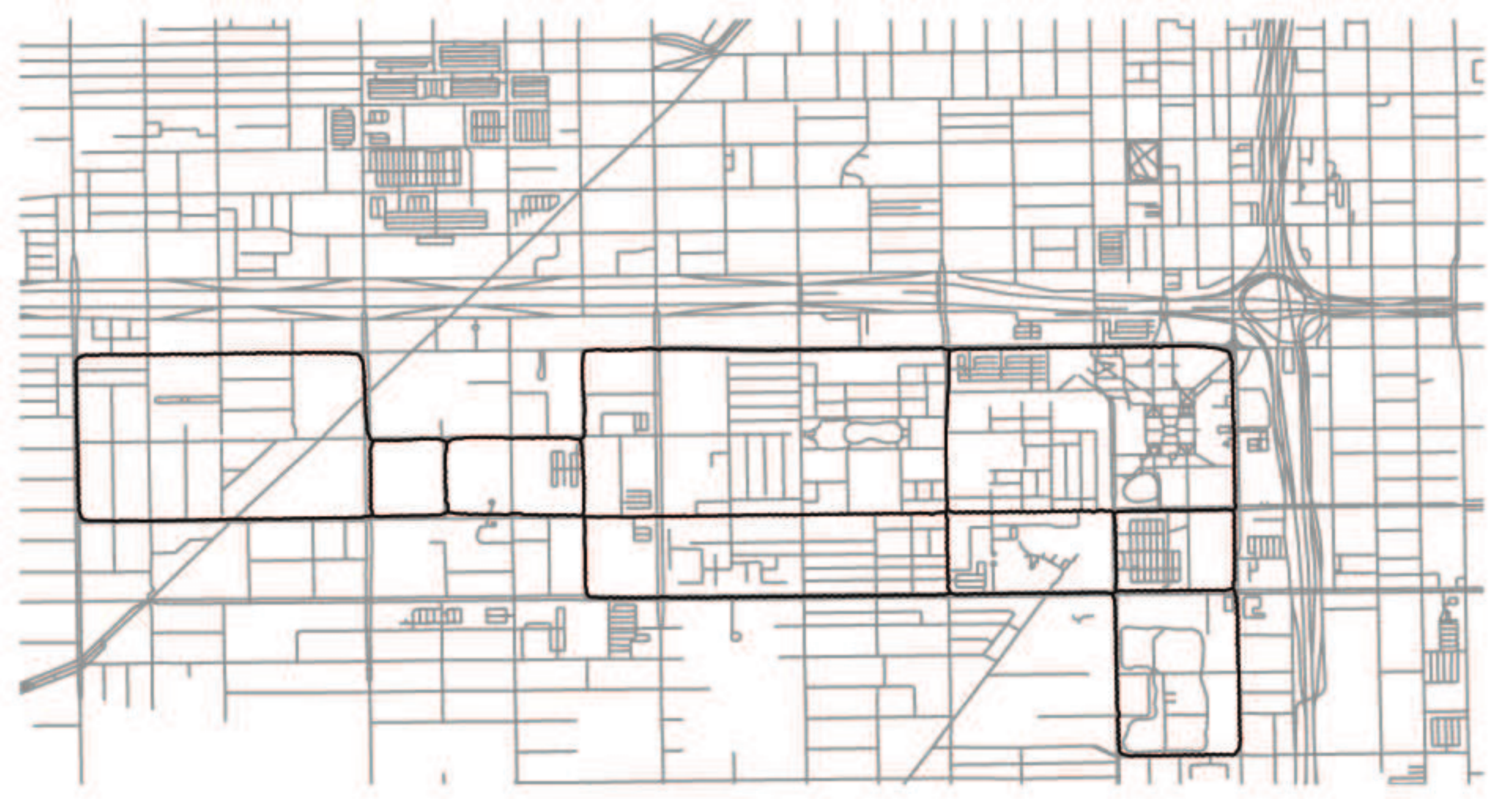}}
\\	 
	 \subfloat[Edelkamp - \emph{Chicago}]{\includegraphics[width=0.47\columnwidth]{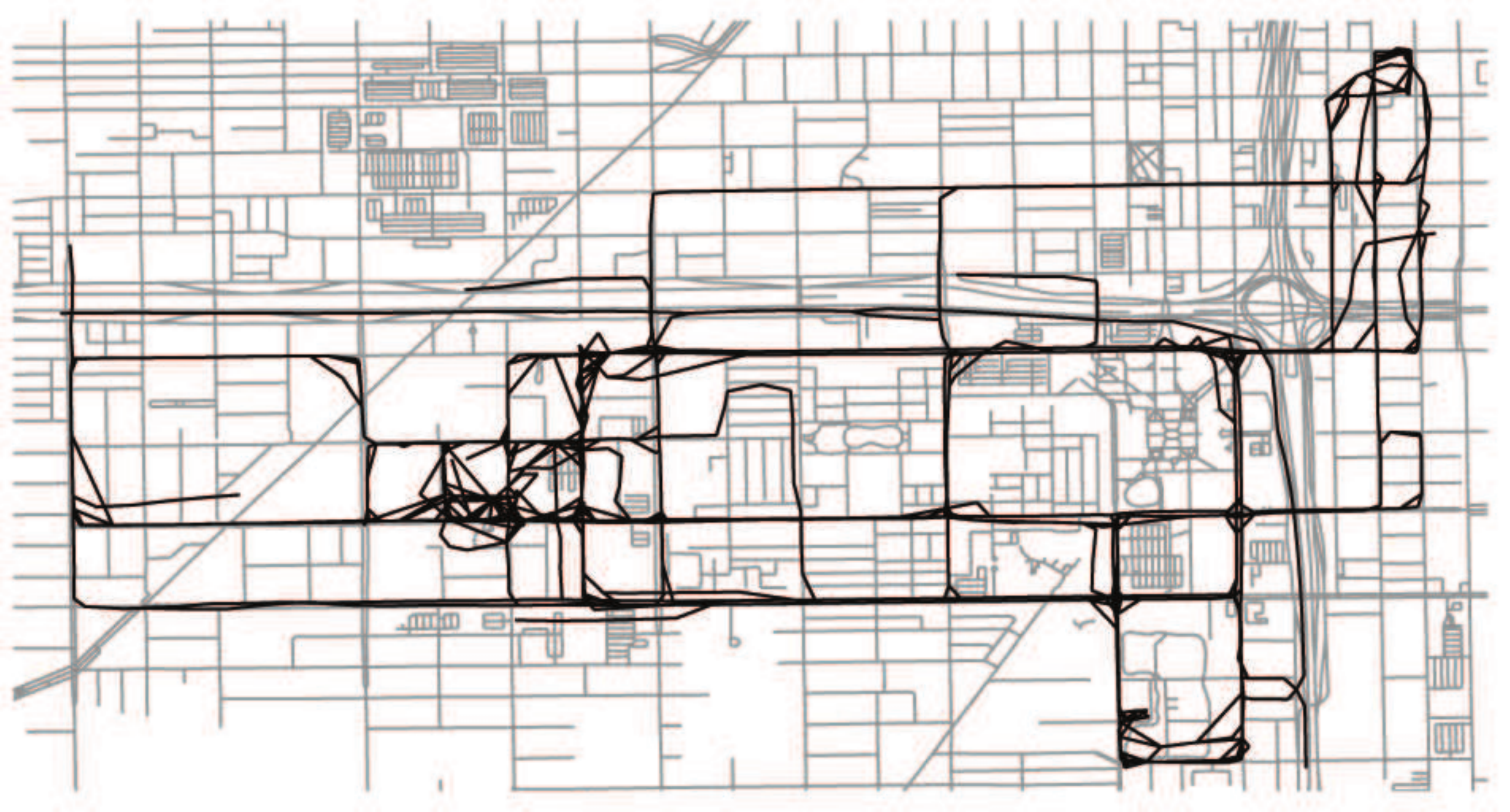}}
	 \subfloat[Ge - \emph{Chicago}]{\includegraphics[width=0.47\columnwidth]{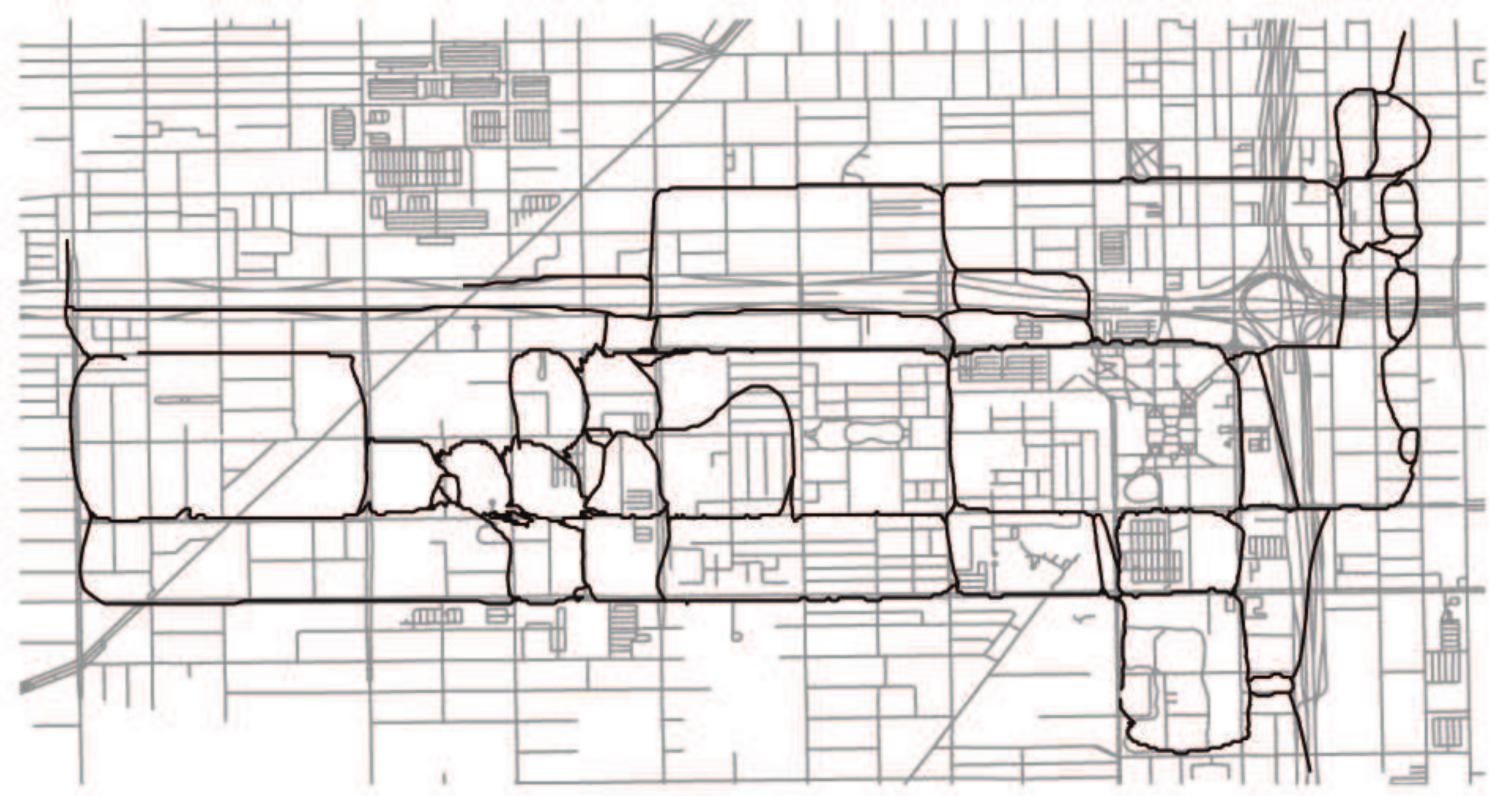}}	 
	 
	 \subfloat[Karagiorgou - \emph{Chicago}]{\includegraphics[width=0.47\columnwidth]{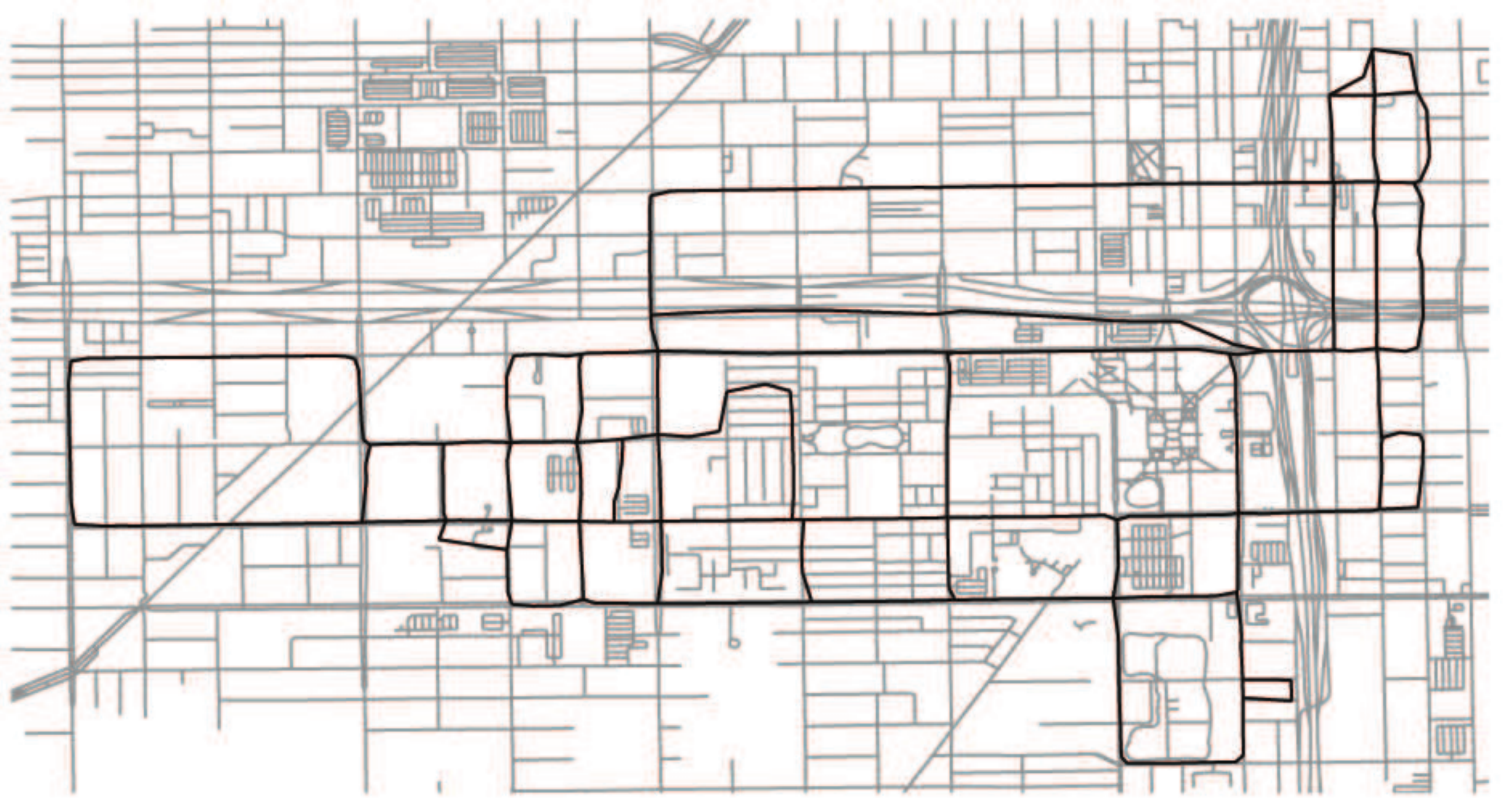}} 
	 \end{center}
\caption{Constructed maps (in black) overlayed on ground-truth map (in gray) (small dataset).}
\label{fig:rn_alls}
\end{figure}

\begin{figure}[htbp]
 \begin{center}	
	 \subfloat[Ahmed - \emph{Berlin}]{\includegraphics[width=0.47\columnwidth]{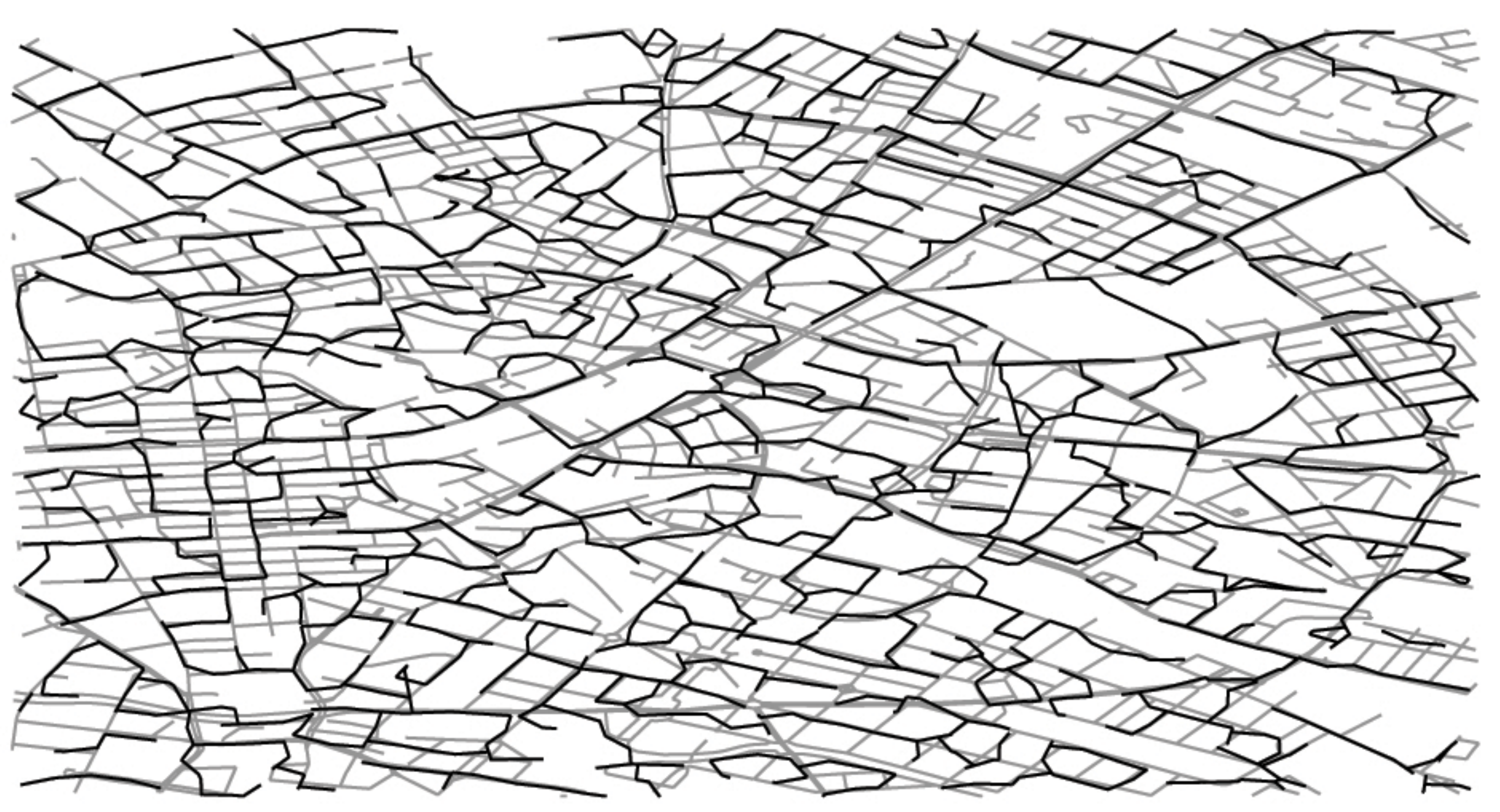}}
	 \subfloat[Ge - \emph{Berlin}]{\includegraphics[width=0.47\columnwidth]{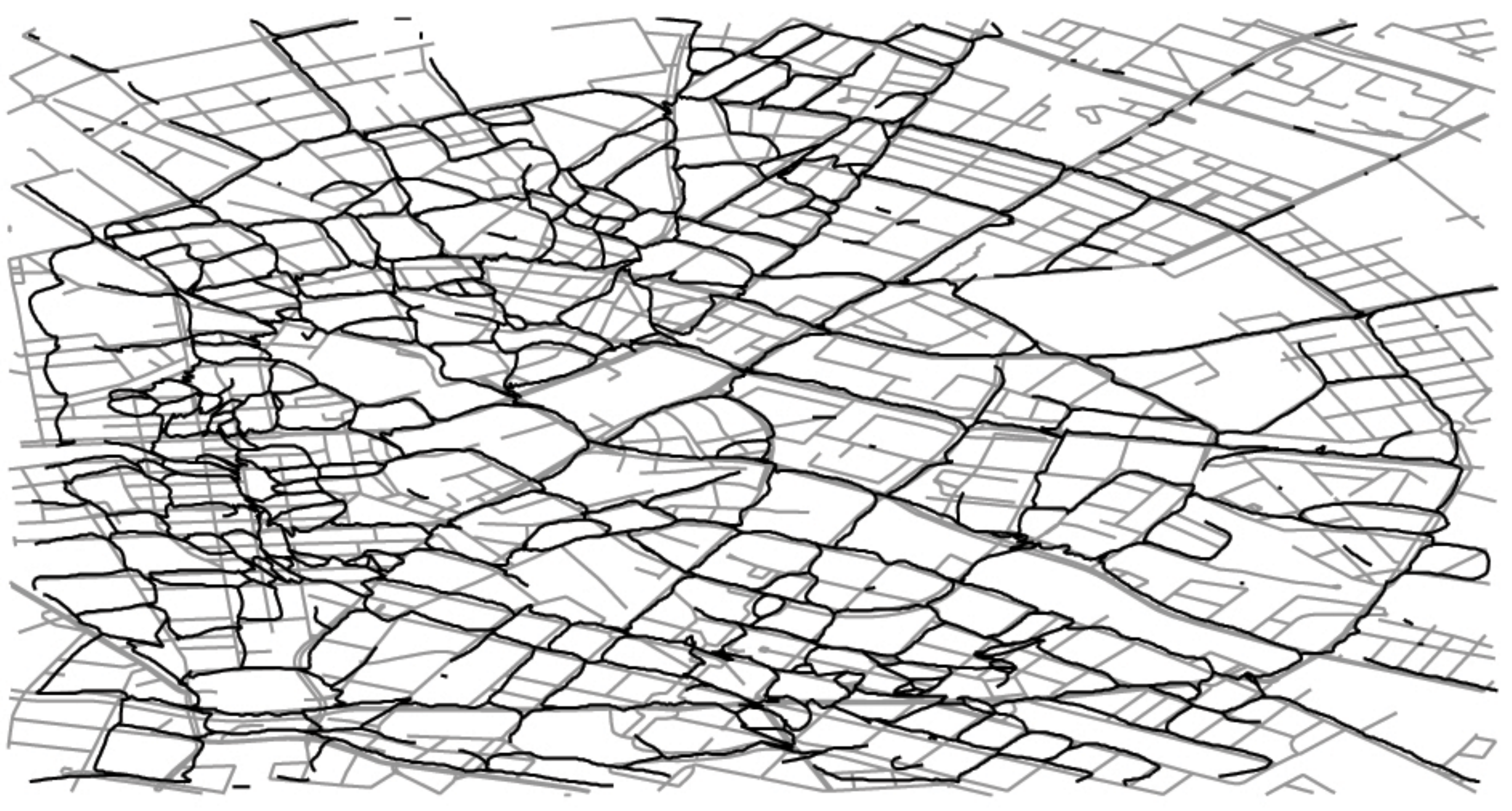}}
	 \\
	 \subfloat[Karagiorgou - \emph{Berlin}]{\includegraphics[width=0.47\columnwidth]{berlin_sophia_rn.pdf}} 
	 \end{center}
\caption{Constructed maps (in black) overlayed on ground-truth map (in gray)  (large dataset).}
\label{fig:rn_allb}
\end{figure}

What follows is an initial overview of the experimentation in terms of constructed maps and the respective result quality.
Figure~\ref{fig:rn_alls} illustrates the ground-truth map (light gray) and the generated maps (black) for the small \emph{Chicago} dataset. On larger datasets, i.e., \emph{Athens large} and \emph{Berlin}, we ran the algorithms described in Subsections~\ref{subsub:subsub_aw},~\ref{subsub:subsub_ge} and ~\ref{subsub:subsub_kp}. Figure~\ref{fig:rn_allb} illustrates the ground-truth map (light gray) and the generated maps (black) for the case of the larger \emph{Berlin} dataset.

Each of the algorithms uses different parameter settings.
For Ahmed and Wenk \cite{csm_esa2012} the values of $\varepsilon$ to cluster subtrajectories are: $180$, $90$, $170$ and $80$ meters for \emph{Athens large}, \emph{Athens small}, \emph{Berlin} and \emph{Chicago}, respectively. 
The respective parameters of \emph{proximity} and \emph{bearing} for the other algorithms are Biagioni 50$m$ \cite{Biagioni:2012:MIF:2424321.2424333}, Cao 20$m$ and $45^{\circ}$ \cite{Cao:2009:GTR:1653771.1653776}, Davies 16$m$ \cite{Davies:2006:SDR:1175887.1176088} and Edelkamp 50$m$ and $45^{\circ}$ \cite{edelkamp:2003:rpmi}. 
For Karagiorgou and Pfoser \cite{Karagiorgou:2012:VTD:2424321.2424334} the values of \emph{direction}, \emph{speed} and \emph{proximity} to extract intersection nodes and to merge trajectories into links are $15^{\circ}$, 40km/h and 25$m$ accordingly. We evaluated all constructed maps using the distance measures described in Subsection~\ref{subsec:subsec_qmu}.

\begin{table}[htbp]\scriptsize
\centering
\begin{tabular}{|c||c c c|}
\hline
Generated&&&\\
Map&\# Vertices&\# Edges& Length (km)\\
\hline
\hline
\emph{Athens large}&&&\\
\hline
Ahmed&7067&7960&1358\\
Ge&20774&21626&9740\\
Karagiorgou&6584&5280&252\\
\hline
\emph{Athens small}&&&\\
\hline
Ahmed&344&378&35\\
Biagioni&391&398&22\\
Cao&20&14&3\\
Davies&209&227&2\\
Edelkamp&526&1037&197\\
Ge&1936&1993&23\\
Karagiorgou&660&637&35\\
\hline
\emph{Berlin}&&&\\
\hline
Ahmed&1322&1567&164\\
Ge&15450&16136&183\\
Karagiorgou&2542&2262&161\\
\hline
\emph{Chicago}&&&\\
\hline
Ahmed&1195&1286&34\\
Biagioni&303&322&24\\
Cao&2092&2948&78\\
Davies&1277&1310&14\\
Edelkamp&828&1247&83\\
Ge&5893&6672&37\\
Karagiorgou&596&558&26\\
\hline
\end{tabular}
\caption{Complexities of the generated maps.}
\label{tab:tab_stat}
\end{table}

A summary of the complexities of the constructed maps is shown in Table \ref{tab:tab_stat}. Here, the number of vertices includes vertices of degree two (which may lie on a polygonal curve describing a single edge), the number of edges refers to the number of undirected line segments between these vertices, and the total length refers to the total length of all undirected line segments.
It appears that the \emph{point clustering algorithms based on kernel density estimation} such as Biagioni et al.\ \cite{be-irmgp-12,Biagioni:2012:MIF:2424321.2424333} and Davies et al.\ \cite{Davies:2006:SDR:1175887.1176088} produce maps with lower complexity (fewer number of vertices and edges) but often \emph{fail to reconstruct streets that are not traversed frequently enough} by the input tracks. 
In particular, the maps reconstructed by Davies et al.'s algorithm are very small. 
On the other hand, the algorithm by Ge et al.\ \cite{DBLP:conf/nips/GeSBW11} subsample all tracks to create a much denser output set, hence the complexity of their constructed maps is always higher.

Map construction algorithms based on \emph{incremental track insertion}, such as Ahmed et al.\ \cite{csm_esa2012} and Cao et al.\ \cite{Cao:2009:GTR:1653771.1653776} fail to cluster tracks together when the variability and error associated with the input tracks is large. As a result, the constructed street maps contain \emph{multiple edges for a single street}, which implies larger values in the total edge length column in Table \ref{tab:tab_stat}.

Several examples of generated maps are shown in Figure~\ref{fig:rn_alls} and Figure~\ref{fig:rn_allb}. Since not all algorithms produced results for all maps, we showcase examples of the smaller \emph{Chicago} map in Figure~\ref{fig:rn_alls}. It can be clearly seen that the coverage and quality of the constructed map varies considerably. Three examples for the \emph{Berlin} map are also given in Figure~\ref{fig:rn_allb}. More examples can be found on the {\tt mapconstruction.org} web site.

\subsection{Path-Based and Hausdorff Distance} 
\label{sub:distance}

For the \emph{path-based distance measure} we generated all paths of link-length $3$ for each generated map. For each path, we computed the \Frd\ between the path and the ground-truth map. We then computed the minimum, maximum, median, average of all the obtained distances. We also computed the $d\%$-distance, as the maximum of the distances after removing the $d\%$ largest distances (``outliers'').
For the \emph{Directed Hausdorff distance}, we computed all link-length $1$ paths and computed the Directed Hausdorff distance of the union of all edges to the ground-truth map. Our results are summarized in Table \ref{tab:tab_pbvsdh}. 
In the case of \emph{Athens small}, the Cao algorithm produced a very small map and thus it was not possible to perform a quantitative evaluation. 

The maps reconstructed using the algorithms by Karagiorgou et al. \cite{Karagiorgou:2012:VTD:2424321.2424334} and by  Biagioni et al.\ \cite{be-irmgp-12,Biagioni:2012:MIF:2424321.2424333} generally have a better path-based distance than the others. Note that Davies et al.'s \cite{Davies:2006:SDR:1175887.1176088} map is unusually small for the \emph{Athens small} dataset. Their idea of averaging trajectories, or computing skeletons, however, seems to help to improve the quality of the edges of the produced map. 

\begin{table}[ht]\scriptsize
\centering
\scalebox{0.7}{
\begin{tabular}{|c||c c c c c c c c|c c c c c c c c|}
\hline
Generated&&&&&&&&&&&&&&&&\\
Map&\multicolumn{8}{|c|}{Path based distance (m)}&\multicolumn{8}{c|}{Directed Hausdorff distance (m)}\\
\hline
\hline
\emph{Athens large}&min&max&median&avg&$2\%$&$5\%$&$10\%$&$15\%$&min&max&median&avg&$2\%$&$5\%$&$10\%$&$15\%$\\
\hline
Ahmed&7&849&70&85&250&164&132&114&1&269&30&33&84&67&56&50\\
Ge&7&956&76&90&237&188&150&116&1&295&35&37&95&74&59&52\\
Karagiorgou&2&175&25&32&109&80&63&53&1&200&10&13&46&35&26&22\\
\hline
\emph{Athens small}&min&max&median&avg&$2\%$&$5\%$&$10\%$&$15\%$&min&max&median&avg&$2\%$&$5\%$&$10\%$&$15\%$\\
\hline
Ahmed&9&224&45&52&101&101&81&72&1&82&25&26&82&54&46&40\\
Biagioni&5&73&35&36&67&66&61&57&3&74&19&20&47&43&31&31\\
Cao&&&&&&&&&5&25&13&13&25&25&25&22\\
Davies&4&38&11&11&38&18&14&14&2&13&7&6&13&13&13&11\\
Edelkamp&2&229&36&39&89&72&68&61&1&86&18&21&63&50&42&37\\
Ge&19&251&52&59&142&113&89&76&3&81&21&23&80&59&39&35\\
Karagiorgou&7&229&32&38&113&68&59&57&2&84&14&17&54&40&33&30\\
\hline
\emph{Berlin}&min&max&median&avg&$2\%$&$5\%$&$10\%$&$15\%$&min&max&median&avg&$2\%$&$5\%$&$10\%$&$15\%$\\
\hline
Ahmed&9&540&66&74&207&147&120&107&1&219&30&33&95&70&60&53\\
Ge&13&808&65&75&214&157&117&103&4&562&36&37&73&62&55&51\\
Karagiorgou&4&306&28&37&120&85&65&52&1&232&14&18&59&42&34&30\\
\hline
\emph{Chicago}&min&max&median&avg&$2\%$&$5\%$&$10\%$&$15\%$&min&max&median&avg&$2\%$&$5\%$&$10\%$&$15\%$\\
\hline
Ahmed&7&201&35&42&127&100&85&76&1&81&14&19&72&59&43&35\\
\cellcolor{gray!25}Biagioni&3&\cellcolor{gray!25}71&15&18&71&\cellcolor{gray!25}38&27&26&2&\cellcolor{gray!25}53&9&11&29&\cellcolor{gray!25}25&23&17\\
Cao&1&126&24&27&79&61&49&42&1&78&9&12&44&35&28&25\\
\cellcolor{gray!25}Davies&2&92&12&14&57&24&22&21&2&20&8&7&20&14&13&12\\
Edelkamp&1&205&29&37&99&84&72&66&1&93&8&13&57&48&35&25\\
Ge&18&346&50&56&158&126&95&75&7&72&26&28&64&61&53&46\\
\cellcolor{gray!25}Karagiorgou&3&89&15&23&72&72&65&51&1&48&7&8&41&23&15&13\\
\hline
\end{tabular}
}
\caption{Path-Based and Directed Hausdorff distance measure evaluation.}
\label{tab:tab_pbvsdh}
\end{table}

For further analysis of the results, we selected the \emph{Chicago} dataset as all map construction algorithms produced results for it.
From Table \ref{tab:tab_pbvsdh} one can see that the path-based distance and the Directed Hausdorff distance are smaller for the generated maps by Biagioni, Davies and Karagiorgou (shaded gray) compared to map generated using other algorithms. A visual inspection of the maps in Figure \ref{fig:rn_alls} justifies the result. Note that Davies et al.'s \cite{Davies:2006:SDR:1175887.1176088} map is comparatively smaller than the other, see Table \ref{tab:tab_stat}.
Although the algorithms by Ahmed et al.\ and by Ge et al.\ produce maps with good coverage, their path-based distances are larger since they employ less aggressive averaging techniques that would help cope with noise in the input tracks.

\begin{figure}[htbp]
\begin{center} 	
\includegraphics[width=0.6\columnwidth]{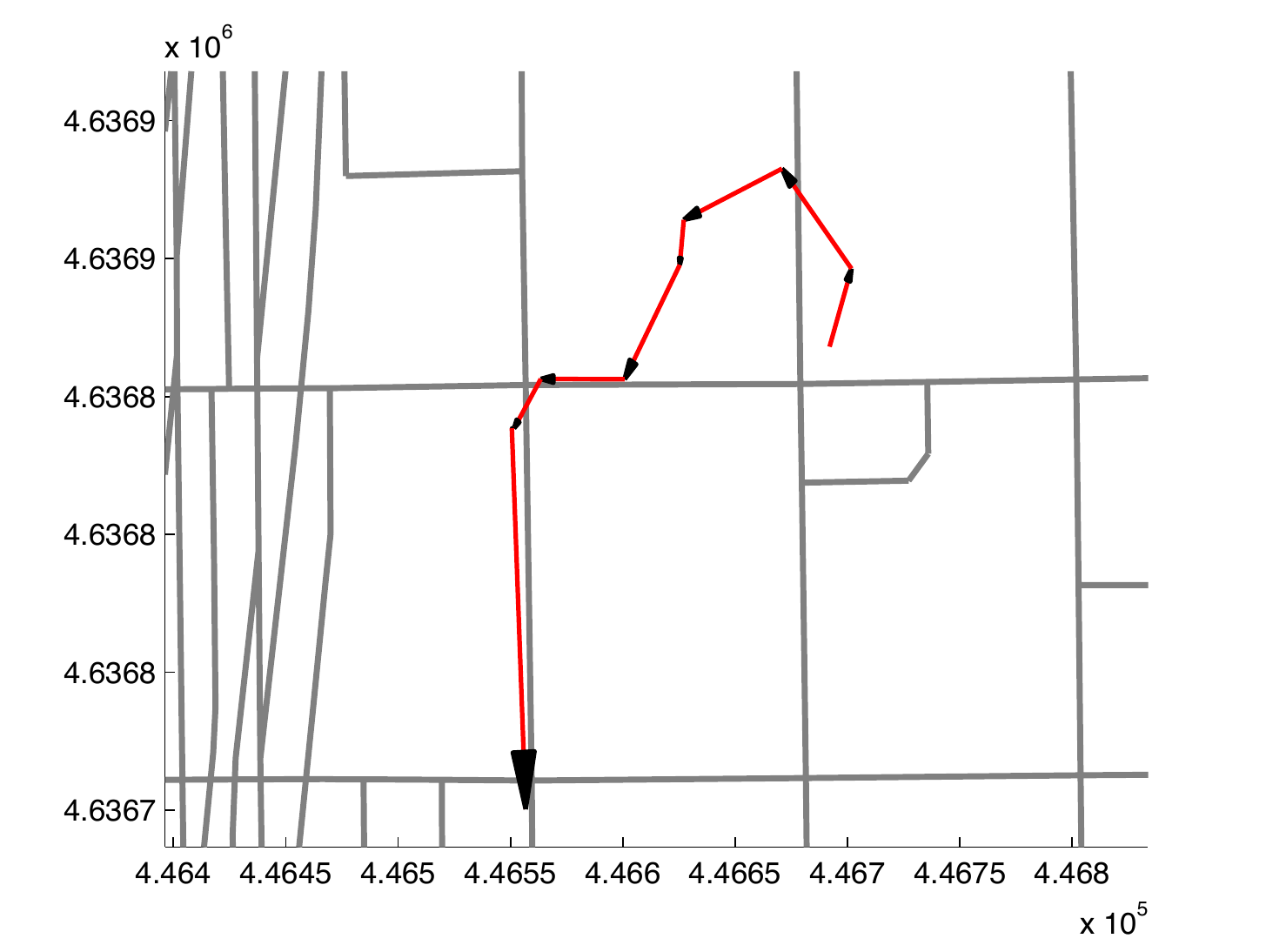}
\caption{A path with \Frd\ greater than Hausdorff distance.}
\label{fig:james_larger_fr}
\end{center}
\vspace*{-2ex}
\end{figure}

To illustrate the appropriateness of the path-based distance, consider the path in Figure~\ref{fig:james_larger_fr} from the map generated by Biagioni et al. This is an example where the Fr\'echet-based distance measure is more effective than any point-based measure. As \Frd\ ensures continuous mapping, the whole path needs to be matched with the bottom horizontal edge of the ground-truth map. The \Frd\ for this path is $71m$. For the same path, the Hausdorff distance is $53m$, as this only requires for each point on the path to have a point on the graph close-by. So, to evaluate the connectivity of a map, the \Frd\ is a more suitable distance measure than any point-based measure.

\begin{figure}[htbp]
\begin{center}
 	\subfloat[Path-Based Distance]{\label{subfig:james_eval_fr}\includegraphics[width=0.50\columnwidth]{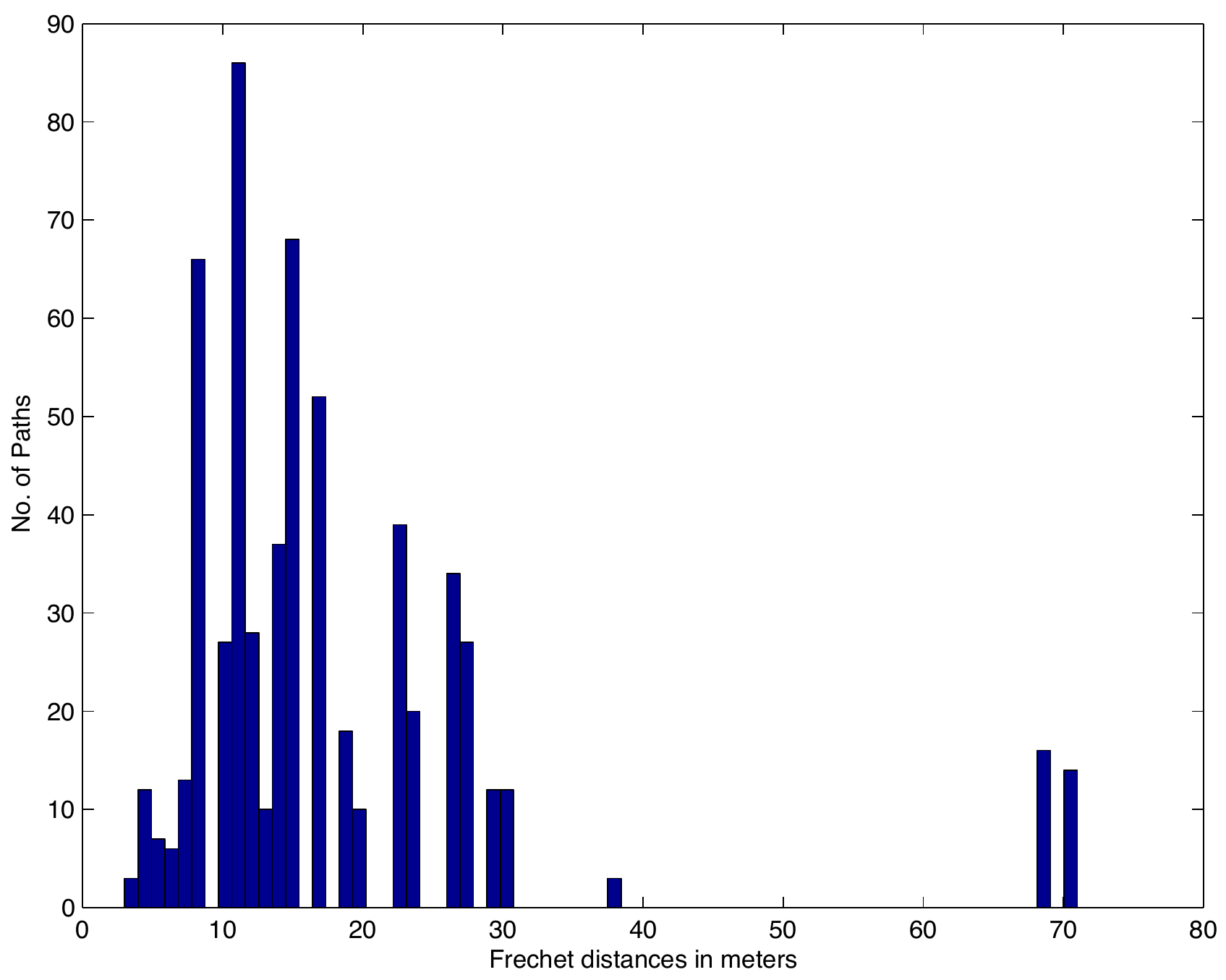}}
 	\subfloat[Directed Hausdorff]{\label{subfig:james_eval_hausdorff}\includegraphics[width=0.50\columnwidth]{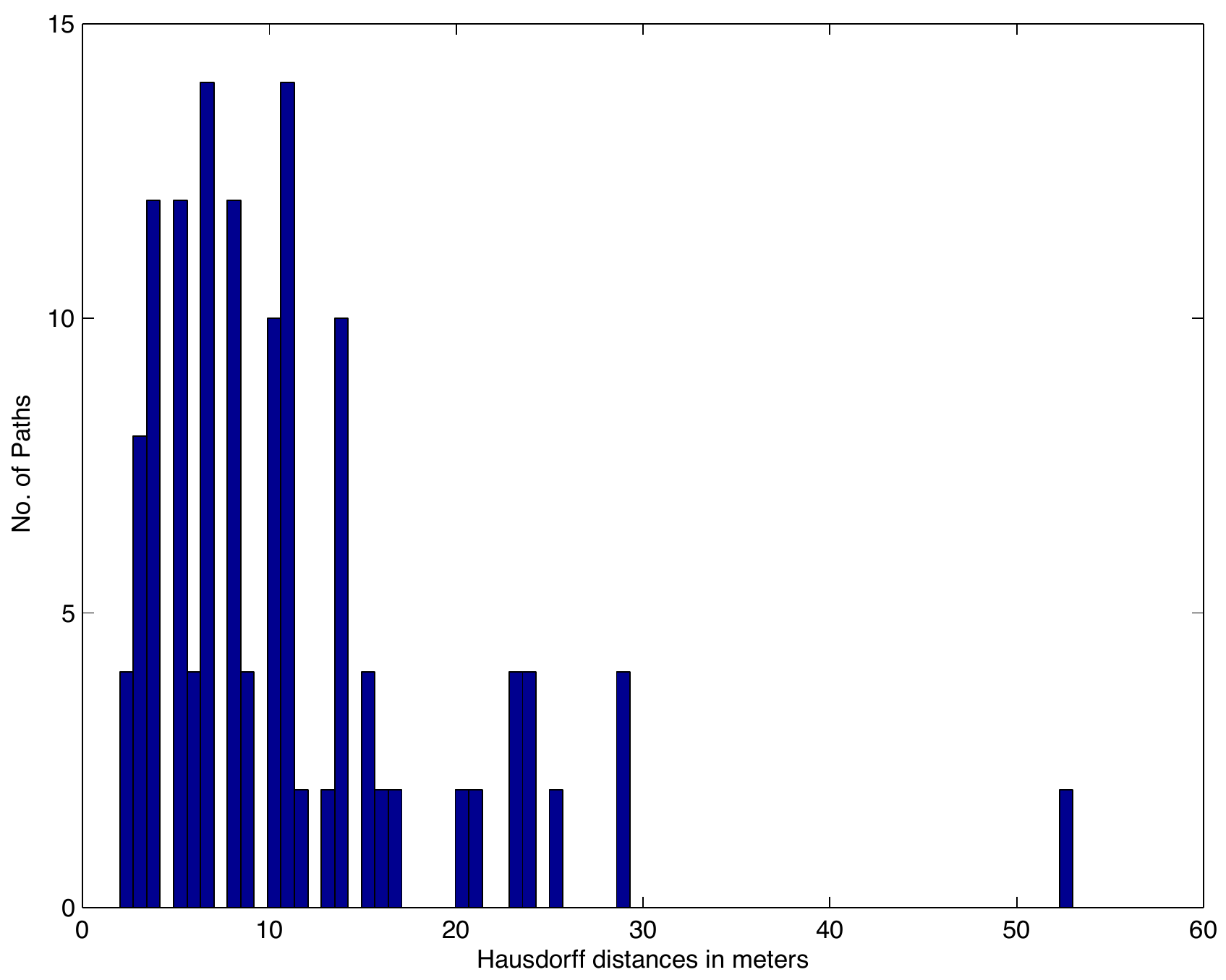}}
 \end{center}
 \caption{Distributions of individual path distances (Biagioni alg. - \emph{Chicago}).}
 \label{fig:distribution}
\vspace*{-2ex}
\end{figure}

\begin{figure}[htbp]
\begin{center}
\subfloat[Biagioni et al. 
Edges in lighter shades indicate smaller distances ($3m$ being the smallest) and darker shades indicate larger distances($71m$ being the largest).
]{\includegraphics[width=0.6\columnwidth]{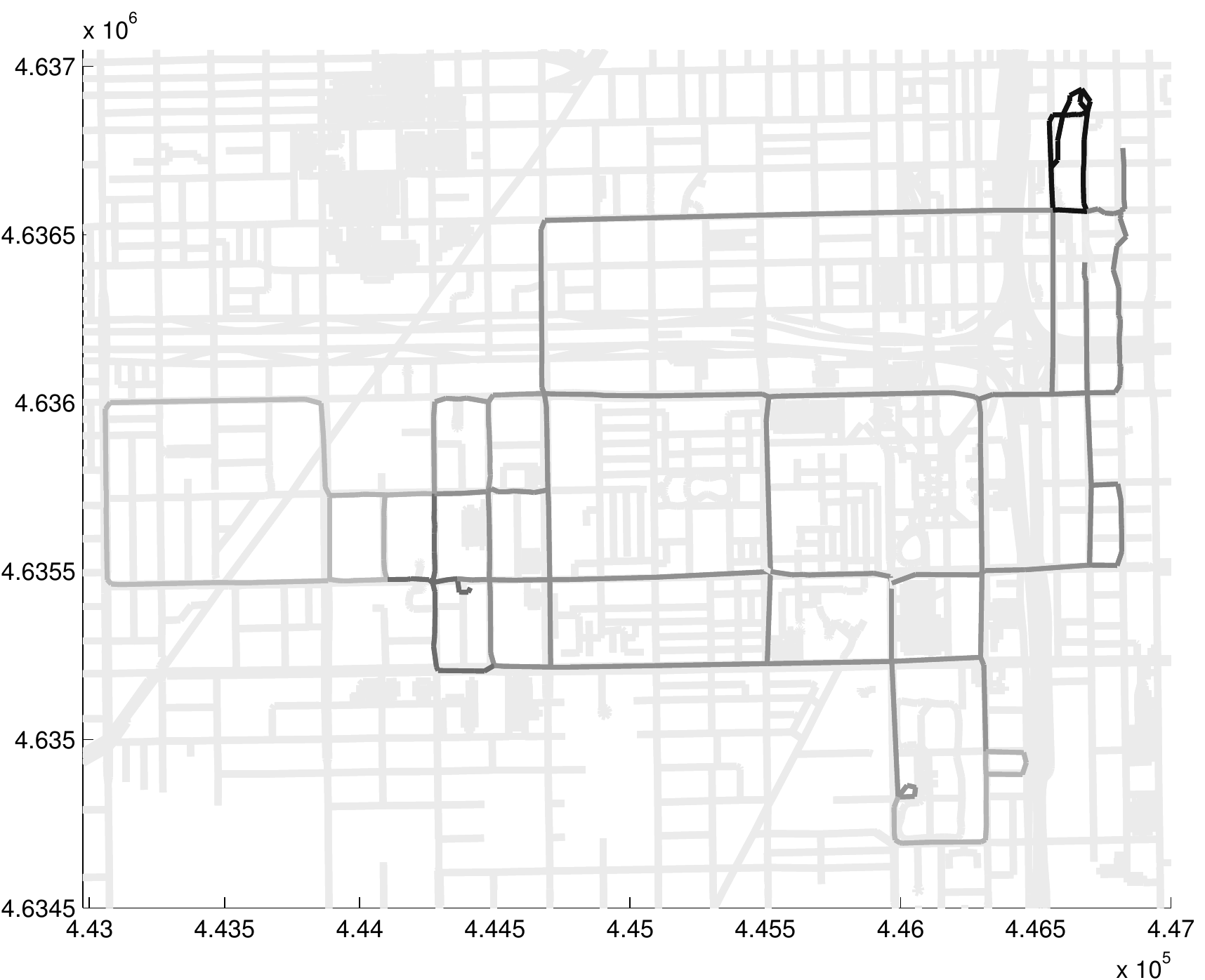}} 

\subfloat[Ahmed et al. 
Edges in lighter shades indicate smaller distances ($7m$ being the smallest) and darker shades indicate larger distances($201m$ being the largest).
]{
\includegraphics[width=0.6\columnwidth]{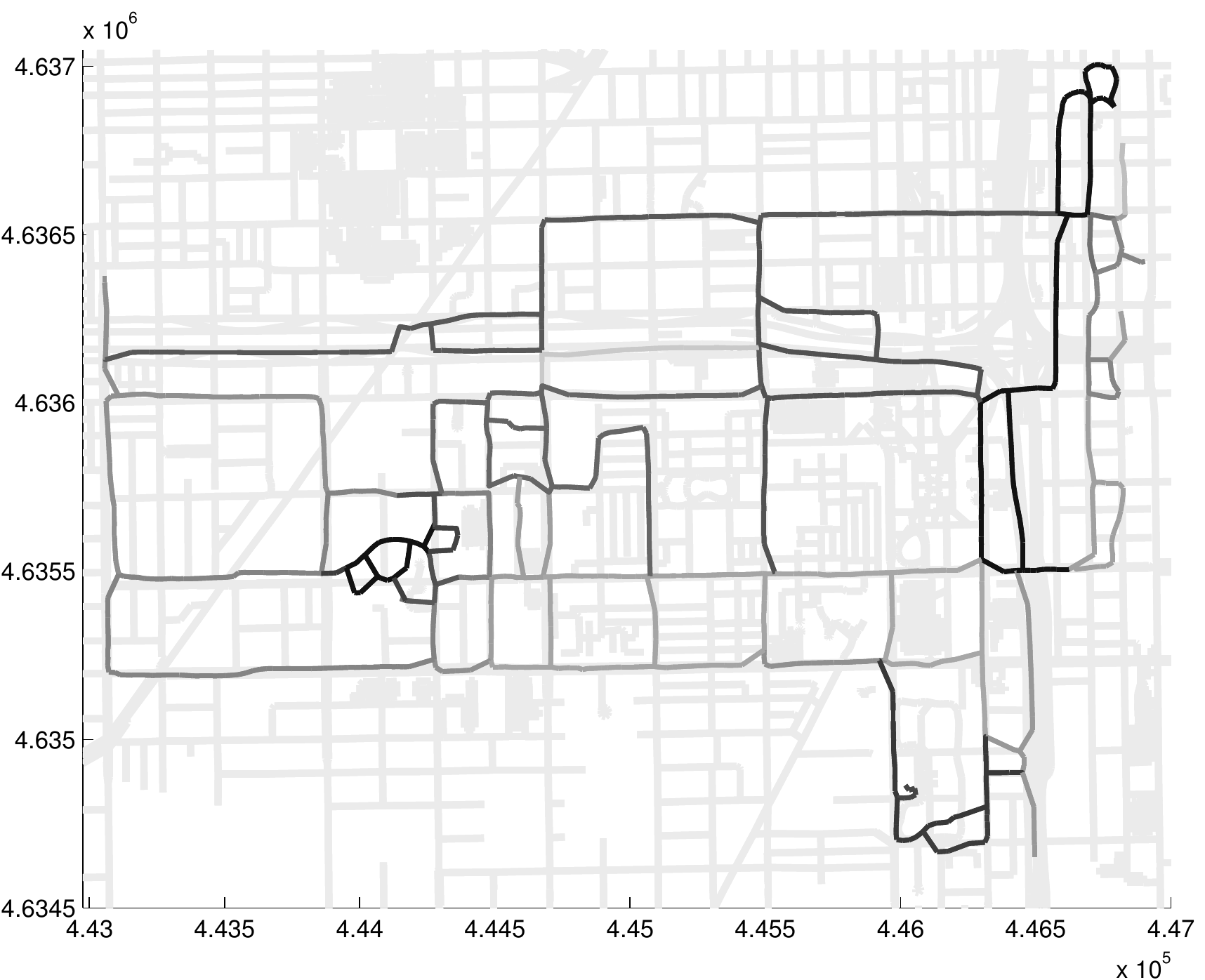}}
\end{center}
\vspace*{-2ex}
\caption{Reconstructed graph overlayed on ground-truth map (light gray). Based on link-length $3$ paths, edges in lighter shades has smaller distance and darker shades has larger distance.}
\label{fig:rmapjames}
\end{figure}

In addition, if desired one can discard outliers by computing the $d\%$-distance. Figure~\ref{fig:distribution} shows the distribution of both the path-based measure and the Directed Hausdorff distance for Biagioni et al. In both cases, a very small number of paths have the maximum distance, and the distances for most of the paths are distributed within a small range. Removing only $5\%$ of the outliers (largest) brings the path-based distance from $71m$ (max) to $38m$ and the Directed Hausdorff distance from $53m$ (max) to $25m$. 
Figure~\ref{fig:rmapjames} shows edges of maps with smaller distances in lighter shades and larger distances in darker shades.
Such visual representation helps to identify areas in the map that have higher distance to the ground-truth map.

\begin{figure}[bthp]
\begin{center}
\subfloat[Ahmed]{\includegraphics[width=0.48\columnwidth]{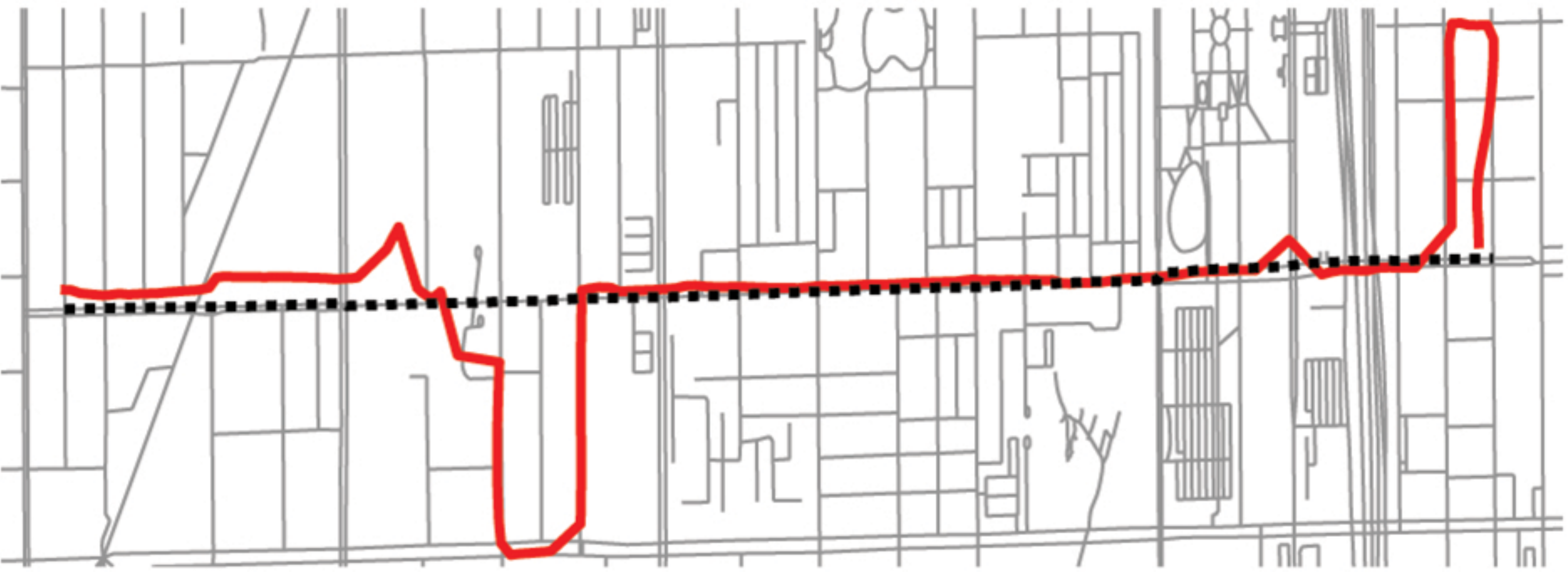}
\label{subfig:ahmed_rn_sp_all}}
\hspace{0.05cm}
\subfloat[Biagioni]{\includegraphics[width=0.48\columnwidth]{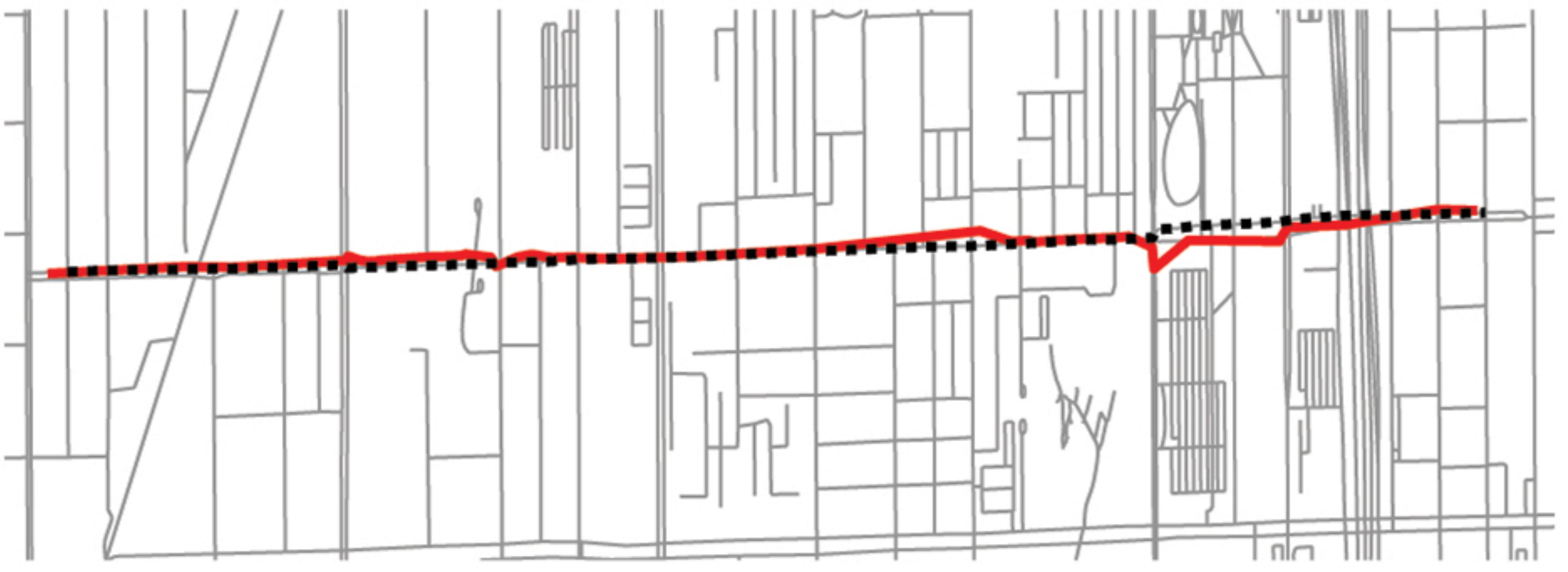}
\label{subfig:bag_rn_sp_all}}
\subfloat[Cao]{\includegraphics[width=0.48\columnwidth]{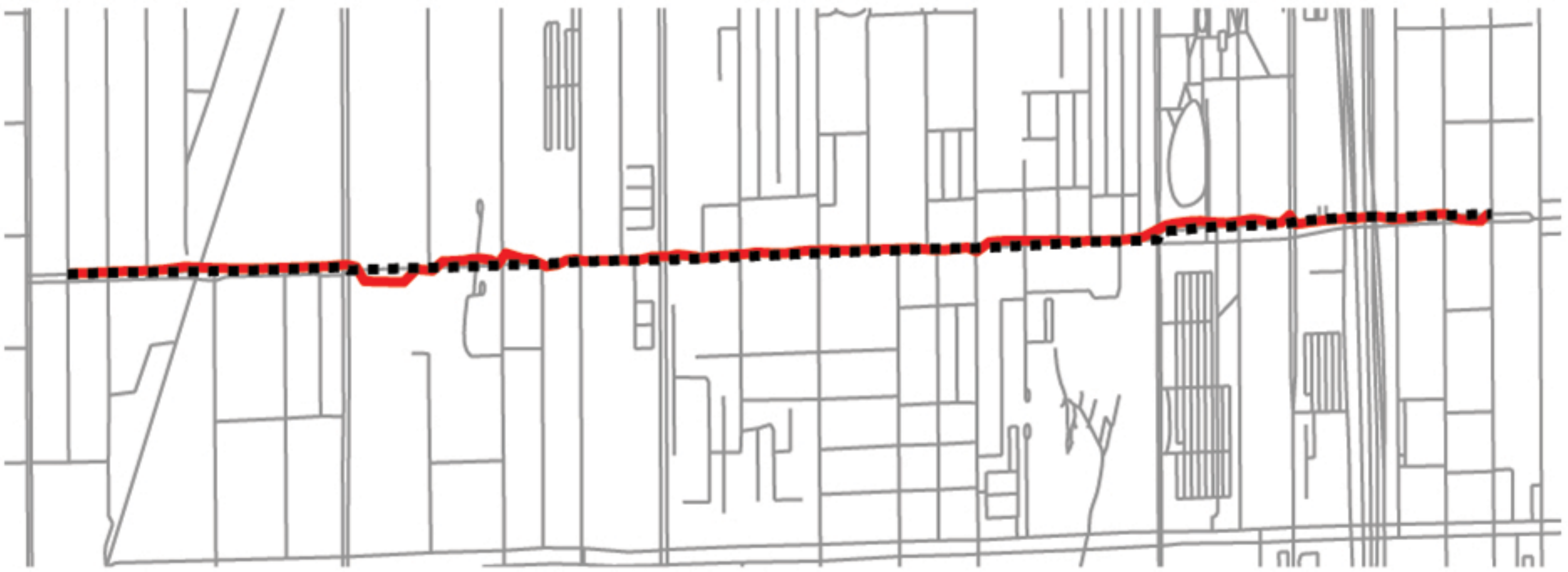}
\label{subfig:cao_rn_sp_all}}
\hspace{0.05cm}
\subfloat[Davies]{\includegraphics[width=0.48\columnwidth]{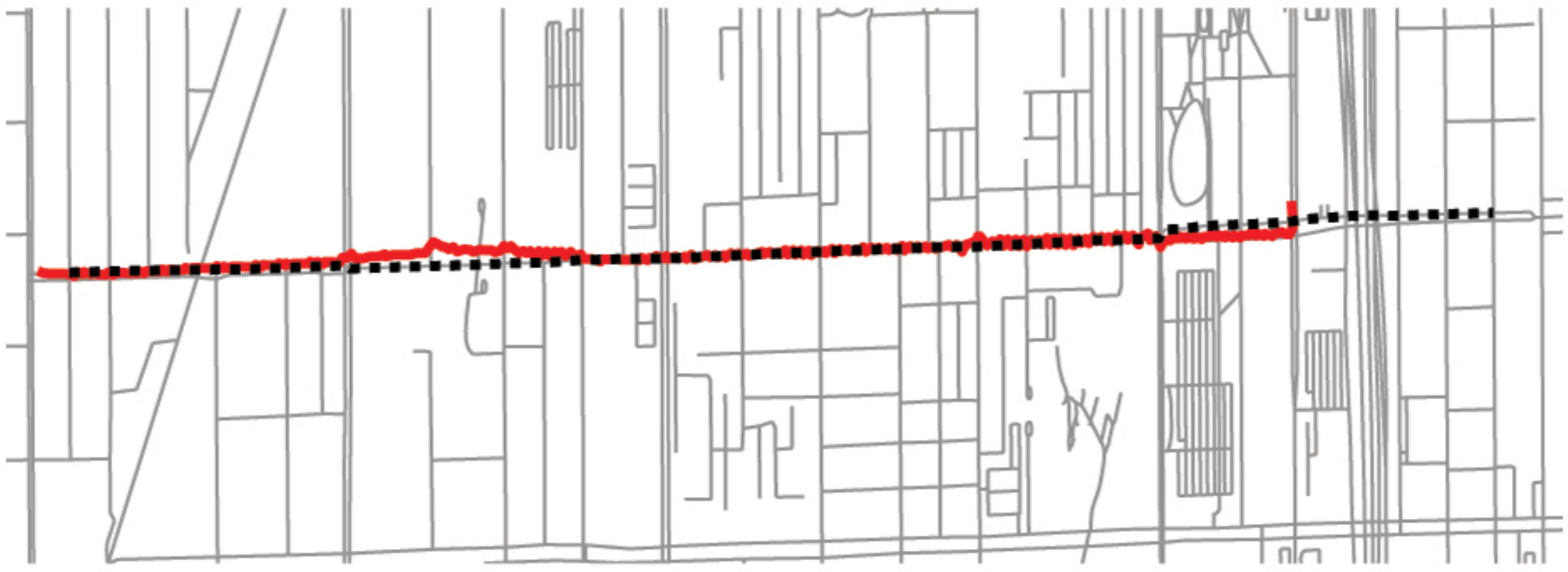}
\label{subfig:dav_rn_sp_all}}
\subfloat[Edelkamp]{\includegraphics[width=0.48\columnwidth]{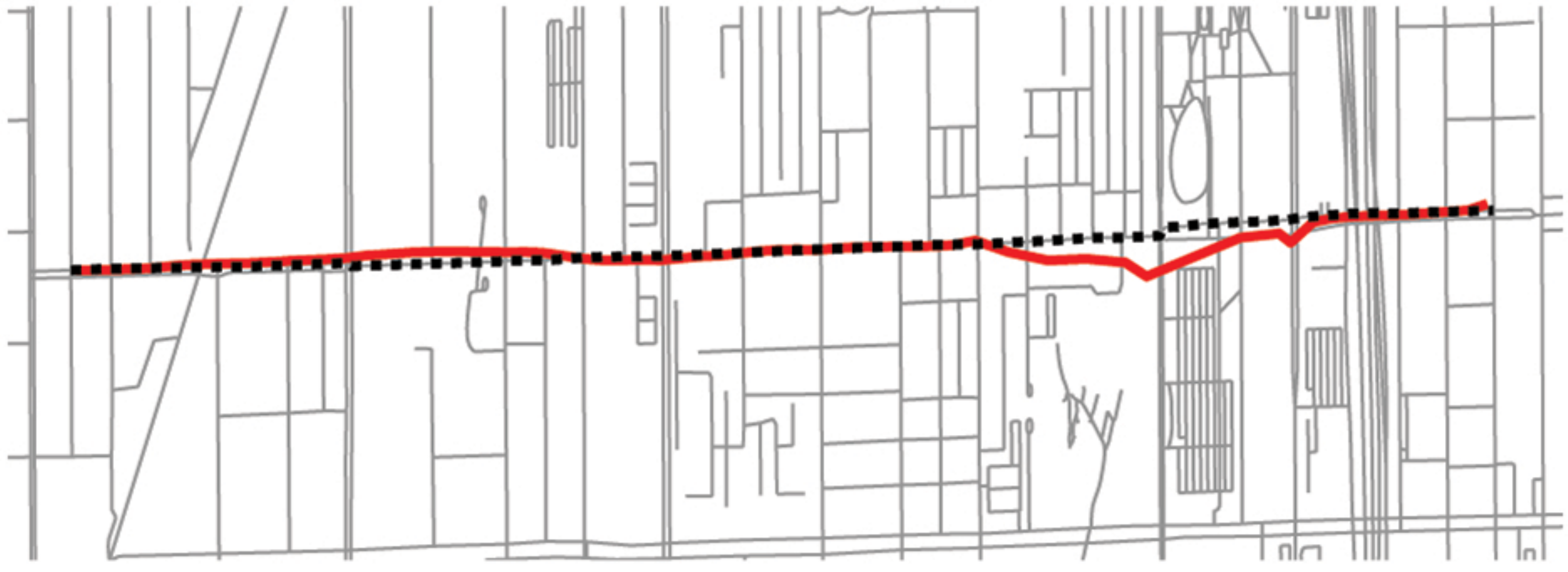}\label{subfig:edel_rn_sp_all}}
\hspace{0.05cm}
\subfloat[Ge]{\includegraphics[width=0.48\columnwidth]{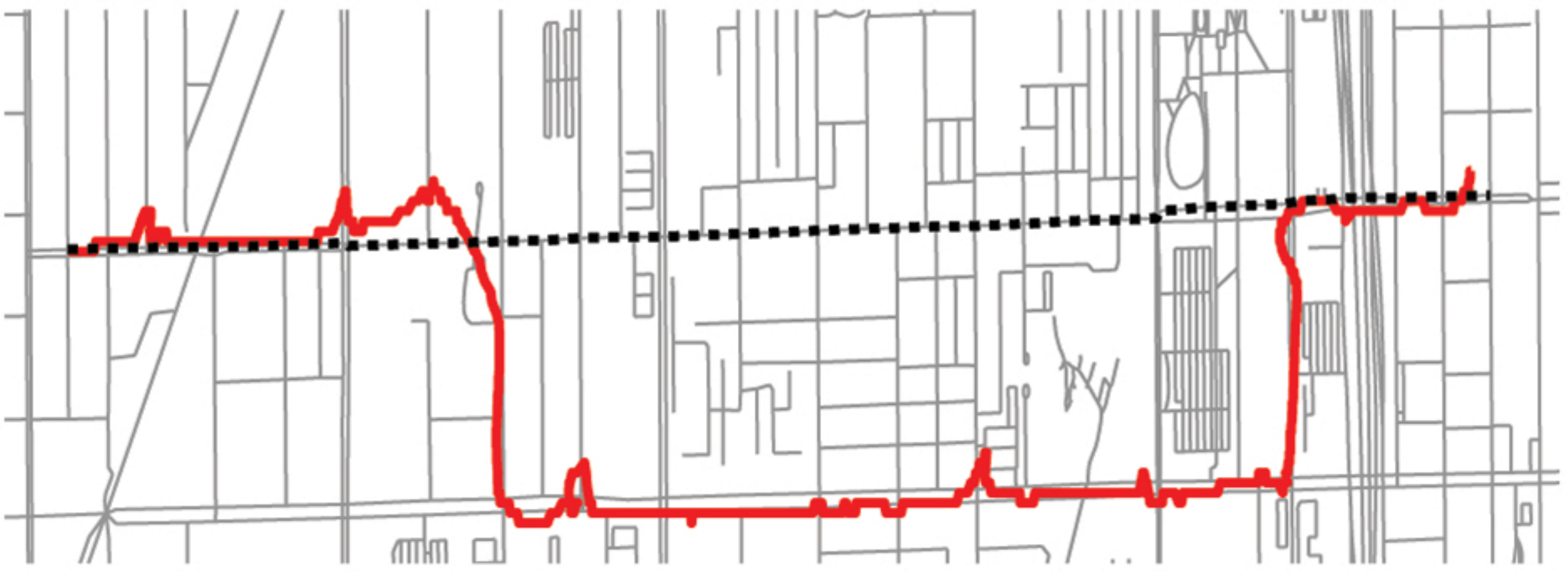}
\label{subfig:ge_rn_sp_all}}
\subfloat[Karagiorgou]{\includegraphics[width=0.48\columnwidth]{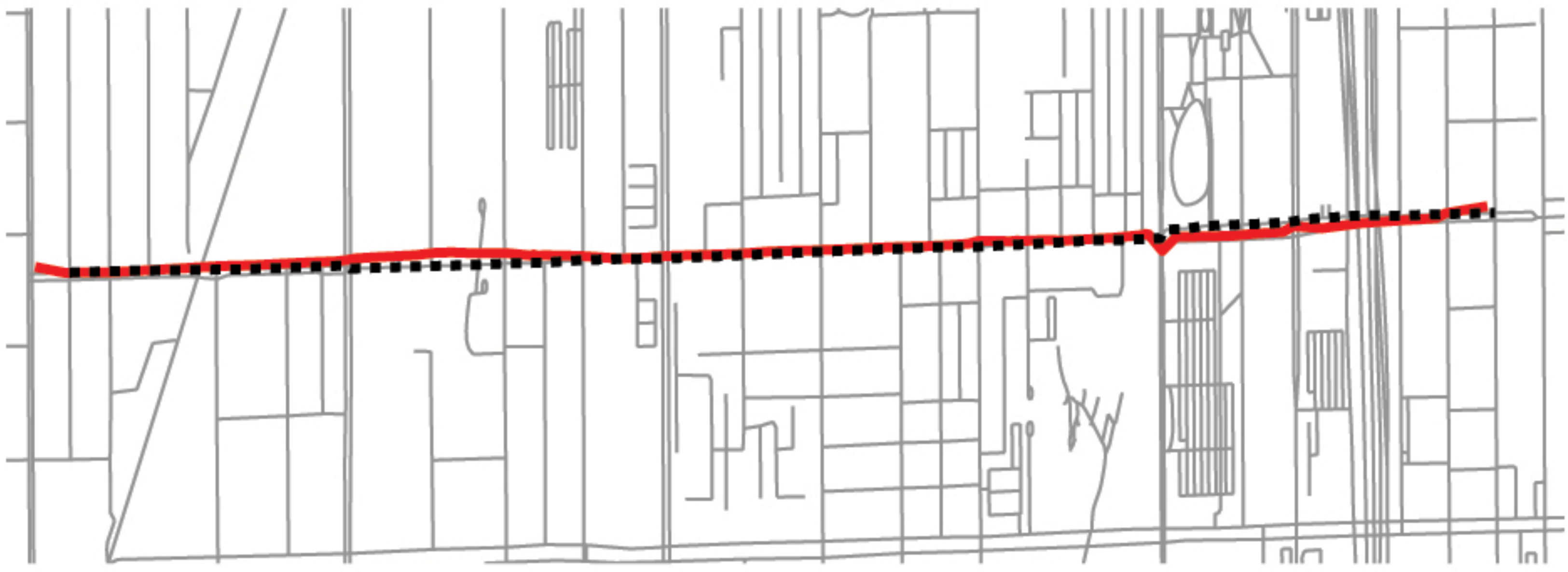}\label{subfig:kar_rn_sp_all}}
\end{center}
\caption{Examples of shortest paths for the \emph{Chicago} dataset.}
\label{fig:rn_sp_all}
\end{figure}

\subsection{Shortest Path Based Measure}

Another means to compare the constructed maps is the shortest path based distance. For each city, we computed a set of 500 random shortest paths with origin and destination nodes uniformly distributed over the maps and compared the paths using the Discrete Fréchet and Average Vertical distance measure.

A first impression on how different constructed maps affect such paths is given in Figure~\ref{fig:rn_sp_all}. Given 
 a specific  origin and destination for the \emph{Chicago} map, the shortest path has length 3.66$km$ in the ground-truth map (black dotted line). 
The computed shortest path for the map generated by each algorithm is shown in red line. In the map generated by Ahmed et al.'s algorithm the shortest path has length 4.67$km$ (a Discrete \Frd\ with respect to the ground-truth map of 65$m$, and an Average Vertical distance of 21$m$). 
The respective results for the other algorithms are Biagioni 3.71$km$ (36$m$, 5$m$), Cao 3.76$km$ (24$m$, 6$m$), Davies 3.39$km$ (35$m$, 4$m$), Edelkamp 3.64$km$, (26$m$, 8$m$), Ge 7.33$km$, (174$m$, 98$m$), and Karagiorgou 3.73$km$ (21$m$, 5$m$).
For most algorithms the resulting paths have small distance to the shortest path in the ground-truth map. However, in the case of Ahmed (Figure~\ref{subfig:ahmed_rn_sp_all}) and Ge (Figure~\ref{subfig:ge_rn_sp_all}), due to significant differences in the generated map, different shortest paths have been computed that have a larger distance to the shortest path in the ground-truth map. This result is in line with the path-based measure of Section~\ref{sub:distance}, where also Biagioni, Davies and Karagiorgou produced the best constructed maps.
 
\begin{figure}[htbp]
 \begin{center}
	 \subfloat[Discrete Fréchet distance - \emph{Athens large}]{\label{subfig:athens_eval_fr} \includegraphics[width=0.7\columnwidth]{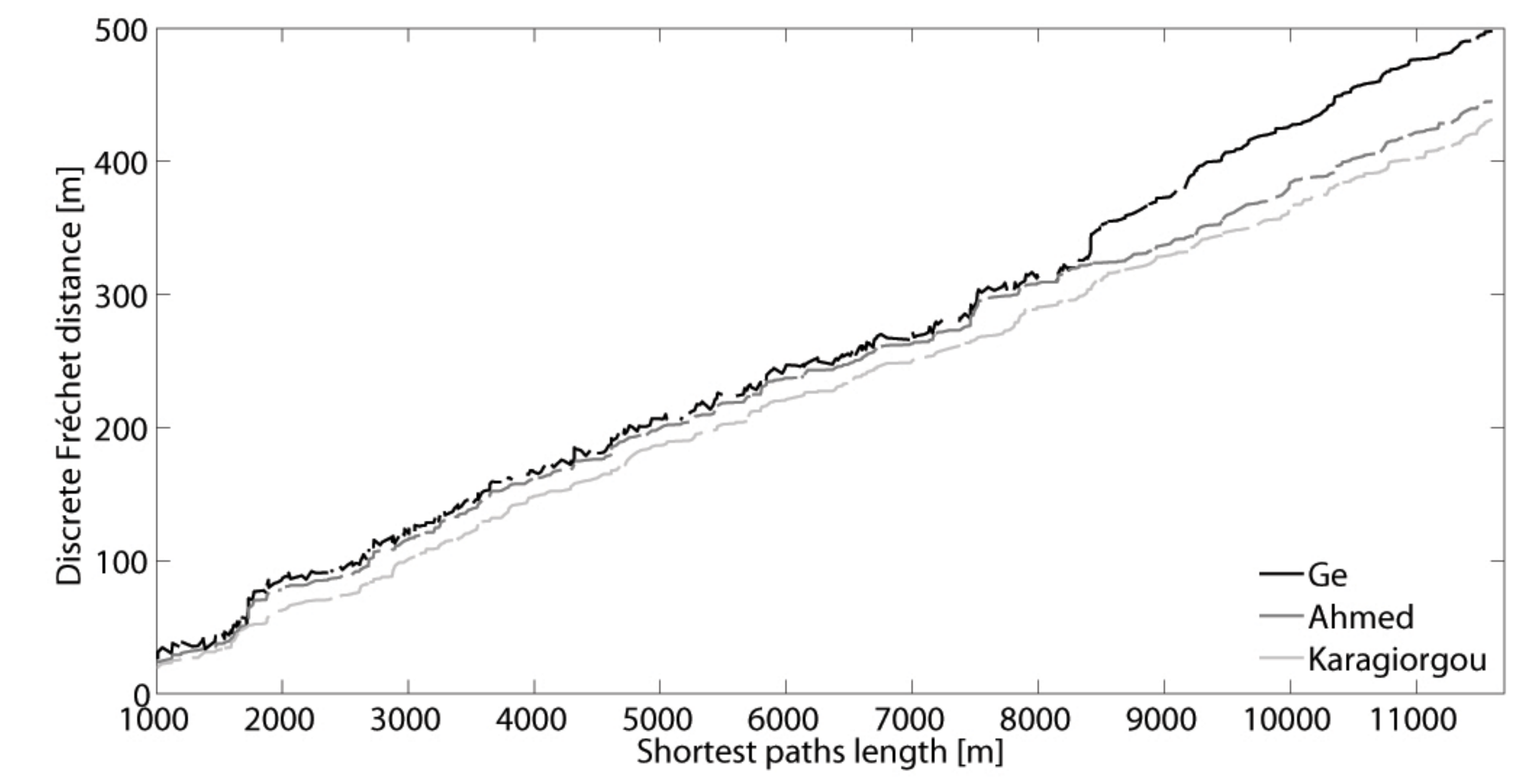}}
\hspace{0.05cm}	 
	 \subfloat[Average Vertical distance - \emph{Athens large}]{\label{subfig:athens_eval_vert} \includegraphics[width=0.7\columnwidth]{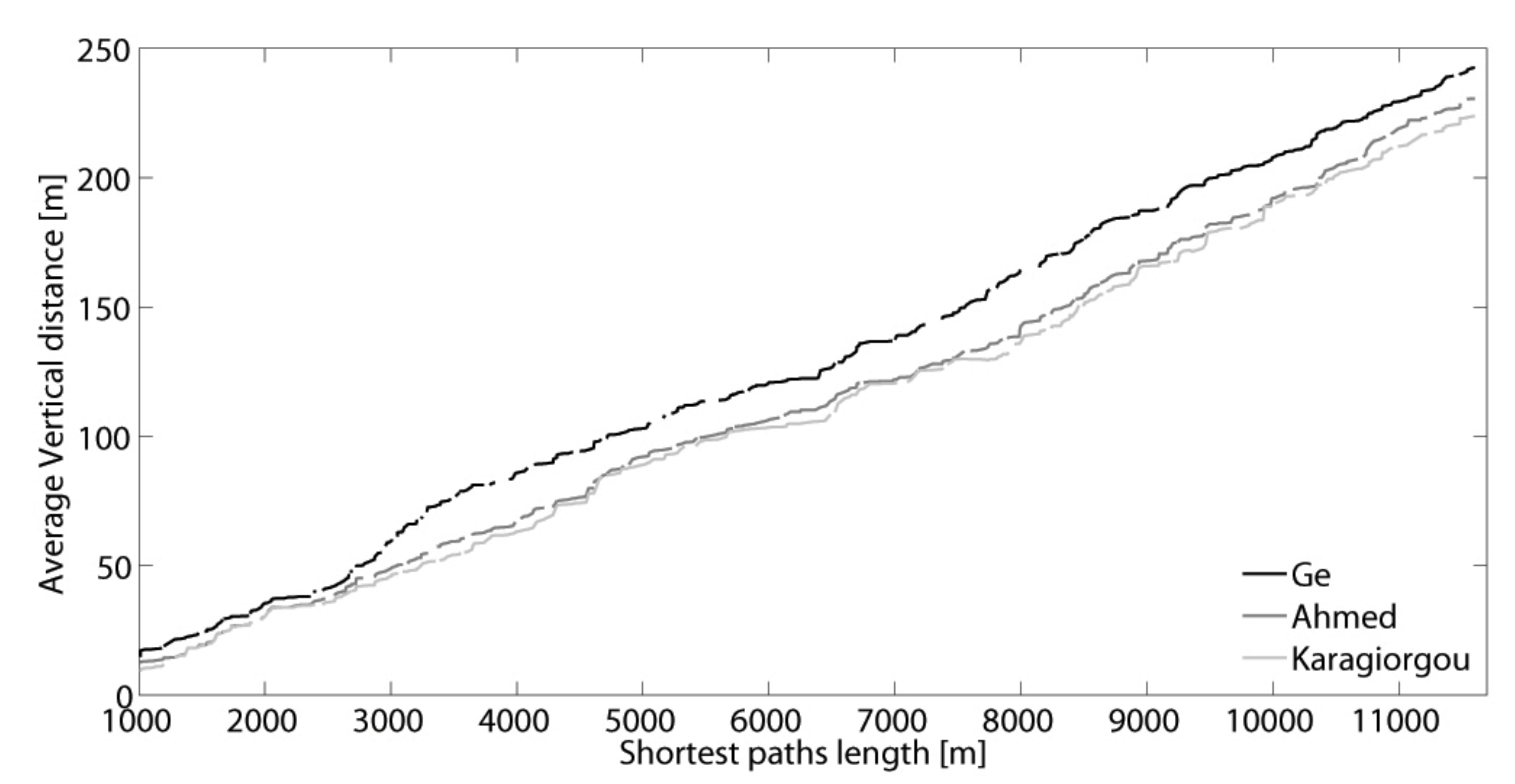}}	
 \end{center}
\caption{Map comparison.}
\end{figure}

Figures~\ref{subfig:athens_eval_fr} and \ref{subfig:athens_eval_vert} show the Discrete Fréchet and the Average Vertical distance measures for each of the 500 paths per algorithm for the \emph{Athens large} map. The paths are ordered by increasing distance of the shortest path length with respect to the ground-truth map. Some paths could not be computed for some maps due to connectivity problems (missing links). Some other paths experience greater distance measures due to spatial accuracy problems. The graph shows that some algorithms produce maps which resemble the actual map more closely, as assessed by this shortest path sampling approach.

Finally, the shortest path based evaluation is summarized in Table~\ref{tab:sp_summary}. The first column shows the percentage ($\%$) of shortest paths that in each case could be computed, i.e., an algorithm might find an accurate, but small map. The second and the third column show the two different distance measures used to compare the resulting paths. The fourth column gives some statistics with respect to the computed shortest paths.
Considering the example of \emph{Berlin} and here the Ahmed algorithm result (shaded light gray) in Table~\ref{tab:sp_summary}, this algorithm produces a map that in turn generates paths that have a min, max, and avg. Discrete \Frd\ of 21$m$, 469$m$, and 192$m$, respectively. 
An aspect not captured by these distances are missing paths due to \emph{limited map coverage}. Consider the case of Davies for \emph{Chicago} and Cao for \emph{Athens small} (shaded light gray in Table~\ref{tab:sp_summary}). In both cases, the distance measures suggest good map quality. However, in both cases the constructed map has a small coverage, as only 92.6$\%$ and 7.0$\%$ of the 500 total paths were computed. In this evaluation, Karagiorgou produces maps that have both good coverage and high path similarity (cf. dark-shaded entry for Berlin - good coverage and small distance measure indicating similar paths between constructed and ground-truth map).

Overall, shortest path sampling provides an effective means for assessing the quality of constructed maps as it not only considers \emph{similarity}, but also the \emph{coverage of the map}.

\begin{table}[htbp]\scriptsize
\centering
\scalebox{0.77}{
\begin{tabular}{|c||c|c c c c|c c c c|c c c c|}
\hline
Generated&&&&&&&&&&&&&\\

Map&\multicolumn{1}{c|}{Found (\%)} & \multicolumn{4}{c|}{Discrete Fréchet dist. (m)}& \multicolumn{4}{c|}{Average Vertical dist. (m)}& \multicolumn{4}{c|}{Shortest path dist. (km)}\\
\hline
\hline
\emph{Athens large}&&min&max&avg&stddev&min&max&avg&stddev&min&max&avg&stddev\\
\hline
Ahmed&92.6&23&445&137&103&12&230&106&62&1.12&11.84&6.93&2.92\\
Ge&92.8&25&497&149&112&14&241&120&65&1.47&11.91&7.13&3.18\\
Karagiorgou&94.2&19&432&125&96&9&225&98&58&1.01&11.62&6.84&2.86\\
\hline
\emph{Athens small}&&min&max&avg&stddev&min&max&avg&stddev&min&max&avg&stddev\\
\hline
Ahmed&97.6&13&234&96&62&6&91&38&24&1.28&5.72&3.11&1.84\\
Biagioni&94.2&7&214&84&50&4&80&28&21&0.79&5.23&2.97&1.41\\
\cellcolor{gray!25}Cao&\cellcolor{gray!25}7.0&\cellcolor{gray!25}7&\cellcolor{gray!25}26&\cellcolor{gray!25}10&\cellcolor{gray!25}11&4&13&6&5&0.17&0.31&0.22&0.21\\
Davies&22.6&9&258&102&69&5&81&31&22&0.85&5.25&2.99&1.47\\
Edelkamp&97.2&15&228&97&64&6&93&40&26&0.93&5.29&3.02&1.51\\
Ge&93.4&21&290&123&75&11&127&63&33&1.43&5.93&3.41&1.92\\
Karagiorgou&96.8&7&212&81&48&3&81&27&20&0.78&5.21&2.95&1.39\\
\hline
\emph{Berlin}&&min&max&avg&stddev&min&max&avg&stddev&min&max&avg&stddev\\
\hline
\cellcolor{gray!25}Ahmed&\cellcolor{gray!25}93.2&\cellcolor{gray!25}21&\cellcolor{gray!25}469&\cellcolor{gray!25}191&\cellcolor{gray!25}123&12&231&121&63&1.56&5.88&3.49&1.96\\
Ge&92.4&25&475&194&128&15&236&127&64&1.85&5.93&3.84&2.03\\
Karagiorgou&93.8&18&428&183&112&8&209&106&58&1.32&5.67&3.27&1.84\\
\hline
\emph{Chicago}&&min&max&avg&stddev&min&max&avg&stddev&min&max&avg&stddev\\
\hline
Ahmed&99.8&13&208&97&56&6&92&43&19&1.21&6.95&4.45&2.04\\
Biagioni&98.6&4&98&40&27&2&49&20&13&0.89&6.03&3.76&1.57\\
Cao&99.2&7&131&67&34&4&76&41&17&1.02&6.87&3.94&1.84\\
\cellcolor{gray!25}Davies&\cellcolor{gray!25}92.6&\cellcolor{gray!25}5&\cellcolor{gray!25}97&\cellcolor{gray!25}41&\cellcolor{gray!25}27&3&51&23&15&0.93&6.08&3.88&1.66\\
Edelkamp&99.0&12&211&98&58&5&89&41&18&1.19&6.88&4.32&1.97\\
Ge&99.8&19&241&127&63&8&94&49&22&1.58&6.98&4.69&2.25\\
\cellcolor{gray!40}Karagiorgou&\cellcolor{gray!40}99.2&\cellcolor{gray!40}4&\cellcolor{gray!40}103&\cellcolor{gray!40}41&\cellcolor{gray!40}28&2&50&21&14&0.90&6.05&3.82&1.59\\
\hline
\end{tabular}
}
\caption{Shortest path measure evaluation summary.} 
\label{tab:sp_summary}
\end{table}

\subsection{Graph-Sampling Based Distance}

For this measure we use the source code obtained from the authors of \cite{be-irmgp-12}. We modified the code to use Euclidean distance as our data uses projected coordinate system. The method that computes this measure has four parameters: 
\begin{inparaenum}
\item{\emph{sampling density}, how densely the map should be sampled (marbles for generated map and holes for ground-truth map), we use $5$ meters.}
\item{\emph{matched distance}, the maximum distance between a matched marble-hole pair, we vary this distance from $10$ to $120m$.}
\item{\emph{maximum distance from root}, the maximum distance from randomly selected start location one will explore, we use $300m$.}
\item{\emph{number of runs}, number of start locations to consider, we use 1000.}
\end{inparaenum}
To make our comparison of all generated maps consistent, we generated a sequence of random locations for each dataset and used the first $1,000$ locations from the same sequence for each algorithm for which both maps (ground-truth and generated) had correspondences within \emph{matched distance}. When two maps are very similar, they should have very few unmatched marbles and holes, which implies the precision, recall and F-score values should be very close to 1. In our case, as we used a superset of the ground-truth map, there should be a large number of unmatched holes, which implies lower recall and F-score values than in \cite{be-irmgp-12}, but still the relative comparison of F-score values should provide an idea of whether an algorithm performs better than another.

\begin{figure}[htbp]
\begin{center}	
\includegraphics[width=0.7\columnwidth, clip=true, trim = 15mm 65mm 2mm 60mm ]{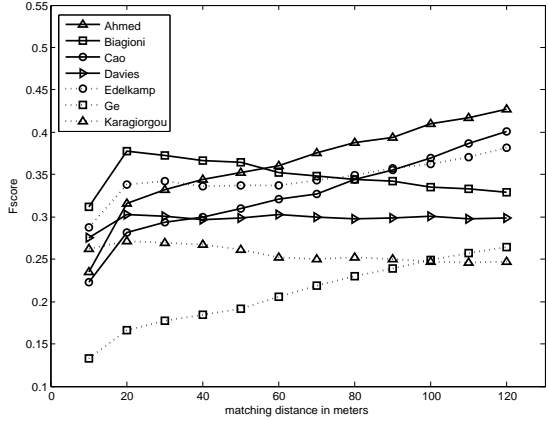}
\end{center}
\vspace{-2pt} 
 \caption{Comparison of F-scores - \emph{Chicago}.}
 \label{fig:biagioniMeasure}
\vspace{-2pt}
\end{figure}


\begin{table}[htbp]\scriptsize
\centering
\begin{tabular}{|c||c c c c|}
\hline
Generated& \multicolumn{4}{c|}{Precision Value}\\
Map&  \multicolumn{4}{c|}{(for {\em matched distance} 10, 40, 70, 100)}\\
\hline
\hline
\multicolumn{1}{|c||}{\emph{Athens large}}&10&40&70&100\\
\hline
Ahmed&0.216&0.407&0.497&0.591\\
Ge&0.149&0.368&0.507&0.635\\
Karagiorgou&0.394&0.559&0.630&0.711\\
\hline
\multicolumn{1}{|c||}{\emph{Athens small}}&10&40&70&100\\
\hline
Ahmed&0.265&0.442&0.503&0.579\\
Biagioni&0.450&0.586&0.662&0.727\\
Cao&0.415&.691&0.722&0.810\\
Davies&0.439&0.574&0.617&0.670\\
Edelkamp&0.106&0.156&0.197&0.232\\
Ge&0.409&0.527&0.624&0.708\\
Karagiorgou&0.343&0.489&0.561&0.647\\
\hline
\multicolumn{1}{|c||}{\emph{Berlin}}&10&40&70&100\\
\hline
Ahmed&0.123&0.326&0.422&0.485\\
Ge&0.142&0.457&0.534&0.584\\
Karagiorgou&0.294&0.590&0.633&0.649\\
\hline
\multicolumn{1}{|c||}{\emph{Chicago}}&10&40&70&100\\
\hline
Ahmed&0.312&0.563&0.658&0.738\\
\cellcolor{gray!40}Biagioni&0.491&0.699&0.730&0.775\\
\cellcolor{gray!25}Cao&0.209&0.321&0.376&0.456\\
\cellcolor{gray!40}Davies&0.488&0.650&0.690&0.739\\
Edelkamp&0.334&0.431&0.473&0.541\\
Ge&0.306&0.487&0.565&0.645\\
\cellcolor{gray!40}Karagiorgou&0.602&0.740&0.751&0.801\\
\hline
\end{tabular}
\caption{Precisions for varying {\em matched distance}.}
\label{tab:biagioniSummary}
\end{table}

Figure \ref{fig:biagioniMeasure} shows F-score values for the \emph{Chicago} dataset for different generated maps. As our ground-truth is essentially a superset of the actual ground-truth represented by the tracking dataset, a larger matching distance creates unexpected results for algorithms that generate extra edges and vertices. For example, Cao and Edelkamp for \emph{Chicago}, the precision is low as there will be lots of unmatched marbles (cf. entry for Cao and Edelkamp for \emph{Chicago} in Table~\ref{tab:biagioniSummary}). However, a larger matching distance decreases the number of unmatched marbles by matching these with available holes that probably are not part of the actual ground-truth. A higher recall value yields a higher F-score, which does not necessarily reflect better-quality maps (cf. Figure~\ref{fig:rn_alls} and Figure~\ref{fig:rn_allb}).

In Figure~\ref{fig:biagioniMeasure} we also see the performance based on F-score declines for Biagioni, Davies and Karagiorgou as the matching distance threshold increases. After investigating the reason of this unexpected behaviour we found, although precision increases with matching distances the recall declines for these three algorithms; and smaller recall indicates larger number of unmatched sample points on ground-truth (empty holes). Figure~\ref{fig:rn_alls} and Table~\ref{tab:tab_stat} show these three algorithms reconstruct less streets than others, which means they produce smaller number of marbles to match with larger number of holes.
 
Hence, in Table \ref{tab:biagioniSummary} we are ignoring F-score and recall values and \emph{showcase only precision values}. According to precision values, the algorithms by Biagioni, Davies and Karagiorgou perform best for dataset \emph{Chicago}, which is consistent with our findings using the other three distance measures.

\subsection{Summary} 
\label{sub:summary}

The best way to characterize the constructed maps is in terms of coverage and accuracy. 
Here, it appears that KDE-based point clustering algorithms such as Biagioni and Davies produce maps with lower complexity (fewer number of vertices and edges) and often fail to reconstruct streets that are not traversed frequently enough by the input tracks. 
On the other hand, the algorithm by Ge subsample all tracks to create a much denser output set, hence the complexity of their constructed maps is always higher. 
A similar observation can be made for algorithms based on incremental track insertion, such as the Ahmed and Cao algorithms. They fail to cluster tracks together when the variability and error associated with the input tracks is large. As a result, the constructed street maps contain multiple edges for a single street, which implies a larger constructed, but not necessarily more accurate road network.

In terms of map quality and accuracy, the maps reconstructed using the algorithms by Karagiorgou, Davis, and  Biagioni generally have smallest path-based and Directed Hausdorff distances and their constructed maps can be considered more accurate. 
Although the algorithms by Ahmed and Ge produce maps with good coverage and provide quality guarantees, their path-based distances are larger, since they employ less aggressive averaging techniques that would help cope with noise in the input tracks.
%
In an effort to assess both, accuracy and coverage, the shortest path based measure shows for the cases of Davies and \emph{Chicago} and Cao and \emph{Athens small} good map quality, but at the same only limited coverage. In this evaluation, Karagiorgou produces maps that have both good coverage and high path similarity.

An overall observation to be made based on our experimentation is that map construction algorithms tend to produce either accurate maps, or maps with good coverage, but not both. The algorithm of Karagiorgou however seems to be a good compromise, in that it produces maps of good coverage and accuracy at the same time.


\section{Conclusions}
\label{sec:conclusions}
This survey has considered the active field of road network construction and has considered a variety of such construction algorithms.
In the past, the lack of benchmark data and quantitative evaluation methods has hindered a cross-comparison between algorithms. 
In this paper, the contribution of benchmark data sets and code for road network construction algorithms and evaluation measures for the first time enables a standardized assessment and comparison of road network construction algorithms. All data, road network construction, and evaluation algorithms are available with detailed execution instructions on the {\tt mapconstruction.org} web site. Directions for future work include the expansion of the web site towards the inclusion of more algorithms and source code. The final goal will be to provide an easy-to-use benchmark suite and automated quality measurements for generated maps.
 
\begin{acknowledgements}
This work has been supported by the National Science Foundation grant CCF-1301911 and the European Union Seventh Framework Programme - Marie Curie Actions, Initial Training Network GEOCROWD (www.geocrowd.eu) under grant agreement No. FP7-PEOPLE-2010-ITN-264994.

We thank James Biagioni for making the source code for the graph sampling-based distance measure \cite{be-irmgp-12} available to us, for implementing the map construction algorithms by \cite{Cao:2009:GTR:1653771.1653776,Davies:2006:SDR:1175887.1176088, edelkamp:2003:rpmi, Biagioni:2012:MIF:2424321.2424333} and for making them publicly available.
We thank Xiaoyin Ge and Yusu Wang for running their map construction algorithm \cite{DBLP:conf/nips/GeSBW11} on our benchmark datasets.
Associated data and software will be made available at {\tt mapconstruction.org}.
\end{acknowledgements}

\bibliographystyle{spmpsci}      
\bibliography{references} 


\end{document}